\documentclass{aa}

\usepackage{graphics}
\usepackage{lscape}

\begin{document}

\thesaurus{04           
             (11.06.2;  
              11.11.1;  
              11.19.2;  
              11.19.6)} 

\title{The Ursa Major Cluster of Galaxies. IV ; HI synthesis observations}
\subtitle{}

\author{Marc A.W. Verheijen \inst{1,2} 
   \and Renzo Sancisi \inst{2,3}}

\offprints{M.A.W. Verheijen}

\institute{NRAO, PO Box O, Socorro, NM 87801, USA 
      \and Kapteyn Institute, Postbus 800, 9700 AV, Groningen, The Netherlands
      \and Osservatorio Astronomico di Bologna, Via Ranzani 1, I-40127 Bologna, Italy }

\date{Received: 20 November 2000 / Accepted: 4 December 2000 }

\maketitle

\begin{abstract}
 In this data paper we present the results of an extensive 21cm-line
synthesis imaging survey of 43 spiral galaxies in the nearby Ursa Major
cluster using the Westerbork Synthesis Radio Telescope. Detailed
kinematic information in the form of position-velocity diagrams and
rotation curves is presented in an atlas together with HI channel maps,
21cm continuum maps, global HI profiles, radial HI surface density
profiles, integrated HI column density maps, and HI velocity fields. The
relation between the corrected global HI linewidth and the rotational
velocities V$_{\mbox{\scriptsize max}}$ and V$_{\mbox{\scriptsize
flat}}$ as derived from the rotation curves is investigated. Inclination
angles obtained from the optical axis ratios are compared to those
derived from the inclined HI disks and the HI velocity fields. The
galaxies were not selected on the basis of their HI content but solely
on the basis of their cluster membership and inclination which should be
suitable for a kinematic analysis. The observed galaxies provide a
well-defined, volume limited and equidistant sample, useful to
investigate in detail the statistical properties of the Tully-Fisher
relation and the dark matter halos around them.
 \end{abstract}

\section{Introduction}

The statistical properties of the Tully-Fisher relation (TFr), like its
scatter and slope, and in more detail the characteristics of dark matter
halos around galaxies, like their core densities and radii, are of great
interest to those who study galaxy formation scenarios.  The relevant
observables for the TFr are total luminosities and rotational velocities
while actual rotation curves and luminosity profiles are required to
obtain constraints on the density profiles of dark matter halos.
However, the available data sets from which these observables can be
obtained are often suffering from incompleteness, distance uncertainties
and inhomogeneous observing and data analysis techniques. 

Furthermore, the interpretation of the observables is not always
unambiguous. For instance, the rotational velocity of a galaxy is
generally inferred from the width of its global HI profile and often
this measured width is being related directly to the dark matter
potential (e.g Navarro \& Steinmetz \cite{navarro}). However, the width
of the HI profile is a complicated convolution of the 2-dimensional
distribution of HI in a galaxy disk and the shape and extent of its
rotation curve as sampled by the HI gas. Needless to say one has to
exercise caution when relating the observed scatter and slope in the TFr
to the outcome of numerical and semi-analytical simulations of galaxy
formation. This is especially the case when the observed TFr was
constructed to serve as an empirical distance estimator and the
selection criteria and applied corrections to the raw observables were
optimized to linearize the relation and minimize its scatter.

To overcome some of the observational issues, we initiated a program to
obtain detailed multi-band photometric and kinematic information on
individual galaxies in a well-defined, complete sample. The nearby Ursa
Major cluster of galaxies provides a particularly suitable sample. In
the first place, the galaxies in the cluster are all at roughly the same
distance. Therefore, there is little doubt about their relative
luminosities, sizes and masses. Furthermore, the Ursa Major cluster
contains overwhelmingly gas-rich systems and the morphological mix of
its galaxy members is close to that of the lower density field. A
detailed discussion on the definition of the Ursa Major cluster is given
by Tully et al. (\cite{tully96}) (Paper I). The characteristics of this cluster
will be discussed in more detail in Section 2.

In this paper we present the results of an extensive 21cm-line synthesis
imaging survey of individual galaxies in the Ursa Major cluster using
the Westerbork Synthesis Radio Telescope (WSRT). The HI data are
presented in the form of an atlas. Multi-band optical and near-infrared
imaging photometry is presented in Paper I together with B-band and
K$^\prime$-band images of all identified cluster members. Forthcoming
papers in this series will use these data to investigate the TFr while
the HI rotation curves will be supplemented in the inner regions with
already obtained high resolution optical rotation curves to derive
constraints on the structural properties of the dark matter halos and
the mass-to-light ratios of the stellar populations.

This data paper is organized as follows. Section 2 describes the Ursa
Major cluster in more detail and contains morphological and photometric
information on the galaxies observed with the WSRT. Data acquisition and
reduction procedures are explained in Section 3. Section 4 explains how
the corrected linewidth can be matched to V$_{\mbox{\scriptsize max}}$
and V$_{\mbox{\scriptsize flat}}$ from the rotation curves through an
appropriate correction for turbulent motion. Inclinations are derived
from the optical images, HI column density maps and HI velocity fields
and Section 5 presents a comparison of these inclinations. The layout of
the HI atlas is described in Section 6 with details about the various
elements of the atlas pages. The HI properties of the Ursa Major cluster
galaxies as a sample are presented in Section 7. Some concluding remarks
are given in Section 8.

\begin{table*}[p]
\begin{center}

\caption[]{All galaxies in the Ursa Major cluster brighter than
M$^{b,i}$(B)=$-$16.8 and more inclined than 45 degrees.}

{\fontsize{9}{10}
\selectfont

\begin{tabular}[t]{lcccclcrcccccc}

\hline
\noalign{\vspace{0.8mm}}
Name     &    R.A.    &   Dec.   &\multicolumn{2}{c}{Galactic}& Type & D$_{25}$(B)& PA & 1$-$b/a &i$_{\mbox{\scriptsize opt}}$ &i$_{\mbox{\scriptsize adopt}}$ & S.B. & \multicolumn{2}{c}{{\scriptsize [BH]} A$^{\mbox{\scriptsize b}}_{\mbox{\scriptsize B}}$ {\scriptsize [SFD]}} \\
\noalign{\vspace{0.8mm}}
         &\multicolumn{2}{c}{(1950)}& Long. & Lat. & &($^\prime$) &($^\circ$)& &($^\circ$)&($^\circ$)& & mag & mag \\
\noalign{\vspace{0.8mm}}
  (1)    &     (2)    &   (3)    &   (4)  &  (5)  & (6) & (7) & (8)  &  (9)  &(10)& (11)& (12) & (13) & (14)\\
\noalign{\vspace{0.8mm}}
\hline
\hline
\noalign{\vspace{0.8mm}}

\multicolumn{13}{l}{\hspace{-2mm}{\em \underline{Galaxies with fully analyzed HI data:}}} \\
\noalign{\vspace{0.8mm}}
U6399    & 11 20 35.9 & 51 10 09 & 152.08 & 60.96 & Sm  & 2.40 & 140 & 0.72 & 79 & 75$\pm$2 & LSB & 0.00 & 0.07 \\
U6446    & 11 23 52.9 & 54 01 21 & 147.56 & 59.14 & Sd  & 2.27 & 200 & 0.38 & 54 & 51$\pm$3 & LSB & 0.00 & 0.07 \\
N3726    & 11 30 38.7 & 47 18 20 & 155.38 & 64.88 & SBc & 5.83 & 194 & 0.38 & 54 & 53$\pm$2 & HSB & 0.01 & 0.07 \\
N3769    & 11 35 02.8 & 48 10 10 & 152.72 & 64.75 & SBb & 2.97 & 150 & 0.69 & 76 & 70$\pm$2 & HSB & 0.01 & 0.10 \\
U6667    & 11 39 45.3 & 51 52 32 & 146.27 & 62.29 & Scd & 3.43 &  88 & 0.88 & 90 & 89$\pm$1 & LSB & 0.00 & 0.07 \\
N3877    & 11 43 29.3 & 47 46 21 & 150.72 & 65.96 & Sc  & 5.40 &  36 & 0.78 & 84 & 76$\pm$1 & HSB & 0.01 & 0.10 \\
N3893    & 11 46 00.2 & 48 59 20 & 148.15 & 65.23 & Sc  & 3.93 & 352 & 0.33 & 49 & 49$\pm$2 & HSB & 0.02 & 0.09 \\
N3917    & 11 48 07.7 & 52 06 09 & 143.65 & 62.79 & Scd & 4.67 & 257 & 0.76 & 82 & 79$\pm$2 & LSB & 0.01 & 0.09 \\
N3949    & 11 51 05.5 & 48 08 14 & 147.63 & 66.40 & Sbc & 2.90 & 297 & 0.38 & 54 & 55$\pm$2 & HSB & 0.03 & 0.09 \\
N3953    & 11 51 12.4 & 52 36 18 & 142.21 & 62.59 & SBbc& 6.10 &  13 & 0.50 & 62 & 62$\pm$1 & HSB & 0.01 & 0.13 \\
N3972    & 11 53 09.0 & 55 35 56 & 138.85 & 60.06 & Sbc & 3.43 & 298 & 0.72 & 79 & 77$\pm$1 & HSB & 0.00 & 0.06 \\
U6917    & 11 53 53.1 & 50 42 27 & 143.46 & 64.45 & SBd & 3.17 & 123 & 0.46 & 59 & 56$\pm$2 & LSB & 0.03 & 0.12 \\
U6923    & 11 54 14.4 & 53 26 19 & 140.51 & 62.06 & Sdm & 1.97 & 354 & 0.58 & 68 & 65$\pm$2 & LSB & 0.00 & 0.12 \\
U6930$^i$& 11 54 42.3 & 49 33 41 & 144.54 & 65.51 & SBd & 3.00 &  39 & 0.14 & 32 & 31$\pm$3 & LSB & 0.05 & 0.13 \\
N3992    & 11 55 00.9 & 53 39 11 & 140.09 & 61.92 & SBbc& 6.93 & 248 & 0.44 & 58 & 56$\pm$2 & HSB & 0.01 & 0.13 \\
U6940$^f$& 11 55 12.4 & 53 30 46 & 140.17 & 62.06 & Scd & 0.83 & 135 & 0.72 & 79 & 75$\pm$3 & LSB & 0.00 & 0.12 \\
U6962$^i$& 11 55 59.5 & 43 00 44 & 154.08 & 71.05 & SBcd& 2.33 & 179 & 0.20 & 38 & 37$\pm$3 & HSB & 0.00 & 0.09 \\
N4010    & 11 56 02.0 & 47 32 16 & 146.68 & 67.36 & SBd & 4.63 &  65 & 0.88 & 90 & 89$\pm$1 & LSB & 0.00 & 0.11 \\
U6969    & 11 56 12.9 & 53 42 11 & 139.70 & 61.96 & Sm  & 1.50 & 330 & 0.69 & 76 & 76$\pm$2 & LSB & 0.01 & 0.13 \\
U6973    & 11 56 17.8 & 43 00 03 & 153.97 & 71.10 & Sab & 2.67 &  40 & 0.61 & 70 & 71$\pm$3 & HSB & 0.00 & 0.09 \\
U6983    & 11 56 34.9 & 52 59 08 & 140.27 & 62.62 & SBcd& 3.20 & 270 & 0.34 & 50 & 49$\pm$1 & LSB & 0.01 & 0.12 \\
N4051    & 12 00 36.4 & 44 48 36 & 148.88 & 70.08 & SBbc& 5.90 & 311 & 0.34 & 50 & 49$\pm$3 & HSB & 0.00 & 0.06 \\
N4085    & 12 02 50.4 & 50 37 54 & 140.59 & 65.17 & Sc  & 2.80 & 255 & 0.76 & 82 & 82$\pm$2 & HSB & 0.01 & 0.08 \\
N4088    & 12 03 02.0 & 50 49 03 & 140.33 & 65.01 & Sbc & 5.37 & 231 & 0.63 & 71 & 69$\pm$2 & HSB & 0.01 & 0.09 \\
N4100    & 12 03 36.4 & 49 51 41 & 141.11 & 65.92 & Sbc & 5.23 & 344 & 0.71 & 77 & 73$\pm$2 & HSB & 0.03 & 0.10 \\
N4102    & 12 03 51.3 & 52 59 22 & 138.08 & 63.07 & SBab& 3.00 &  38 & 0.44 & 58 & 56$\pm$2 & HSB & 0.01 & 0.09 \\
N4157    & 12 08 34.2 & 50 45 47 & 138.47 & 65.41 & Sb  & 6.73 &  63 & 0.83 & 90 & 82$\pm$3 & HSB & 0.02 & 0.09 \\
N4183    & 12 10 46.5 & 43 58 33 & 145.39 & 71.73 & Scd & 4.77 & 346 & 0.86 & 90 & 82$\pm$2 & LSB & 0.00 & 0.06 \\
N4217    & 12 13 21.6 & 47 22 11 & 139.90 & 68.85 & Sb  & 5.67 & 230 & 0.74 & 80 & 86$\pm$2 & HSB & 0.00 & 0.08 \\
N4389    & 12 23 08.8 & 45 57 41 & 136.73 & 70.74 & SBbc& 2.50 & 276 & 0.34 & 50 & 50$\pm$4 & HSB & 0.00 & 0.06 \\

\noalign{\vspace{0.8mm}}
\multicolumn{13}{l}{\hspace{-2mm}{\em \underline{Galaxies with partially analyzed HI data:}}} \\
\noalign{\vspace{0.8mm}}
N3718    & 11 29 49.9 & 53 20 39 & 147.01 & 60.22 & Sa  & 7.53 & 195 & 0.58 & 68 & 69$\pm$3 & HSB & 0.00 & 0.06 \\
N3729    & 11 31 04.9 & 53 24 08 & 146.64 & 60.28 & SBab& 2.80 & 164 & 0.32 & 48 & 49$\pm$3 & HSB & 0.00 & 0.05 \\
U6773    & 11 45 22.1 & 50 05 12 & 146.89 & 64.27 & Sm  & 1.53 & 341 & 0.47 & 60 & 58$\pm$3 & LSB & 0.00 & 0.07 \\
U6818    & 11 48 10.1 & 46 05 09 & 151.76 & 67.78 & Sd  & 2.20 &  77 & 0.72 & 79 & 75$\pm$3 & LSB & 0.00 & 0.09 \\
U6894    & 11 52 47.3 & 54 56 08 & 139.52 & 60.63 & Scd & 1.67 & 269 & 0.84 & 90 & 83$\pm$3 & LSB & 0.00 & 0.06 \\
N3985    & 11 54 06.4 & 48 36 48 & 145.94 & 66.27 & Sm  & 1.40 &  70 & 0.37 & 53 & 51$\pm$3 & HSB & 0.05 & 0.11 \\
N4013    & 11 55 56.8 & 44 13 31 & 151.86 & 70.09 & Sb  & 4.87 & 245 & 0.76 & 88 & 90$\pm$1 & HSB & 0.00 & 0.07 \\
U7089    & 12 03 25.4 & 43 25 18 & 149.90 & 71.52 & Sdm & 3.50 & 215 & 0.81 & 90 & 80$\pm$3 & LSB & 0.00 & 0.07 \\
U7094    & 12 03 38.5 & 43 14 05 & 150.14 & 71.70 & Sdm & 1.60 &  39 & 0.64 & 72 & 70$\pm$3 & LSB & 0.00 & 0.06 \\
N4117    & 12 05 14.2 & 43 24 17 & 149.07 & 71.72 & S0  & 1.53 &  21 & 0.56 & 67 & 68$\pm$3 & LSB & 0.00 & 0.06 \\
N4138    & 12 06 58.6 & 43 57 49 & 147.29 & 71.40 & Sa  & 2.43 & 151 & 0.37 & 53 & 53$\pm$3 & HSB & 0.00 & 0.06 \\
N4218    & 12 13 17.4 & 48 24 36 & 138.88 & 67.88 & Sm  & 1.17 & 316 & 0.40 & 55 & 53$\pm$3 & HSB & 0.00 & 0.07 \\
N4220    & 12 13 42.8 & 48 09 41 & 138.94 & 68.13 & Sa  & 3.63 & 140 & 0.69 & 76 & 78$\pm$3 & HSB & 0.00 & 0.08 \\
 
\noalign{\vspace{0.8mm}}
\multicolumn{13}{l}{\hspace{-2mm}{\em \underline{Galaxies with confused HI data:}}} \\
\noalign{\vspace{0.8mm}}
1135+48  & 11 35 09.2 & 48 09 31 & 152.71 & 64.77 & Sm  & 1.23 & 114 & 0.69 & 76 & 73$\pm$3 & LSB & 0.01 & 0.10 \\
N3896    & 11 46 18.6 & 48 57 10 & 148.10 & 65.29 & Sm  & 1.60 & 308 & 0.33 & 49 & 48$\pm$3 & LSB & 0.02 & 0.09 \\

\noalign{\vspace{0.8mm}}
\multicolumn{13}{l}{\hspace{-2mm}{\em \underline{Not observed or too little HI content:}}} \\
\noalign{\vspace{0.8mm}}
N3870    & 11 43 17.5 & 50 28 40 & 147.02 & 63.75 & S0a & 1.13 &  17 & 0.31 & 47 & 48$\pm$3 & HSB & 0.00 & 0.07 \\
N3990    & 11 55 00.3 & 55 44 13 & 138.25 & 60.04 & S0  & 1.47 &  40 & 0.50 & 62 & 63$\pm$3 & HSB & 0.00 & 0.07 \\
N4026    & 11 56 50.7 & 51 14 24 & 141.94 & 64.20 & S0  & 4.37 & 177 & 0.74 & 80 & 84$\pm$3 & HSB & 0.04 & 0.10 \\
N4111    & 12 04 31.0 & 43 20 40 & 149.53 & 71.69 & S0  & 4.47 & 150 & 0.78 & 84 & 90$\pm$3 & HSB & 0.00 & 0.06 \\
U7129    & 12 06 23.6 & 42 01 08 & 151.00 & 72.99 & Sa  & 1.27 &  72 & 0.31 & 47 & 48$\pm$3 & HSB & 0.00 & 0.06 \\
N4143    & 12 07 04.6 & 42 48 44 & 149.18 & 72.40 & S0  & 2.60 & 143 & 0.46 & 59 & 60$\pm$3 & HSB & 0.00 & 0.06 \\
N4346    & 12 21 01.2 & 47 16 15 & 136.57 & 69.39 & S0  & 3.47 &  98 & 0.67 & 75 & 77$\pm$3 & HSB & 0.00 & 0.06 \\
\noalign{\vspace{0.8mm}}
\hline
\hline

\end{tabular}
}
\end{center}
\end{table*}

\begin{table*}[p]
\begin{center}

\caption[]{Photometrics of all galaxies in the UMa cluster
brighter than M$^{b,i}$(B)=$-$16.8 and more inclined than 45 degrees.}

{\fontsize{9}{10}
\selectfont

\begin{tabular}[t]{lrrrrrccccrrrrr}

\hline
\noalign{\vspace{0.8mm}}
Name &
 m$^{tot}_{\mbox{\scriptsize B}}$ & 
 m$^{tot}_{\mbox{\scriptsize R}}$ & 
 m$^{tot}_{\mbox{\scriptsize I}}$ & 
 m$^{tot}_{\mbox{\scriptsize K$^\prime$}}$ &
 W$_{R,I}^i$ &
 A$^i_{\mbox{\scriptsize B}}$ &
 A$^i_{\mbox{\scriptsize R}}$ &
 A$^i_{\mbox{\scriptsize I}}$ &
 A$^i_{\mbox{\scriptsize K$^\prime$}}$ &
 M$^{b,i}_{\mbox{\scriptsize B}}$ &
 M$^{b,i}_{\mbox{\scriptsize R}}$ &
 M$^{b,i}_{\mbox{\scriptsize I}}$ &
 M$^{b,i}_{\mbox{\scriptsize K$^\prime$}}$ &
 D$^{b,i}_{25}$ \\

\noalign{\vspace{0.8mm}}
 & mag & mag & mag & mag & km/s & 
   mag & mag & mag & mag &
   mag & mag & mag & mag &
   ($^\prime$) \\
\noalign{\vspace{0.8mm}}
  (1)    &  (2)  &  (3)  &  (4)  &  (5)  &  (6)  & (7) & (8) &  (9)  &  (10)  & (11)  &  (12) & (13) & (14) & (15) \\
\noalign{\vspace{0.8mm}}
\hline
\hline
\noalign{\vspace{0.8mm}}

\multicolumn{15}{l}{\hspace{-2mm}{\em \underline{Galaxies with fully analyzed HI data:}}} \\
\noalign{\vspace{0.8mm}}
 U6399    & 14.33 & 13.31 & 12.88 & 11.09 & 172 & 0.47 & 0.36 & 0.27 & 0.06 & $-$17.56 & $-$18.44 & $-$18.77 & $-$20.33 & 1.84 \\ 
 U6446    & 13.52 & 12.81 & 12.58 & 11.50 & 174 & 0.18 & 0.14 & 0.10 & 0.02 & $-$18.08 & $-$18.72 & $-$18.90 & $-$19.88 & 2.07 \\ 
 N3726    & 11.00 &  9.97 &  9.51 &  7.96 & 331 & 0.34 & 0.25 & 0.20 & 0.05 & $-$20.76 & $-$21.67 & $-$22.07 & $-$23.45 & 5.32 \\ 
 N3769    & 12.80 & 11.56 & 10.99 &  9.10 & 256 & 0.67 & 0.50 & 0.39 & 0.09 & $-$19.32 & $-$20.35 & $-$20.80 & $-$22.35 & 2.34 \\ 
 U6667    & 14.33 & 13.11 & 12.63 & 10.81 & 167 & 0.74 & 0.58 & 0.43 & 0.10 & $-$17.83 & $-$18.87 & $-$19.18 & $-$20.65 & 2.18 \\ 
 N3877    & 11.91 & 10.46 &  9.72 &  7.75 & 335 & 1.06 & 0.78 & 0.62 & 0.15 & $-$20.60 & $-$21.73 & $-$22.29 & $-$23.76 & 3.95 \\ 
 N3893    & 11.20 & 10.19 &  9.71 &  7.84 & 382 & 0.31 & 0.23 & 0.18 & 0.04 & $-$20.55 & $-$21.45 & $-$21.86 & $-$23.56 & 3.67 \\ 
 N3917    & 12.66 & 11.42 & 10.85 &  9.08 & 276 & 0.87 & 0.64 & 0.51 & 0.12 & $-$19.65 & $-$20.63 & $-$21.05 & $-$22.40 & 3.48 \\ 
 N3949    & 11.55 & 10.69 & 10.28 &  8.43 & 321 & 0.33 & 0.24 & 0.20 & 0.05 & $-$20.22 & $-$20.96 & $-$21.31 & $-$22.98 & 2.66 \\ 
 N3953    & 11.03 &  9.66 &  9.02 &  7.03 & 446 & 0.60 & 0.43 & 0.35 & 0.08 & $-$21.05 & $-$22.20 & $-$22.74 & $-$24.41 & 5.38 \\ 
 N3972    & 13.09 & 11.90 & 11.34 &  9.39 & 264 & 0.76 & 0.56 & 0.44 & 0.11 & $-$19.08 & $-$20.05 & $-$20.48 & $-$22.08 & 2.62 \\ 
 U6917    & 13.15 & 12.16 & 11.74 & 10.30 & 224 & 0.31 & 0.23 & 0.18 & 0.04 & $-$18.63 & $-$19.49 & $-$19.84 & $-$21.10 & 2.84 \\ 
 U6923    & 13.91 & 12.97 & 12.36 & 11.04 & 160 & 0.28 & 0.22 & 0.16 & 0.04 & $-$17.84 & $-$18.67 & $-$19.20 & $-$20.36 & 1.67 \\ 
 U6930$^i$& 12.70 & 11.71 & 11.39 & 10.33 & 231 & 0.08 & 0.06 & 0.05 & 0.01 & $-$18.86 & $-$19.78 & $-$20.07 & $-$21.04 & 2.98 \\ 
 N3992    & 10.86 &  9.55 &  8.94 &  7.23 & 547 & 0.56 & 0.40 & 0.33 & 0.08 & $-$21.18 & $-$22.28 & $-$22.80 & $-$24.21 & 6.27 \\ 
 U6940$^f$& 16.45 & 15.65 & 15.44 & 13.99 &  50 & 0.00 & 0.00 & 0.00 & 0.00 & $-$15.02 & $-$15.77 & $-$15.96 & $-$17.37 & 0.64 \\ 
 U6962$^i$& 12.88 & 11.88 & 11.42 & 10.11 & 327 & 0.16 & 0.11 & 0.09 & 0.02 & $-$18.72 & $-$19.64 & $-$20.06 & $-$21.27 & 2.26 \\ 
 N4010    & 13.36 & 12.14 & 11.55 &  9.22 & 254 & 1.20 & 0.89 & 0.70 & 0.17 & $-$19.30 & $-$20.17 & $-$20.55 & $-$22.31 & 2.97 \\ 
 U6969    & 15.12 & 14.32 & 14.04 & 12.58 & 117 & 0.20 & 0.17 & 0.11 & 0.02 & $-$16.56 & $-$17.28 & $-$17.48 & $-$18.80 & 1.19 \\ 
 U6973    & 12.94 & 11.26 & 10.53 &  8.23 & 364 & 0.71 & 0.52 & 0.42 & 0.10 & $-$19.21 & $-$20.67 & $-$21.28 & $-$23.23 & 2.21 \\ 
 U6983    & 13.10 & 12.27 & 11.91 & 10.52 & 221 & 0.21 & 0.16 & 0.12 & 0.03 & $-$18.58 & $-$19.31 & $-$19.61 & $-$20.87 & 2.99 \\ 
 N4051    & 10.98 &  9.88 &  9.37 &  7.86 & 308 & 0.28 & 0.21 & 0.17 & 0.04 & $-$20.71 & $-$21.72 & $-$22.18 & $-$23.54 & 5.45 \\ 
 N4085    & 13.09 & 11.87 & 11.28 &  9.20 & 247 & 0.78 & 0.58 & 0.46 & 0.11 & $-$19.12 & $-$20.11 & $-$20.57 & $-$22.27 & 2.08 \\ 
 N4088    & 11.23 & 10.00 &  9.37 &  7.46 & 362 & 0.74 & 0.54 & 0.43 & 0.10 & $-$20.95 & $-$21.94 & $-$22.45 & $-$24.00 & 4.40 \\ 
 N4100    & 11.91 & 10.62 & 10.00 &  8.02 & 386 & 0.97 & 0.70 & 0.57 & 0.14 & $-$20.51 & $-$21.49 & $-$21.97 & $-$23.48 & 4.07 \\ 
 N4102    & 12.04 & 10.54 &  9.93 &  7.86 & 393 & 0.46 & 0.34 & 0.27 & 0.07 & $-$19.86 & $-$21.20 & $-$21.73 & $-$23.57 & 2.69 \\ 
 N4157    & 12.12 & 10.60 &  9.88 &  7.52 & 399 & 1.40 & 1.02 & 0.82 & 0.20 & $-$20.72 & $-$21.83 & $-$22.33 & $-$24.04 & 4.64 \\ 
 N4183    & 12.96 & 11.99 & 11.51 &  9.76 & 228 & 1.01 & 0.76 & 0.59 & 0.14 & $-$19.46 & $-$20.16 & $-$20.46 & $-$21.74 & 3.13 \\ 
 N4217    & 12.15 & 10.62 &  9.84 &  7.61 & 381 & 1.05 & 0.76 & 0.62 & 0.15 & $-$20.33 & $-$21.54 & $-$22.16 & $-$23.90 & 4.29 \\ 
 N4389    & 12.56 & 11.33 & 10.87 &  9.12 & 212 & 0.20 & 0.15 & 0.12 & 0.03 & $-$19.05 & $-$20.21 & $-$20.63 & $-$22.27 & 2.31 \\ 
\noalign{\vspace{0.8mm}}
\multicolumn{15}{l}{\hspace{-2mm}{\em \underline{Galaxies with partially analyzed HI data:}}} \\  
\noalign{\vspace{0.8mm}}
 N3718    & 11.28 &  9.95 &  9.29 &  7.47 & 476 & 0.77 & 0.55 & 0.45 & 0.11 & $-$20.90 & $-$21.99 & $-$22.54 & $-$24.00 & 6.30 \\ 
 N3729    & 12.31 & 10.94 & 10.30 &  8.60 & 296 & 0.25 & 0.18 & 0.15 & 0.03 & $-$19.34 & $-$20.62 & $-$21.22 & $-$22.78 & 2.60 \\ 
 U6773    & 14.42 & 13.61 & 13.15 & 11.23 & 112 & 0.09 & 0.08 & 0.05 & 0.01 & $-$17.09 & $-$17.87 & $-$18.28 & $-$20.14 & 1.35 \\ 
 U6818    & 14.43 & 13.62 & 13.15 & 11.70 & 151 & 0.39 & 0.31 & 0.22 & 0.05 & $-$17.40 & $-$18.10 & $-$18.46 & $-$19.71 & 1.69 \\ 
 U6894    & 15.27 & 14.31 & 14.00 & 12.40 & 124 & 0.37 & 0.31 & 0.21 & 0.05 & $-$16.51 & $-$17.39 & $-$17.59 & $-$19.01 & 1.13 \\ 
 N3985    & 13.25 & 12.26 & 11.81 & 10.19 & 180 & 0.18 & 0.14 & 0.10 & 0.02 & $-$18.39 & $-$19.30 & $-$19.69 & $-$21.19 & 1.29 \\ 
 N4013    & 12.44 & 10.79 &  9.95 &  7.68 & 377 & 1.10 & 0.80 & 0.64 & 0.15 & $-$20.08 & $-$21.40 & $-$22.07 & $-$23.83 & 3.61 \\ 
 U7089    & 13.73 & 12.77 & 12.36 & 11.11 & 138 & 0.42 & 0.34 & 0.24 & 0.05 & $-$18.11 & $-$18.96 & $-$19.26 & $-$20.30 & 2.46 \\ 
 U7094    & 14.74 & 13.70 & 13.22 & 11.58 &  76 & 0.00 & 0.00 & 0.00 & 0.00 & $-$16.67 & $-$17.69 & $-$18.16 & $-$19.78 & 1.29 \\ 
 N4117    & 14.05 & 12.47 & 11.81 &  9.98 & 285 & 0.00 & 0.00 & 0.00 & 0.00 & $-$17.36 & $-$18.92 & $-$19.57 & $-$21.38 & 1.29 \\
 N4138    & 12.27 & 10.72 & 10.09 &  8.19 & 374 & 0.36 & 0.26 & 0.21 & 0.05 & $-$19.50 & $-$20.93 & $-$21.50 & $-$23.22 & 2.22 \\ 
 N4218    & 13.69 & 12.83 & 12.41 & 10.83 & 150 & 0.15 & 0.12 & 0.09 & 0.02 & $-$17.88 & $-$18.68 & $-$19.06 & $-$20.55 & 1.06 \\ 
 N4220    & 12.34 & 10.79 & 10.03 &  8.36 & 399 & 0.94 & 0.68 & 0.55 & 0.13 & $-$20.03 & $-$21.29 & $-$21.91 & $-$23.13 & 2.85 \\ 
\noalign{\vspace{0.8mm}}
\multicolumn{15}{l}{\hspace{-2mm}{\em \underline{Galaxies with confused HI data:}}} \\
\noalign{\vspace{0.8mm}}
 1135+48  & 14.95 & 14.05 & 13.61 & 11.98 & 111 & 0.17 & 0.15 & 0.09 & 0.02 & $-$16.67 & $-$17.51 & $-$17.88 & $-$19.40 & 0.97 \\
 N3896    & 13.75 & 12.96 & 12.47 & 11.35 &  83 & 0.00 & 0.00 & 0.00 & 0.00 & $-$17.69 & $-$18.45 & $-$18.92 & $-$20.01 & 1.49 \\
\noalign{\vspace{0.8mm}}
\multicolumn{15}{l}{\hspace{-2mm}{\em \underline{Not observed or too little HI content:}}} \\
\noalign{\vspace{0.8mm}}
 N3870    & 13.67 & 12.71 & 12.16 & 10.73 & 127 & 0.08 & 0.06 & 0.04 & 0.01 & $-$17.83 & $-$18.74 & $-$19.26 & $-$20.64 & 1.06 \\ 
 N3990    & 13.53 & 12.08 & 11.36 &  9.54 & ... & 0.00 & 0.00 & 0.00 & 0.00 & $-$17.89 & $-$19.31 & $-$20.02 & $-$21.82 & 1.28 \\ 
 N4026    & 11.71 & 10.25 &  9.57 &  7.65 & ... & 0.00 & 0.00 & 0.00 & 0.00 & $-$19.74 & $-$21.16 & $-$21.82 & $-$23.71 & 3.32 \\ 
 N4111    & 11.40 &  9.95 &  9.25 &  7.60 & ... & 0.00 & 0.00 & 0.00 & 0.00 & $-$20.01 & $-$21.44 & $-$22.13 & $-$23.76 & 3.24 \\ 
 U7129    & 14.13 & 12.80 & 12.19 &  ...  & ... & 0.08 & 0.06 & 0.04 & 0.01 & $-$17.36 & $-$18.65 & $-$19.23 &     ...  & 1.19 \\ 
 N4143    & 12.06 & 10.55 &  9.84 &  7.86 & ... & 0.00 & 0.00 & 0.00 & 0.00 & $-$19.35 & $-$20.83 & $-$21.54 & $-$23.50 & 2.30 \\ 
 N4346    & 12.14 & 10.69 &  9.96 &  8.21 & ... & 0.00 & 0.00 & 0.00 & 0.00 & $-$19.27 & $-$20.70 & $-$21.42 & $-$23.15 & 2.75 \\ 
\noalign{\vspace{0.8mm}}
\hline
\hline

\end{tabular}
}
\end{center}
\end{table*}

\section{The sample}

The nearby Ursa Major cluster as defined in Paper I has 79 identified
members.  It is located in the Supergalactic plane at an angular
distance of 38 degrees from the core of the Virgo cluster.  It has a
recession velocity of 950 km$\;$s$^{-1}$ and a velocity dispersion of
only $\approx$150 km$\;$s$^{-1}$.  In Paper~I and in Tully \& Verheijen
(\cite{tully97}) (Paper II) a distance of 15.5 Mpc was adopted. However,
new HST cepheid distances to local TFr calibrators (e.g. Sakai et al.
\cite{sakai}) and a new correction formalism for internal extinction
(Tully et al. \cite{tully98}) now place the Ursa Major cluster galaxies
at a mean distance of 18.6 Mpc (Tully \& Pierce \cite{tully00}). At this
distance, 1 arcmin corresponds to 5.4 kpc.  The morphological mix of the
cluster members is made up overwhelmingly by late type systems and only
a dozen lenticulars are known members.  Morphological and photometric
properties in the optical and near-infrared of individual galaxies are
described in detail in Paper I. The galaxy distribution shows no
concentration toward any core and no X-ray emitting intra-cluster gas
has been detected. 

It should be noted that individual galaxies in the nearby Virgo (Warmels
\cite{warmels88a}, \cite{warmels88b}; Cayatte et al. \cite{cayatte90}),
A1367 (Dickey \& Gavazzi \cite{dickey91}), Hercules (Dickey
\cite{dickey97}) and Coma (Bravo-Alfaro et al. \cite{bravo}) clusters
have also been studied in detail and the effect of the dense environment
on the properties of the HI disks in these clusters has been clearly
demonstrated. The HI disks in the cores of these rich clusters are in
general very small and often offset from the optical galaxy. On the
other hand, the volume limited survey of the Hydra cluster (McMahon
1993) does not show such an HI deficiency although some interesting
dynamical substructure has been revealed in this system. It should be
stressed that the Ursa Major cluster is markedly different from these
more massive and denser clusters. In selecting the Ursa Major sample,
these environmental effects are carefully avoided as well as fore- and
background contamination caused by high velocity dispersions and complex
dynamical and spatial substructures.

Since the Ursa Major galaxies are all at the same distance, the effects
of incompleteness and uncertain relative distances are minimized.  A
complete sample of 62 galaxies brighter than M$_B$$\approx$$-$16.8, i.e.
roughly twice the luminosity of the Small Magellanic Cloud, was
constructed and nearly all cluster members were observed with the WSRT. 
In this paper, however, only those 49 galaxies which are more inclined
than 45 degrees, as derived from the optical axis ratio, will be
considered for a detailed kinematic study. 

Table~1 gives a summary of the positional and morphological properties
of these 49 galaxies while photometrics are presented in Table~2, based
on a 18.6 Mpc distance.  There are 3 additional galaxies in the tables
which do not meet the luminosity ($^f$) and inclination ($^i$) criteria
but happened to be in the same WSRT fields as galaxies from the complete
sample.  Of all those 52 galaxies, the HI synthesis data of 30 were
fully analyzed. Thirteen systems were observed and detected but the HI
data of these galaxies are presented in an abbreviated form comprising 
only the channel maps, global profiles and position-velocity diagrams. 
Two of the smaller galaxies were detected in HI but they are confused
with the HI emission from their more massive companions.  Finally, there
are 7 galaxies in the complete sample which have not been observed or
detected because of their low HI content known from single dish
observations.  These are in general S0 or Sa systems.

Table~1 presents the following positional and morphological information:\\
 \noindent{\it Column} (1) gives the NGC or UGC numbers. \\  
 \noindent{\it Columns} (2) and (3) provide the equatorial coordinates
(B1950) derived from the optical images. \\
 \noindent{\it Columns} (4) and (5) give the Galactic coordinates. \\
 \noindent{\it Column} (6) provides the morphological type.\\
 \noindent{\it Column} (7) gives the observed major axis diameter of the
25$^{th}$ mag$\;$arcsec$^{-2}$ blue isophote. \\
 \noindent{\it Column} (8) contains the position angle of the receding side
of the galaxy.  For galaxies which are not observed or not detected in
HI, this is the smallest position angle of the major axis measured
eastward from the north. \\
 \noindent{\it Column} (9) contains the observed ellipticity of the
optical galaxy image. \\
 \noindent{\it Column} (10) gives the inclination i$_{\mbox{\scriptsize
opt}}$ as derived from the observed axis ratio (b/a). See Section 5.1
for further details.\\
 \noindent{\it Column} (11) gives the adopted inclination angle as
derived from several methods described in Section 5. \\
 \noindent{\it Column} (12) indicates whether a galaxy has a low (LSB)
or high surface brightness (HSB) according to paper II. \\
 \noindent{\it Columns} (13) and (14) provide the galactic extinction in
the B-band according to Burstein \& Heiles (1984) (BH) and Schlegel et
al. (1998) (SFD) as reported by the NASA Extragalactic Database.

Table~2 presents the following photometric information: \\
  \noindent{\it Column} (1) gives the NGC or UGC numbers. \\
 \noindent{\it Columns} (2)$-$(5) give the observed total magnitudes
in the B, R, I and  K$^\prime$ passbands from Paper I. \\
 \noindent{\it Column} (6) contains the corrected HI line widths at the
20\% level, used to calculate the internal extinction as explained below. \\
 \noindent{\it Columns} (7)$-$(10) present the calculated internal
extinction corrections in thee B, R, I and K$^\prime$ passbands toward
face-on A$_\lambda^{i\rightarrow 0}$, calculated according to Tully et
al. (1998):

 $$ \mbox{A}_\lambda^{i\rightarrow 0} = \gamma_\lambda \; \mbox{\rm log (a/b)} $$

\noindent where a/b is the observed axis ratio of the galaxy as an
indication of inclination while $\gamma_\lambda$ depends on the
luminosity and is calculated according to

\begin{center}
\begin{tabular}{lcl}
 $\gamma_{\mbox{\tiny B}}$          & = & 1.57 + 2.75 ( log W$^i_{R,I}$ $-$ 2.5 ) \\
\noalign{\vspace{0.8mm}}
 $\gamma_{\mbox{\tiny R}}$          & = & 1.15 + 1.88 ( log W$^i_{R,I}$ $-$ 2.5 ) \\
\noalign{\vspace{0.8mm}}
 $\gamma_{\mbox{\tiny I}}$          & = & 0.92 + 1.63 ( log W$^i_{R,I}$ $-$ 2.5 ) \\
\noalign{\vspace{0.8mm}}
 $\gamma_{\mbox{\tiny K$^\prime$}}$ & = & 0.22 + 0.40 ( log W$^i_{R,I}$ $-$ 2.5 ) \\
\end{tabular}
\end{center}

\noindent where W$^i_{R,I}$ is the distance independent HI line width
corrected for instrumental resolution as described in Section 3.2,
corrected for turbulent motion according to Tully \& Fouqu\'e
(\cite{tully85}) (TFq hereafter) with W$_{t,20}$=22 km/s as motivated in
Section~4 and corrected for inclination using i$_{\mbox{\scriptsize
adopt}}$ from Table~1. For dwarf galaxies with W$^i_{R,I}~<~85$ km/s and
for lenticulars with no dust features, the value of $\gamma_\lambda$ is
set to zero at all passbands.\\
 \noindent{\it Columns} (11)$-$(14) give the total absolute B, R, I and
K$^\prime$ magnitudes corrected for Galactic and internal extinction and
a distance modulus of 31.35 corresponding to a distance to the Ursa
Major cluster of 18.6~Mpc:

  $$\mbox{M}^{b,i}_\lambda = \mbox{m}^{tot}_\lambda - \mbox{A}_\lambda^b - \mbox{A}_\lambda^{i\rightarrow 0} - 31.35$$

\noindent where the Galactic extinction A$_{\mbox{\scriptsize B}}^b$ is
taken from SFD as listed in Table~1. Extinction corrections in the other
passbands are made according to the Galactic reddening law given by
Cardelli et al. (\cite{cardelli}) as summarized by SFD under the Landolt
filters in their Table~6. It's given by
 A$^b_{\mbox{\scriptsize R}}$/A$^b_{\mbox{\scriptsize B}}$=~0.62,
 A$^b_{\mbox{\scriptsize I}}$/A$^b_{\mbox{\scriptsize B}}$=~0.45 and
 A$^b_{\mbox{\scriptsize K$^\prime$}}$/A$^b_{\mbox{\scriptsize B}}$=~0.08. \\
 \noindent{\it Column} (15) gives the diameter of the 25$^{th}$
mag$\;$arcsec$^{-2}$ blue isophote corrected for both galactic and
internal extinction and projection according to TFq:

  $$\mbox{Log}(\mbox{D}^{b,i}_{25}) = \mbox{Log}(\mbox{D}_{25}) - 0.22\;\mbox{Log}(\mbox{D}_{25}/\mbox{d}_{25}) + 0.09\;\mbox{A}^b_{\mbox{\scriptsize B}}$$

\noindent where $\mbox{d}_{25}$ is the minor axis diameter at the
25$^{th}$ mag$\;$arcsec$^{-2}$ blue isophote and A$^b_{\mbox{\scriptsize
B}}$ is taken from SFD.  \\

\section{Data acquisition and reduction}

The HI data presented in this paper were obtained with the Westerbork
Synthesis Radio Telescope (WSRT) between 1991 and 1996.  The integration
times varied between 1$\times$12$^{\mbox{\scriptsize h}}$ and
5$\times$12$^{\mbox{\scriptsize h}}$ depending on the required
signal-to-noise.  The angular resolution at the center of the cluster is
$12^{\prime\prime}\times16^{\prime\prime}$ or 1.08$\times$1.44 kpc at
the adopted distance of 18.6 Mpc.  The FWHM of the primary beam is 37.4
arcminutes or 202 kpc.  As a result, often more than one galaxy was
mapped in a single field of view.  The observed bandwidth was either 2.5
or 5 MHz, depending on the width of the global profiles.  The
observations of the NGC3992-group and the NGC4111-group required a
broad frequency band of 5 MHz and at the same time also sufficient
velocity resolution for the dwarf systems.  To comply with the
correlator restrictions, those two fields were observed only in one
polarization (XX) which allowed for a velocity resolution of 10
km$\;$s$^{-1}$ but resulted in less sensitivity.  During the earlier
measurements an on-line Hanning taper was applied but this tapering was
abandoned later to obtain the highest possible velocity resolution.  The
various obtained velocity resolutions (dependent on the correlator
restrictions) were 5, 8, 10, 20 or 33 km/s, corresponding to typical
rms-noise levels of respectively 3.1, 1.9, 2.9, 1.6 and 1.0
mJy$\;$beam$^{-1}$ for a single 12$^{\mbox{\scriptsize h}}$ observation
at the highest angular resolution.  The data of NGC4013 were kindly made
avaible by R.~Bottema who studied this system in great detail (Bottema
\cite{bottema96} and references therein). 

More details on the observational parameters for each field are
tabulated in the atlas along with the data. What follows is a brief
description of the reduction procedures.

\begin{table*}[p]

\caption[]{A comparison of the widths and integrated fluxes from the
present WSRT survey and from the literature.}

\begin{center}

{\fontsize{8.6}{9.6}
\selectfont

\begin{tabular}[t]{lrrrrrrrrrrr@{}l}
\hline
\noalign{\vspace{0.8mm}}
 & \multicolumn{5}{|c}{This study} & \multicolumn{7}{|c}{Literature} \\

\multicolumn{1}{c}{Name}      &
\multicolumn{1}{|c}{W$_{20}$}  & 
\multicolumn{1}{c}{$\pm$}     &
\multicolumn{1}{c}{Res.}      &
\multicolumn{1}{c}{$\int$Sdv}       & 
\multicolumn{1}{c}{$\pm$}     &
\multicolumn{1}{|c}{W$_{20}$}  &
\multicolumn{1}{c}{$\pm$}     & 
\multicolumn{1}{c}{Res.}      & 
\multicolumn{1}{c}{$\int$Sdv}       & 
\multicolumn{1}{c}{$\pm$}     & 
\multicolumn{1}{c}{} &
\multicolumn{1}{@{}l}{Ref.}      \\
 &
\multicolumn{3}{|c}{-- -- -- km$\;$s$^{-1}$ -- -- --} &
\multicolumn{2}{c}{-- Jykm$\;$s$^{-1}$ --} &
\multicolumn{3}{|c}{-- -- -- km$\;$s$^{-1}$ -- -- --} &
\multicolumn{2}{c}{-- Jykm$\;$s$^{-1}$ --} & \\

\multicolumn{1}{c}{(1)} &
\multicolumn{1}{|c}{(2)} &
\multicolumn{1}{c}{(3)} &
\multicolumn{1}{c}{(4)} &
\multicolumn{1}{c}{(5)} &
\multicolumn{1}{c}{(6)} &
\multicolumn{1}{|c}{(7)} &
\multicolumn{1}{c}{(8)} &
\multicolumn{1}{c}{(9)} &
\multicolumn{1}{c}{(10)} &
\multicolumn{1}{c}{(11)} &
\multicolumn{1}{c}{} &
\multicolumn{1}{@{}l}{(12)} \\
\noalign{\vspace{0.8mm}}
\hline
\hline
\noalign{\vspace{0.8mm}}

U6399           & 188.1 & 1.4 &  8.3 &  10.5 & 0.3 & 178\phantom{$^m$} & 20 & 22\phantom{.4} &  10.1           & 1.9 &  1 & \\
U6446           & 154.1 & 1.0 &  5.0 &  40.6 & 0.5 & 162\phantom{$^m$} & 10 & 22\phantom{.4} &  45.9           & 4.1 &  1 & \\
N3718$^{(c)}$   & 492.8 & 1.0 & 33.2 & 140.9 & 0.9 & 480\phantom{$^m$} & 10 &  5.5           &  84.9           &26.4 &  1 & \\
                &       &     &      &       &     & 508$^m$           & .. & 33\phantom{.4} & 120\phantom{.4} & ... &  8 &$^{WSRT}$ \\
N3726           & 286.5 & 1.6 &  5.0 &  89.8 & 0.8 & 290\phantom{$^m$} & 10 &  5.5           &  83.9           &10.8 &  1 & \\
N3729$^{noSD}$  & 270.8 & 1.5 & 33.2 &   5.5 & 0.3 & ...\phantom{$^m$} & .. & ..\phantom{.4} &   ...           & ... & .. & \\
                &       &     &      &       &     & 279$^m$           & .. & 33\phantom{.4} & 25$^?$          & ... &  8 &$^{WSRT}$ \\
N3769$^i$       & 265.3 & 6.7 &  8.3 &  62.3 & 0.6 & 276\phantom{$^m$} & 20 &  7.4           &  44.1           & 4.2 &  2 & \\
U6667           & 187.5 & 1.4 &  5.0 &  11.0 & 0.4 & 210\phantom{$^m$} & 20 & 22\phantom{.4} &  11.6           & 2.2 &  1 & \\
N3877           & 373.4 & 5.0 & 33.2 &  19.5 & 0.6 & 352\phantom{$^m$} & 10 & 22\phantom{.4} &  24.8           & 5.6 &  1 & \\
U6773           & 110.4 & 2.3 &  8.3 &   5.6 & 0.4 & 118\phantom{$^m$} &  8 & 22\phantom{.4} &   5.6           & 0.7 &  6 & \\
N3893$^{(c)}$   & 310.9 & 1.0 &  5.0 &  69.9 & 0.5 & 313\phantom{$^m$} &  8 & 22\phantom{.4} &  85.3           & 5.1 &  1 & \\
N3917           & 294.5 & 1.9 &  8.3 &  24.9 & 0.6 & 284\phantom{$^m$} & 10 & 22\phantom{.4} &  21.9           & 4.7 &  1 & \\
U6818           & 166.9 & 2.3 &  8.3 &  13.9 & 0.2 & 168\phantom{$^m$} & 15 & 22\phantom{.4} &  14.8           & 2.1 &  1 & \\
N3949           & 286.5 & 1.4 &  8.3 &  44.8 & 0.4 & 289\phantom{$^m$} & 10 & 22\phantom{.4} &  42.7           & 5.4 &  1 & \\
N3953$^l$       & 441.9 & 2.4 & 33.1 &  39.3 & 0.8 & 423\phantom{$^m$} & 10 & 22\phantom{.4} &  41.0           & 3.9 &  1 & \\
U6894           & 141.8 & 1.1 &  8.3 &   5.8 & 0.2 & 159\phantom{$^m$} & 20 &  7.4           &   5.1           & 1.7 &  2 & \\
N3972           & 281.2 & 1.4 &  8.3 &  16.6 & 0.4 & 270\phantom{$^m$} & 15 & 22\phantom{.4} &  14.0           & 2.6 &  1 & \\
U6917           & 208.9 & 3.2 &  8.3 &  26.2 & 0.3 & 211\phantom{$^m$} & 10 & 22\phantom{.4} &  31.5           & 4.1 &  1 & \\
N3985           & 160.2 & 3.7 &  8.3 &  15.7 & 0.6 & 168\phantom{$^m$} & .. & 22\phantom{.4} &  14.1           & 0.9 &  5 & \\
U6923           & 166.8 & 2.4 & 10.0 &  10.7 & 0.6 & 175\phantom{$^m$} & 15 & 22\phantom{.4} &   8.2           & 2.9 &  1 & \\
                &       &     &      &       &     & 189$^m$           & 15 & 41.4           &  15.6           & ... & 12 &$^{VLA}$ \\
U6930           & 136.5 & 0.5 &  8.3 &  42.7 & 0.3 & 145\phantom{$^m$} &  8 & 22\phantom{.4} &  38.2           & 3.5 &  1 & \\
N3992$^l$       & 478.5 & 1.4 & 10.0 &  74.6 & 1.5 & 480\phantom{$^m$} & 10 & 22\phantom{.4} &  81.2           & 5.3 &  1 & \\
                &       &     &      &       &     & 507$^m$           & 15 & 41.4           &  79.9           & ... & 12 &$^{VLA}$ \\
U6940           &  59.3 & 3.8 & 10.0 &   2.1 & 0.3 & 226\phantom{$^m$} & .. & 22\phantom{.4} &   7.0           & 1.0 &  3 & \\
                &       &     &      &       &     & 121$^m$           & 15 & 41.4           &   2.7           & ... & 12 &$^{VLA}$ \\
N4013           & 425.0 & 0.9 & 33.0 &  41.5 & 0.2 & 403\phantom{$^m$} & 10 & 22\phantom{.4} &  33.8           & 3.7 &  1 & \\
U6962$^{(c)}$   & 220.3 & 6.6 &  8.3 &  10.0 & 0.3 & ...\phantom{$^m$} & .. & 22\phantom{.4} &  21.6           & 4.4 &  1 & \\
                &       &     &      &       &     & 235\phantom{$^m$} & .. & 33\phantom{.4} &   9.2           & 1.0 &  4 &$^{WSRT}$ \\
N4010           & 277.7 & 1.0 &  8.3 &  38.2 & 0.3 & 281\phantom{$^m$} & 10 & 22\phantom{.4} &  38.1           & 3.4 &  1 & \\
U6969$^c$       & 132.1 & 6.4 & 10.0 &   6.1 & 0.5 & 146\phantom{$^m$} & .. & 13.2           &   6.0           & 1.4 &  3 & \\
                &       &     &      &       &     & 159$^m$           & 15 & 41.4           &   6.9           & ... & 12 &$^{VLA}$ \\
U6973$^{noSD}$  & 367.8 & 1.8 &  8.3 &  22.9 & 0.2 & ...\phantom{$^m$} & .. & ..\phantom{.4} &   ...           & ... & .. & \\
                &       &     &      &       &     & 408\phantom{$^m$} & .. & 33\phantom{.4} &  18.3           & 1.2 &  4 &$^{WSRT}$ \\
U6983           & 188.4 & 1.3 &  5.0 &  38.5 & 0.6 & 205\phantom{$^m$} & 10 & 22\phantom{.4} &  36.2           & 4.4 &  1 & \\
N4051$^l$       & 255.4 & 1.8 &  5.0 &  35.6 & 0.8 & 274\phantom{$^m$} & 15 & 22\phantom{.4} &  43.4           & 3.3 &  1 & \\
N4085$^c$       & 277.4 & 6.6 & 19.8 &  14.6 & 0.9 & 299\phantom{$^m$} & 20 &  7.4           &  23.3           & 2.5 &  2 & \\
                &       &     &      &       &     & 311$^m$           & .. & 33\phantom{.4} &  24$^!$         & ... & 13 &$^{WSRT}$ \\
N4088$^{(c)}$   & 371.4 & 1.7 & 19.8 & 102.9 & 1.1 & 381\phantom{$^m$} &  8 & 22\phantom{.4} & 109.2           & 6.4 &  1 & \\
                &       &     &      &       &     & 378$^m$           & .. & 33\phantom{.4} & 128$^!$         & ... & 13 &$^{WSRT}$ \\
U7089$^{(c)}$   & 156.7 & 1.7 & 10.0 &  17.0 & 0.6 & 162\phantom{$^m$} & 10 & 22\phantom{.4} &  17.8           & 2.2 &  1 & \\
                &       &     &      &       &     & 176\phantom{$^m$} & .. & 33.4           &  18.9           & ... & 11 &$^{WSRT}$ \\
N4100           & 401.8 & 2.0 & 19.9 &  41.6 & 0.7 & 420\phantom{$^m$} & 20 & 22\phantom{.4} &  54.0           & 7.3 &  1 & \\
U7094$^{c}$     &  83.7 & 1.7 & 10.0 &   2.9 & 0.2 & 112\phantom{$^m$} &  8 & 22\phantom{.4} &   6.0           & 0.6 &  6 & \\
                &       &     &      &       &     & 153$^?$           & .. & 33.4           &   2.5           & ... & 11 &$^{WSRT}$ \\
N4102           & 349.8 & 2.0 &  8.3 &   8.0 & 0.2 & 327\phantom{$^m$} & 20 &  7.4           &  10.3           & 2.1 &  2 & \\
N4117$^{noSD}$  & 289.4 & 7.5 & 10.0 &   6.9 & 1.1 & ...\phantom{$^m$} & .. & ..\phantom{.4} &   ...           & ... & .. & \\
                &       &     &      &       &     & 314\phantom{$^m$} & .. & 33.4           &   5.3           & ... & 11 &$^{WSRT}$ \\
N4138           & 331.6 & 4.5 & 19.9 &  19.2 & 0.7 & 354$^m$           & 30 &  6.8           &  16\phantom{.4} & ... & 14 & \\
                &       &     &      &       &     & 340\phantom{$^m$} &  5 &  5.2           &  20.6           & 0.3 &  7 &$^{VLA}$ \\
N4157$^{(c),l}$ & 427.6 & 2.2 & 19.9 & 107.4 & 1.6 & 436\phantom{$^m$} & 10 & 22\phantom{.4} & 123.9           & 9.5 &  1 & \\
N4183		& 249.6 & 1.2 &  8.3 &	48.9 & 0.7 & 258\phantom{$^m$} & 10 & 22\phantom{.4} &	49.6           & 5.3 &	1 & \\
N4218		& 138.0 & 5.0 &  8.3 &	 7.8 & 0.2 & 160\phantom{$^m$} & 20 & 13\phantom{.4} &	 5.7           & 0.9 &	9 & \\
N4217           & 428.1 & 5.1 & 33.2 &  33.8 & 0.7 & 426\phantom{$^m$} & 20 & 22\phantom{.4} &  51.8           & 7.2 &  1 & \\
N4220           & 438.1 & 1.3 & 33.1 &   4.4 & 0.3 & 382$^m$           & .. & 11\phantom{.4} &   3.3           & ... & 10 & \\
N4389           & 184.0 & 1.5 &  8.3 &   7.6 & 0.2 & 174\phantom{$^m$} & 20 &  7.4           &   7.6           & 0.8 &  2 & \\

\hline
\hline
\end{tabular}
}
\end{center}
\end{table*}

\addtocounter{table}{-1}
\begin{table*}[t]
\begin{center}
\caption[]{(cont.) Notes}

\begin{tabular}[t]{rll}
\hline
\noalign{\vspace{0.8mm}}
  $^{(c)}$ : & \multicolumn{2}{l}{the authors suggest possible confusion with a dwarf companion.} \\
      $^c$ : & \multicolumn{2}{l}{flagged by the authors as confused with near companion.} \\
      $^l$ : & \multicolumn{2}{l}{large correction factor ($>$1.20) applied for primary beam flux attenuation.} \\
      $^i$ : & \multicolumn{2}{l}{flagged by the authors as possibly interacting.} \\
 $^{noSD}$ : & \multicolumn{2}{l}{no useful single dish profile available due to obvious confusion.} \\
      $^m$ : & \multicolumn{2}{l}{line width directly measured from the published HI profile.} \\
      $^!$ : & \multicolumn{2}{l}{the integrated flux as quoted by the author is a factor 2 larger than is quoted by} \\
             & \multicolumn{2}{l}{any other source. Therefore, half the integrated flux was adopted from this source.} \\
 $^{WSRT}$ : & \multicolumn{2}{l}{synthesis observation with the WSRT.} \\
  $^{VLA}$ : & \multicolumn{2}{l}{synthesis observation with the VLA.} \\
References : & 1: Fisher \& Tully (\cite{fisher})           &  8: Schwarz (\cite{schwarz})            \\
             & 2: Appleton \& Davies (\cite{appleton})      &  9: Thuan \& Martin (\cite{thuan})      \\
             & 3: Richter \& Huchtmeier (\cite{richter91})  & 10: Magri (\cite{magri})                \\
             & 4: Oosterloo \& Shostak (\cite{oosterloo})   & 11: Van der Burg (\cite{vanderburg})    \\
             & 5: Huchtmeier \& Richter (\cite{huchtmeier}) & 12: Gottesman et al. (\cite{gottesman}) \\
             & 6: Schneider et al. (\cite{schneider})       & 13: Van Moorsel (\cite{vanmoorsel})     \\
             & 7: Jore et al. (\cite{jore})                 & 14: Grewing \& Mebold (\cite{grewing})  \\
\noalign{\vspace{0.8mm}}
\hline

\end{tabular}
\end{center}
\end{table*}


The raw UV$-$data were calibrated, interactively flagged and Fast
Fourier Transformed (FFT) using the NEWSTAR software developed at the
NFRA in \penalty-10000 Dwingeloo. The UV points were weighted according
to the local density of points in the UV plane and a Gaussian baseline
taper was applied with a FWHM of 2293 (m) which attenuates the longest
baseline by 50\%. To deal with the frequency dependent antenna pattern,
five antenna patterns were calculated for each data cube at a regular
frequency separation thoughout the bandpass.  Pixel sizes of 5 arcsec in
RA and $\frac{5}{\mbox{sin}(\delta)}$ arcsec in declination ensure an
adequate sampling of the synthesized beam, $12^{\prime\prime}\times
12^{\prime\prime}/\mbox{sin}(\delta)$. 

After the FFT, the datacubes and antenna patterns were further processed
using the Groningen Image Processing SYstem (GIPSY).  Several channels
at the low and high velocity end of the bandpass were discarded because
of their higher noise.  As a result, there are 110 or 53 usable channels
for a bandpass of 2.5 or 5 MHz respectively, except for the N3992 and
N4111 fields which had 110 channels across a 5 MHz bandpass.  All
datacubes were smoothed to lower angular resolutions of
$30^{\prime\prime}\times30^{\prime\prime}$ and
$60^{\prime\prime}\times60^{\prime\prime}$.  This facilitates the
detection of extended low level HI emission and the identification of
the `continuum' channels which are free from line emission.

\subsection{The radio continuum emission}

The channels free from HI emission were averaged and the resulting
continuum map was subtracted from all channels in the data cube.  The
residuals of the frequency dependent grating rings were only a minor
fraction of the noise in the channels containing the line emission.  The
dirty continuum maps were cleaned (H\"ogbom \cite{hogbom}) down to
0.3$\sigma$.  The clean-components were restored with a gaussian beam of
similar FWHM as the synthesized beam.  When radio continuum emission
from a galaxy was detected, its continuum flux was determined from the
cleaned continuum map. In cases of no detection, an upper limit for
extended emission was derived by calculating the rms scatter in the flux
values obtained by integrating the noise in each of fifteen elliptical
areas enclosed by the 25$^{\mbox{\scriptsize th}}$mag blue isophote and
positioned at various emission-free regions in the map.

\subsection{The HI channel maps and the global HI profiles}

At all three spatial resolutions, the regions of HI emission were
defined by the areas enclosed by the 2$\sigma$ contours in the `dirty'
60$^{\prime\prime}$ resolution maps.  Grating rings and noise peaks
above this level were removed manually.  The selected regions were
enlarged by moving their boundaries 1 armin outwards to account for
possible emission in the sidelobes.  The resulting masks vary from
channel to channel in shape, size and position due to the rotation of
the HI disk.  These masks defined the regions that were cleaned down to
0.3$\sigma$ at all three spatial resolutions. 

The clean-components were restored with a Gaussian beam of similar FWHM
as the synthesized beam. The global HI profiles were derived by
determining the primary beam corrected flux in each cleaned region. 
Since the size and shape of the clean masks vary as a function of
velocity, the uncertainty in the flux densities at each velocity in the
global HI profile varies as well.  The noise on the global HI profile
was determined by projecting each clean mask at nine different line-free
positions in a channel map and integrating over each of them. 

For further analysis, each profile was divided up in three equal
velocity bins in which the peak fluxes $F^{\mbox{\scriptsize
peak}}_{\mbox{\scriptsize low}}$, $F^{\mbox{\scriptsize
peak}}_{\mbox{\scriptsize mid}}$ and $F^{\mbox{\scriptsize
peak}}_{\mbox{\scriptsize high}}$ were determined for the low, middle
and high velocity bin respectively.  These three peak fluxes were then
used to classify a global profile shape according to:

\smallskip
\begin{tabular}{lr}
Double peaked & : \phantom{or} $F^{\mbox{\scriptsize peak}}_{\mbox{\scriptsize low}}$ $>$ $F^{\mbox{\scriptsize peak}}_{\mbox{\scriptsize mid}}$ $<$ $F^{\mbox{\scriptsize peak}}_{\mbox{\scriptsize high}}$ \\
\noalign{\vspace{0.8mm}}
Gaussian      & : \phantom{or} $F^{\mbox{\scriptsize peak}}_{\mbox{\scriptsize low}}$ $<$ $F^{\mbox{\scriptsize peak}}_{\mbox{\scriptsize mid}}$ $>$ $F^{\mbox{\scriptsize peak}}_{\mbox{\scriptsize high}}$ \\
\noalign{\vspace{0.8mm}}
Distorted     & : \phantom{or} $F^{\mbox{\scriptsize peak}}_{\mbox{\scriptsize low}}$ $<$ $F^{\mbox{\scriptsize peak}}_{\mbox{\scriptsize mid}}$ $<$ $F^{\mbox{\scriptsize peak}}_{\mbox{\scriptsize high}}$ \\
              &            or  $F^{\mbox{\scriptsize peak}}_{\mbox{\scriptsize low}}$ $>$ $F^{\mbox{\scriptsize peak}}_{\mbox{\scriptsize mid}}$ $>$ $F^{\mbox{\scriptsize peak}}_{\mbox{\scriptsize high}}$ \\
\noalign{\vspace{0.8mm}}
Boxy          & : \phantom{or} $F^{\mbox{\scriptsize peak}}_{\mbox{\scriptsize low}}$ $\approx$ $F^{\mbox{\scriptsize peak}}_{\mbox{\scriptsize mid}}$ $\approx$ $F^{\mbox{\scriptsize peak}}_{\mbox{\scriptsize high}}$ \\
\end{tabular}
\smallskip

\noindent In case of a double peaked profile, the peak fluxes on both
sides were considered separately when calculating the 20\% and 50\%
levels.  In all other cases, the overal peak flux was used.  The four
velocities V$_{low}^{20\%}$, V$_{low}^{50\%}$, V$_{high}^{50\%}$ and
V$_{high}^{20\%}$ corresponding to these 20\% and 50\% levels were
determined by linear interpolation between the data points, going from
the center outward.  In the few cases of non-monotonically decreasing
edges, this procedure tends to slightly underestimate the widths.  The
widths are calculated according to \\

W$_{20}$ = V$_{\mbox{\scriptsize high}}^{20\%}$ $-$ V$_{\mbox{\scriptsize low}}^{20\%}$ and
W$_{50}$ = V$_{\mbox{\scriptsize high}}^{50\%}$ $-$ V$_{\mbox{\scriptsize low}}^{50\%}$.\\

\noindent The systemic velocity is calculated according to \\

V$_{\mbox{\scriptsize sys}}$ = 0.25 ( V$_{\mbox{\scriptsize low}}^{20\%}$+V$_{\mbox{\scriptsize low}}^{50\%}$+V$_{\mbox{\scriptsize high}}^{50\%}$+V$_{\mbox{\scriptsize high}}^{20\%}$ )\\

Because in interferometric measurements some flux may be lost due to the
missing short baselines, it is useful to compare the widths and flux
densities from the WSRT profiles with those from published single dish
observations.  However, a meaningful comparison requires that the
profile widths are all corrected in the same way for instrumental
broadening.  In general, the widths that are published by various
authors were corrected for instrumental broadening using nearly as many
different methods.  Therefore, the published line widths had to be
de-corrected first to ensure a uniformly applied correction.  The
de-corrected widths and integrated HI fluxes from the literature are
compiled in columns (7)$-$(11) of Table 3 along with the results from
this study in columns (2)$-$(6).  \\
 \noindent{\it Column} (1) gives the NGC or UGC numbers.  \\ 
 \noindent{\it Columns} (2,3) and (7,8) give the widths of the global
profiles at the 20\% levels and the formal uncertainties.  \\ 
 \noindent{\it Columns} (4) and (9) give the velocity resolutions of the
observations.  \\ 
 \noindent{\it Columns} (5,6) and (10,11) contain the integrated HI
fluxes derived from the global profiles. \\
 \noindent{\it Column} (12) provides the references to the literature
sources.  \\ 
 \noindent In case the authors suggest that the single dish profile of a
particular galaxy may be confused and synthesis data on that galaxy do
exist, these synthesis data are included as well and used in the
following comparison. However, first it will be explained how the
observed linewidths are corrected for the different instrumental
resolutions.

\subsubsection{Correcting W$_{20}$ for instrumental broadening}

The most widely used method to correct for broadening of the global HI
profiles due to a finite instrumental velocity resolution was provided
by Bottinelli et al. (\cite{bottinelli90}).  For the widths at the 20\%
and 50\% levels of the peak flux they advocate the following linear
relations:

\begin{center}
\setlength{\tabcolsep}{1.2mm}
\begin{tabular}{lclcl}
 W$_{20,R}$ & $=$ & W$_{20}$ $-$ $\delta$W$_{20}$ & $=$ & W$_{20}$ $-$ 0.55 $R$ \\
 W$_{50,R}$ & $=$ & W$_{50}$ $-$ $\delta$W$_{50}$ & $=$ & W$_{50}$ $-$ 0.13 $R$ \\
\end{tabular}
\end{center}

\noindent where W$_{20}$ is the observed linewidth and W$_{20,R}$ is
the linewidth corrected for the instrumental velocity resolution $R$ in
km/s. This empirical prescription is based on comparing
linewidths obtained at different resolutions.

However, the correction method applied here deviates from Bottinelli et
al.'s method and is based on more analytic considerations. It is easy to
imagine that both edges of an intrinsic global profile, when chopped off
at their peaks and glued together, approximate a Gaussian with
dispersion $\sigma_0$.  The width at the 20\% level of this `true'
Gaussian is then given by

    $$\mbox{W}_{20,R} = \sigma_0\sqrt{8 \mbox{ln}(5)}$$

\noindent A spectral Hanning smoothing was applied to most of the WSRT
observations presented in this paper. This smoothing function can also
be approximated by a Gaussian with a FWHM equal to the instrumental
velocity resolution $R$ and has a dispersion $\sigma_R$ 

    $$\sigma_{\mbox{\scriptsize R}} = \frac{R}{\sqrt{8 \mbox{ln}(2)}}$$

\noindent The dispersion $\sigma_c$ of the convolved observed Gaussian
is then given by

    $$\sigma_c = \sqrt{ \sigma_0^2 + \sigma_R^2 }$$

\noindent and the 20\% line width of this convolved or observed Gaussian
is given by

\begin{eqnarray*}
 \mbox{W}_{20} & = & \sigma_c \sqrt{8\mbox{ln}(5)} \\
               & = & \sqrt{8\mbox{ln}(5)}\cdot\sqrt{\sigma_0^2 + \sigma_R^2 } \\
               & = & \sqrt{8\mbox{ln}(5)}\cdot\sqrt{\sigma_0^2 + \frac{R^2}{8\mbox{ln}(2)} }
\end{eqnarray*}

\noindent So, at the 20\% level, the intrinsic width $\mbox{W}_{20,R}$ is
broadened to $\mbox{W}_{20}$ by $\delta\mbox{W}$ given by

\begin{eqnarray*}
 \delta\mbox{W}_{20} & = & \mbox{W}_{20} - \mbox{W}_{20,R} \\
                     & = & \sqrt{8\mbox{ln}(5)}\cdot\sqrt{\sigma_0^2 + \frac{R^2}{8\mbox{ln}(2)} } - \sigma_0 \sqrt{8\mbox{ln}(5)} \\
                     & = & \sigma_0 \sqrt{8\mbox{ln}(5)}\cdot\left[\sqrt{1 + \frac{(R/\sigma_0)^2}{8\mbox{ln}(2)} } - 1 \right]
\end{eqnarray*}

\noindent The broadening $\delta$W$_{20}$ does not only depend on the
instrumental resolution $R$ but also on the steepness of the slopes of the
edges of the profile, expressed by $\sigma_0$.  Fitting Gaussians to the
edges of a profile yields $\sigma_c$ from which $\sigma_0$ can be
calculated given the known value of $\sigma_{\mbox{\scriptsize R}}$. 
The equation above can be rewritten using $\sigma_c$ instead which
results in

    $$\delta\mbox{W}_{20} = \sigma_c\sqrt{8\mbox{ln}(2)}\left(\frac{\mbox{ln}(5)}{\mbox{ln}(2)}\right)^2\left[1-\sqrt{1-\frac{(R/\sigma_c)^2}{8\mbox{ln}(2)}}\;\right]$$

\noindent However, no Gaussians were fitted to the edges of the new WSRT
profiles.  Instead it is assumed that the slopes of the edges of the
profiles are more or less determined by the turbulent motion of the gas
with a canonical velocity dispersion of $\sigma_0$=10 km$\;$s$^{-1}$. 
This results in

    $$ \delta\mbox{W}_{20} = 35.8\cdot\left[\sqrt{1 + \left(\frac{R}{23.5}\right)^2} - 1 \right] $$

\noindent and similarly for the 50\% level

    $$ \delta\mbox{W}_{50} = 23.5\cdot\left[\sqrt{1 + \left(\frac{R}{23.5}\right)^2} - 1 \right] $$

The differences between Bottinelli et al.'s linear presciption and our
corrections
($\Delta\delta\mbox{W}=\delta\mbox{W}^{Bot}-\delta\mbox{W}^{our}$) are
only minor and tabulated below for typical instrumental resolutions of
the WSRT.

\begin{center}
\begin{tabular}{c|rrrrr}
level      & \multicolumn{5}{c}{$\Delta\delta$W} \\
\hline
           & \multicolumn{5}{c}{-- -- -- -- -- R (km/s) -- -- -- -- -- } \\
           &  5.0 &  8.3 & 16.5 & 19.9 &  33.1 \\
\hline
20\%       &  2.0 &  2.4 &  1.2 &  0.2 &  -7.8 \\
50\%       &  0.2 & -0.3 & -3.1 & -4.7 & -12.8 \\
\end{tabular}
\end{center}

\noindent The larger differences occur for the poorest resolutions at
which only the broadest profiles were observed. Consequently, the
differences are a negligible fraction of the line widths.

Figure 1 shows the comparison of the widths and integrated fluxes
derived from the new WSRT global profiles and those from the literature. 
There are no significant systematic differences.  The unweighted average
difference in widths is -0.9 $\pm$ 2.1 km$\;$s$^{-1}$ with a rms scatter
of 14 km$\;$s$^{-1}$.  The unweighted average difference in integrated
flux is 4.7 $\pm$ 3.5 percent with a rms scatter of 25 percent.  It can
therefore be concluded that on average the WSRT results are in excellent
agreement with the results from single dish observations. 

\begin{figure}
  \resizebox{\hsize}{!}{\includegraphics{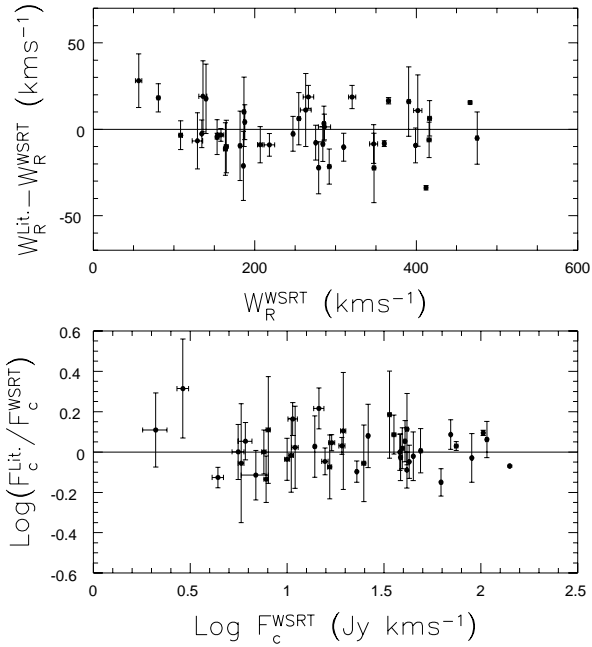}}
  \caption{A comparison of the present WSRT results with pre-existing
single dish and synthesis data from the literature.}
  \label{UMaIVfig1}
\end{figure}

\subsection{The total HI maps}

As a next step, the total integrated HI maps were constructed from the
{\it clean}ed datacubes. The {\it clean}-masks were used to define the
regions with HI emission. Outside these regions, the pixels were set to
zero and all the channels containing a non-zero area were added to build
up the integrated column density map. This was then corrected for
attenuation by the primary beam. Although the advantage of this
procedure is a higher signal-to-noise ratio at a certain pixel in the HI
map, the disadvantage is that the noise is no longer uniform across the
map. As a result, the 3$\sigma$-contour level in an integrated HI map is
not defined. Signal-to-noise maps have been made, however, using the
prescription outlined in Appendix A and the average pixel value of all
pixels with $2.75<(\frac{S}{N})<3.25$ was determined. This average value
was adopted as the `3$\sigma$' level for the column density. 

\subsection{The radial HI surface density profiles}

The integrated column density maps were used to derive the radial HI
surface density profiles by azimuthally averaging in concentric
ellipses. The orientations and widths of the ellipses were the same as
those of the projected tilted rings fitted to the HI velocity field (see
Section~3.6.1). In the case of a warp with overlapping ellipses, the
flux in the overlapping regions was proportionally assigned to each
ellipse. The azimuthal averaging was done separately for the receding
and approaching halves of each tilted ring to reveal possible
asymmetries. Pixels in the HI map without any measured signal were set
to zero. Finally, the entire radial profile was scaled by the total HI
mass as derived from the global HI profile. No attempt was made to
correct the profiles for the effect of beam smearing. 

This method for extracting the surface density profiles from integrated
HI maps breaks down for nearly edge-on systems; the highly inclined
annuli with large major axis diameters could still pick up some flux
along the minor axis due to beam smearing. In such cases, Lucy's
(\cite{lucy}) iterative deprojection scheme as adapted and developed by
Warmels (\cite{warmels88b}) might be preferable. 

Due to the complex noise structure of the integrated HI map, no attempt
was made to estimate the errors on the radial HI surface density
profiles.

\subsection{The HI velocity fields}

The HI data cubes were smoothed to velocity resolution of $\approx$19
km$\;$s$^{-1}$ in order to obtain a good spectral signal-to-noise ratio.
 HI velocity fields were then constructed by fitting single Gaussians to
the velocity profiles at each pixel. Initial estimates for the fits were
given by the various moments of the profiles determined over the
velocity range covered by the masks. Only those fits were accepted for
which 1) the central velocity of the fitted Gaussian lies inside the
masked volume, 2) the amplitude is larger than five times the rms noise
in the profile and 3) the uncertainty in the central velocity is smaller
than $\frac{1}{3}$ the velocity resolution. 

Due to projection effects and beam smearing, the velocity profiles in
highly inclined systems and in the central regions of galaxies may
deviate strongly from a Gaussian shape. The exact shape depends on the
spatial and kinematic distribution of the gas within a synthesized beam.
Fitting single Gaussians to these usually skewed profiles results in an
underestimate of the rotational velocity at that position. As a
consequence, the gradients in the velocity field become shallower. 
There are several methods to correct for the effects of beam smearing. 
In the present cases, however, the signal-to-noise was in general too
low to allow a useful application of these methods, and, since only a
small number of systems were recognized as seriously affected, the HI
velocity fields were not corrected for the effects of beam smearing.

\subsection{Rotation curves}

The rotation curves were derived in two ways; 1) by fitting tilted-rings
to the velocity fields (Begeman \cite{begeman87}) and 2) by estimating
the rotational velocities by eye from the position-velocity diagrams. 

\subsubsection{using the velocity fields}

The determination of the rotation curves from the velocity fields was
done in three steps by fitting tilted rings to the velocity field (see
Begeman \cite{begeman87}, \cite{begeman89}). The widths of the rings
were set at $\frac{2}{3}$ of the width of the synthesized beam (i.e.
10$^{\prime\prime}$, 20$^{\prime\prime}$ or 40$^{\prime\prime}$). 

First, the systemic velocity and the dynamical center were determined. 
In this first step the inclination and position angles were the same for
each ring and kept fixed at the values derived from the optical images. 
The systemic velocity, center and rotational velocity were fitted for
each ring. All the points along the tilted ring were considered and
weighted uniformly.  In general, no significant trend as a function of
radius could be detected for the systemic velocity and center.  The
adopted values were calculated as the average of all rings. 

Second, the systemic velocity and center of rotation were kept fixed for
each ring while the position angle, inclination and rotational velocity
were fitted. All the points along the tilted ring were considered but
weighted with cos($\theta$) where $\theta$ is the angle in the plane of
the galaxy measured from the receding side. Hence, points along the
minor axis have zero weight. While the position angle can be determined
accurately, the inclination and rotational velocity are rather strongly
correlated for inclinations below 60 degrees and above 80 degrees
(Begeman \cite{begeman89}). As a result, the fitted inclinations can
vary by a large amount from one ring to another. However, a possible
trend in the inclination with radius due to a central bar or a warp can
be detected. A change in inclination angle often goes together with a
change in the more accurately determined position angle. 

Third, the rotational velocity was fitted again for each ring while
keeping the systemic velocity, center of rotation, inclination and
position angle fixed. Again, all the points along the tilted ring were
considered but weighted with cos($\theta$). The fixed values for the
inclination and position angles were determined in the second step by
averaging the solutions over all the rings or fixing a clear trend. For
nearly edge-on galaxies, the inclinations determined in the second step
were often overruled by higher values based on the clear presence of a
dust lane (e.g. N4010, N4157, N4217) or the very thin distribution of
gas in the column density maps (e.g U6667). However, uncertainty in the
inclinations of nearly edge-on systems does not significantly influence
the amplitude of the rotational velocity. 

The results of this 3-step procedure were used to construct a model
velocity field. This model was subtracted from the actual observed
velocity field to yield a map of the residual velocities. In some cases
(e.g. N3769, N4051, N4088) this residual map shows significant
systematic residuals, indicative of non-circular motion or a bad model
fit due to a noisy observed velocity field. As a further check, the
derived rotation curve is projected onto the position-velocity maps
along the major and minor axis. 

The errors on the inclination and position angles and the rotational
velocity are formal errors. They do not include possible systematic
uncertainties due to, for instance, the beam smearing.

\subsubsection{using the position-velocity diagrams}

It has already been remarked (see Section 3.5) that beam smearing
affects the determination of the velocity fields, especially in the
central regions of galaxies and in highly inclined disks.  As a
consequence, the rotation curves derived from such velocity fields are
underestimated as one can see from their projection on the XV-maps.  In
order to overcome this problem, the rotation curves were derived
directly from the major axis XV-maps in a manner similar to that used
for edge-on systems (cf. Sancisi \& Allen \cite{sancisi}). This was done
by two independent human neural networks trained to estimate the maximum
rotational velocity from the asymmetric velocity profiles, taking into
account the instrumental band- and beam-widths and the random gas
motions. This was done for both the receding and approaching side of a
galaxy. The rotation curves were then deprojected (also accounting for
possible warps) by using the same position and inclination angles as
fixed in the third step described in the previous section. In general,
the average rotation curves derived from the XV-diagrams are in
reasonable agreement with those obtained by the tilted ring fits. As
expected, significant differences can only be noted for galaxies which
are highly inclined or have a steeply rising rotation curve. 

From the XV-diagrams it is clear that many galaxies have kinematic
asymmetries in the sense that the rotation curve often rises more
steeply on one side of a galaxy than on the other side (e.g.~N3877,
N3949). The rotation curves as derived from the velocity fields and
XV-diagrams are tabulated in Table~4 for the approaching and receding
parts separately. The adopted changes in inclination and position angles
of N3718 and N4138 are motivated in the notes on the atlas pages of
these galaxies. The uncertainties quoted in Table~4 are not 1-sigma
Gaussian errors but rather reflect fiducial velocity ranges, based on
the position-velocity diagrams.

\begin{table}[pht]
\caption[]{Rotation curves derived from velocity fields and \\
XV-diagrams.}
{\fontsize{9}{10}
\selectfont
\setlength{\tabcolsep}{1.7mm}
\begin{tabular}[t]{rrrrrrrrrr}
\hline
\hline
\noalign{\vspace{0.8mm}}
\multicolumn{1}{c}{Rad.} & V$^{\mbox{\scriptsize app}}_{\mbox{\scriptsize rot}}$ & \multicolumn{2}{c}{+/$-$} & V$^{\mbox{\scriptsize rec}}_{\mbox{\scriptsize rot}}$ & \multicolumn{2}{c}{+/$-$} & \multicolumn{1}{c}{V$^{\mbox{\scriptsize ave}}_{\mbox{\scriptsize rot}}$} & \multicolumn{1}{c}{i} & PA \\
\multicolumn{1}{c}{($^{\prime\prime}$)} & \multicolumn{3}{c}{ - - - km/s - - - } & \multicolumn{3}{c}{ - - - km/s - - - } & km/s & ($^\circ$) & ($^\circ$) \\
\noalign{\vspace{0.8mm}}
\hline
\noalign{\vspace{0.8mm}}

\multicolumn{10}{l}{\it \underline{U6399}} \\
  10 &  25 & 12 &  7 &  25 & 10 &  7 &  25 & 75 & 141 \\
  20 &  44 & 10 &  7 &  49 &  7 &  7 &  46 & 75 & 141 \\
  30 &  61 & 12 &  7 &  61 &  7 &  5 &  61 & 75 & 141 \\
  40 &  70 &  7 &  5 &  69 &  5 &  5 &  70 & 75 & 141 \\
  50 &  77 &  7 &  5 &  78 &  3 &  5 &  78 & 75 & 141 \\
  60 &  82 & 10 &  5 &  84 &  5 &  5 &  83 & 75 & 141 \\
  70 &  84 &  5 &  5 &   - &  - &  - &  84 & 75 & 141 \\
  80 &  86 &  5 &  5 &   - &  - &  - &  86 & 75 & 141 \\
  90 &  88 &  5 &  5 &   - &  - &  - &  88 & 75 & 141 \\

\multicolumn{10}{l}{\it \underline{U6446}} \\
  10 &  39 &  8 &  8 &  23 & 10 & 10 &  31 & 51 & 188 \\
  20 &  49 &  8 &  8 &  61 &  8 & 10 &  55 & 51 & 188 \\
  30 &  57 &  5 &  5 &  65 &  5 &  8 &  61 & 51 & 188 \\
  40 &  63 &  5 &  5 &  65 &  5 &  8 &  64 & 51 & 188 \\
  50 &  69 &  8 &  5 &  65 &  5 &  5 &  67 & 51 & 188 \\
  60 &  71 &  5 &  5 &  70 &  5 &  5 &  70 & 51 & 188 \\
  70 &  75 &  5 &  5 &  72 &  5 &  5 &  74 & 51 & 188 \\
  80 &  79 &  8 &  5 &  77 &  5 &  5 &  78 & 51 & 188 \\
  90 &  81 &  8 &  5 &  81 &  5 &  5 &  81 & 51 & 188 \\
 100 &  81 &  5 &  5 &  80 &  5 &  5 &  81 & 51 & 188 \\
 110 &  81 &  5 &  5 &  82 &  5 &  5 &  81 & 51 & 189 \\
 120 &  82 &  8 &  5 &  83 &  5 &  8 &  82 & 51 & 191 \\
 131 &  82 &  8 &  5 &  84 &  5 &  8 &  83 & 51 & 193 \\
 142 &  83 &  8 &  8 &  86 &  8 &  8 &  85 & 51 & 195 \\
 153 &  83 &  8 &  8 &  86 &  8 &  8 &  84 & 51 & 197 \\
 164 &  82 & 11 & 11 &  85 &  8 & 11 &  83 & 51 & 199 \\
 176 &  80 & 11 & 11 &   - &  - &  - &  80 & 51 & 201 \\
 
\multicolumn{10}{l}{\it \underline{N3726}} \\
  40 & 112 & 10 &  7 &  92 & 12 & 10 & 102 & 53 & 195 \\
  60 & 131 &  7 &  7 & 119 & 10 & 10 & 125 & 53 & 195 \\
  80 & 144 &  5 &  5 & 146 &  7 & 10 & 145 & 53 & 195 \\
 100 & 156 &  5 &  5 & 172 &  5 &  7 & 164 & 53 & 195 \\
 120 & 154 &  7 &  7 & 171 &  5 &  7 & 162 & 53 & 195 \\
 140 & 155 &  7 &  5 & 166 &  7 &  7 & 160 & 53 & 195 \\
 160 & 152 &  7 &  7 & 159 &  5 &  7 & 156 & 53 & 195 \\
 183 & 145 & 10 &  7 & 148 &  5 &  5 & 147 & 57 & 188 \\
 256 & 157 &  8 &  8 & 159 &  6 &  6 & 158 & 72 & 180 \\
 316 & 169 &  9 & 12 &   - &  - &  - & 169 & 75 & 179 \\
 344 & 169 &  9 & 12 &   - &  - &  - & 169 & 75 & 179 \\
 373 & 167 & 15 & 15 &   - &  - &  - & 167 & 75 & 179 \\

\multicolumn{10}{l}{\it \underline{N3769}} \\
  20 &  89 & 13 & 10 &  86 & 20 & 10 &  88 & 70 & 149 \\
  40 & 103 & 10 &  8 & 109 & 13 & 10 & 106 & 70 & 149 \\
  60 & 112 &  8 &  8 & 119 &  8 &  8 & 116 & 70 & 149 \\
  80 & 120 &  8 &  8 & 130 &  8 & 10 & 125 & 70 & 149 \\
 100 & 123 &  5 &  8 & 129 &  5 &  8 & 126 & 70 & 149 \\
 120 & 124 &  5 &  8 & 122 &  5 &  8 & 123 & 70 & 150 \\
 141 & 120 &  5 &  5 & 115 &  8 &  8 & 118 & 70 & 152 \\
 166 & 120 &  8 & 10 & 110 &  8 & 10 & 115 & 70 & 155 \\
 196 & 122 & 14 & 17 &   - &  - &  - & 122 & 70 & 158 \\
 364 & 121 & 10 & 10 &   - &  - &  - & 121 & 70 & 167 \\
 396 & 118 & 10 & 10 &   - &  - &  - & 118 & 70 & 167 \\
 426 & 113 & 11 & 11 &   - &  - &  - & 113 & 70 & 168 \\

\multicolumn{10}{l}{\it \underline{U6667}} \\
  10 &  27 &  5 &  5 &  27 &  7 &  7 &  27 & 89 &  89 \\
  20 &  43 &  2 &  2 &  47 &  5 &  5 &  45 & 89 &  89 \\
  30 &  55 &  5 &  5 &  59 &  7 &  5 &  57 & 89 &  89 \\
  40 &  64 &  2 &  5 &  74 &  5 &  5 &  69 & 89 &  89 \\

\hline
\hline
\end{tabular}}
\end{table}

\addtocounter{table}{-1}
\begin{table}[pht]
\caption[]{continued}
{\fontsize{9}{10}
\selectfont
\setlength{\tabcolsep}{1.7mm}
\begin{tabular}[t]{rrrrrrrrrr}
\hline
\hline
\noalign{\vspace{0.8mm}}
\multicolumn{1}{c}{Rad.} & V$^{\mbox{\scriptsize app}}_{\mbox{\scriptsize rot}}$ & \multicolumn{2}{c}{+/$-$} & V$^{\mbox{\scriptsize rec}}_{\mbox{\scriptsize rot}}$ & \multicolumn{2}{c}{+/$-$} & \multicolumn{1}{c}{V$^{\mbox{\scriptsize ave}}_{\mbox{\scriptsize rot}}$} & \multicolumn{1}{c}{i} & PA \\
\multicolumn{1}{c}{($^{\prime\prime}$)} & \multicolumn{3}{c}{ - - - km/s - - - } & \multicolumn{3}{c}{ - - - km/s - - - } & km/s & ($^\circ$) & ($^\circ$) \\
\noalign{\vspace{0.8mm}}
\hline
\noalign{\vspace{0.8mm}}

\multicolumn{10}{l}{\it \underline{U6667 (cont.)}} \\
  50 &  73 &  5 &  5 &  82 &  5 &  5 &  77 & 89 &  89 \\
  60 &  78 &  5 &  5 &  84 &  5 &  7 &  81 & 89 &  89 \\
  70 &  82 &  2 &  5 &  87 &  5 &  5 &  84 & 89 &  89 \\
  80 &  83 &  2 &  5 &  87 &  5 &  5 &  85 & 89 &  89 \\
  90 &  83 &  5 &  5 &  89 &  5 &  5 &  86 & 89 &  89 \\

\multicolumn{10}{l}{\it \underline{N3877}} \\
  10 &  35 & 10 & 10 &  40 & 15 & 15 &  38 & 76 &  37 \\
  20 &  81 & 10 & 10 &  80 & 15 & 15 &  80 & 76 &  37 \\
  30 & 129 & 10 & 10 & 113 & 15 & 12 & 121 & 76 &  37 \\
  40 & 150 &  8 &  8 & 134 & 12 & 12 & 142 & 76 &  37 \\
  50 & 157 &  8 &  8 & 149 & 10 & 10 & 153 & 76 &  37 \\
  60 & 161 &  8 &  8 & 159 &  8 &  8 & 160 & 76 &  37 \\
  70 & 163 &  8 &  8 & 163 &  8 &  8 & 162 & 76 &  37 \\
  80 & 164 &  8 &  8 & 171 &  5 &  5 & 167 & 76 &  37 \\
  90 & 163 &  8 &  8 & 174 &  8 &  8 & 169 & 76 &  37 \\
 100 & 164 &  8 &  8 & 177 &  8 &  5 & 171 & 76 &  37 \\
 110 & 165 &  8 &  8 & 176 &  8 &  5 & 171 & 76 &  37 \\
 120 & 165 &  8 &  8 & 174 & 10 &  8 & 170 & 76 &  37 \\
 130 & 166 & 10 & 10 & 171 & 10 & 10 & 169 & 76 &  37 \\

\multicolumn{10}{l}{\it \underline{N3893}} \\
  20 & 140 & 10 & 10 & 150 & 10 & 10 & 145 & 49 & 345 \\
  40 & 175 & 10 & 10 & 173 & 10 & 10 & 174 & 49 & 345 \\
  60 & 191 & 10 & 10 & 197 & 10 &  7 & 194 & 49 & 345 \\
  80 & 192 &  7 &  7 & 189 & 10 &  7 & 191 & 49 & 346 \\
 101 & 192 & 10 &  8 & 181 &  8 &  8 & 186 & 47 & 351 \\
 125 & 194 &  8 & 11 & 182 &  8 & 11 & 188 & 45 & 362 \\
 151 & 190 & 10 & 13 & 184 & 10 & 16 & 187 & 43 & 371 \\
 176 & 179 & 11 & 15 &   - &  - &  - & 179 & 41 & 377 \\
 198 & 161 & 15 & 12 & 190 & 19 & 19 & 176 & 39 & 379 \\
 218 & 153 & 12 & 12 & 181 & 24 & 24 & 167 & 37 & 380 \\
 233 & 148 & 21 & 17 &   - &  - &  - & 148 & 36 & 381 \\

\multicolumn{10}{l}{\it \underline{N3917}} \\
  10 &  21 &  5 &  5 &  28 &  7 &  5 &  24 & 80 & 257 \\
  20 &  45 &  5 &  5 &  54 &  7 &  5 &  50 & 80 & 257 \\
  30 &  69 &  7 &  7 &  78 &  7 &  7 &  74 & 80 & 257 \\
  40 &  99 &  7 &  5 & 103 &  5 &  7 & 101 & 80 & 257 \\
  50 & 102 &  5 &  5 & 113 &  5 &  5 & 107 & 80 & 257 \\
  60 & 108 &  5 &  5 & 122 &  5 &  5 & 115 & 80 & 257 \\
  70 & 118 &  5 &  5 & 128 &  5 &  7 & 123 & 80 & 257 \\
  80 & 127 &  5 &  5 & 134 &  5 &  7 & 130 & 80 & 257 \\
  90 & 133 &  5 &  5 & 134 &  5 &  5 & 134 & 79 & 257 \\
 100 & 137 &  7 &  5 & 136 &  5 &  5 & 136 & 78 & 257 \\
 110 & 137 &  7 &  5 & 136 &  5 &  5 & 137 & 77 & 257 \\
 120 & 137 &  5 &  5 & 136 &  5 &  5 & 136 & 77 & 257 \\
 130 & 137 &  5 &  5 & 137 &  5 &  5 & 137 & 76 & 257 \\
 140 & 137 &  5 &  5 & 138 &  5 &  5 & 137 & 75 & 257 \\
 150 & 137 &  5 &  5 &   - &  - &  - & 137 & 75 & 257 \\
 160 & 138 &  5 &  5 &   - &  - &  - & 138 & 74 & 257 \\
 170 & 137 &  8 &  8 &   - &  - &  - & 137 & 73 & 257 \\

\multicolumn{10}{l}{\it \underline{N3949}} \\
  10 &  58 & 10 & 10 &  79 & 14 & 14 &  68 & 55 & 298 \\
  20 & 106 & 10 &  7 & 141 &  7 &  7 & 123 & 55 & 298 \\
  30 & 138 &  7 & 10 & 152 &  7 &  7 & 145 & 55 & 298 \\
  40 & 150 &  5 &  7 & 155 & 10 & 12 & 152 & 55 & 298 \\
  50 & 156 &  7 &  7 & 159 &  7 & 14 & 157 & 55 & 297 \\
  60 & 161 &  5 &  5 & 161 & 10 & 24 & 161 & 55 & 295 \\
  70 & 165 &  7 &  7 & 165 &  7 & 34 & 165 & 55 & 294 \\
  81 &   - &  - &  - & 169 &  7 & 44 & 169 & 55 & 293 \\

\hline
\hline
\end{tabular}}
\end{table}

\addtocounter{table}{-1}
\begin{table}[pht]
\caption[]{continued}
{\fontsize{9}{10}
\selectfont
\setlength{\tabcolsep}{1.7mm}
\begin{tabular}[t]{rrrrrrrrrr}
\hline
\hline
\noalign{\vspace{0.8mm}}
\multicolumn{1}{c}{Rad.} & V$^{\mbox{\scriptsize app}}_{\mbox{\scriptsize rot}}$ & \multicolumn{2}{c}{+/$-$} & V$^{\mbox{\scriptsize rec}}_{\mbox{\scriptsize rot}}$ & \multicolumn{2}{c}{+/$-$} & \multicolumn{1}{c}{V$^{\mbox{\scriptsize ave}}_{\mbox{\scriptsize rot}}$} & \multicolumn{1}{c}{i} & PA \\
\multicolumn{1}{c}{($^{\prime\prime}$)} & \multicolumn{3}{c}{ - - - km/s - - - } & \multicolumn{3}{c}{ - - - km/s - - - } & km/s & ($^\circ$) & ($^\circ$) \\
\noalign{\vspace{0.8mm}}
\hline
\noalign{\vspace{0.8mm}}

\multicolumn{10}{l}{\it \underline{N3953}} \\
  40 & 178 &  7 & 10 & 184 & 15 & 10 & 181 & 62 &  13 \\
  60 & 200 & 10 & 10 & 207 & 10 &  7 & 203 & 62 &  13 \\
  80 & 214 &  7 & 10 & 219 &  7 &  7 & 217 & 62 &  13 \\
 100 & 218 &  7 & 10 & 227 & 10 &  7 & 223 & 62 &  13 \\
 120 & 219 &  7 &  7 & 229 & 10 &  7 & 224 & 62 &  13 \\
 140 & 222 &  7 &  7 & 224 & 10 & 10 & 223 & 62 &  13 \\
 160 & 229 & 10 &  7 & 218 &  7 & 10 & 224 & 62 &  13 \\
 175 & 234 & 10 &  7 &   - &  - &  - & 234 & 62 &  13 \\
 180 &   - &  - &  - & 215 & 10 & 10 & 215 & 62 &  13 \\

\multicolumn{10}{l}{\it \underline{N3972}} \\
  10 &  24 &  5 &  5 &  55 & 10 &  7 &  40 & 77 & 297 \\
  20 &  68 &  7 &  7 &  78 &  7 &  7 &  73 & 77 & 297 \\
  30 &  86 & 12 & 10 &  93 &  7 &  7 &  89 & 77 & 297 \\
  40 & 101 & 10 &  7 & 103 &  5 &  7 & 102 & 77 & 297 \\
  50 & 111 &  7 &  5 & 110 &  5 &  7 & 110 & 77 & 297 \\
  60 & 117 &  7 &  5 & 116 &  5 &  7 & 116 & 77 & 297 \\
  70 & 122 &  5 &  5 & 124 &  7 &  7 & 123 & 77 & 297 \\
  80 & 129 &  5 &  5 & 131 &  5 &  7 & 130 & 77 & 297 \\
  86 & 131 &  7 &  7 &   - &  - &  - & 131 & 77 & 297 \\
  90 &   - &  - &  - & 134 &  5 &  7 & 134 & 77 & 297 \\
 100 &   - &  - &  - & 134 &  5 &  5 & 134 & 77 & 297 \\

\multicolumn{10}{l}{\it \underline{U6917}} \\
  20 &  60 &  5 &  5 &  59 &  5 &  5 &  59 & 56 & 123 \\
  30 &  71 &  5 &  8 &  72 &  5 &  5 &  71 & 56 & 123 \\
  40 &  86 &  8 &  5 &  83 &  8 &  5 &  85 & 56 & 123 \\
  50 &  96 &  8 &  8 &  91 &  5 &  5 &  94 & 56 & 123 \\
  60 &  98 &  5 &  5 &  97 &  5 &  5 &  98 & 56 & 123 \\
  70 & 100 &  5 &  5 & 100 &  5 &  5 & 100 & 56 & 123 \\
  80 & 101 &  8 &  5 & 101 &  5 &  5 & 101 & 56 & 123 \\
  90 & 105 &  5 &  5 & 102 &  5 &  5 & 103 & 56 & 123 \\
 100 & 110 &  5 &  5 & 101 &  5 &  8 & 105 & 56 & 123 \\
 110 & 116 &  7 &  7 & 104 &  5 &  7 & 110 & 57 & 123 \\
 120 &   - &  - &  - & 111 &  5 &  7 & 111 & 60 & 124 \\

\multicolumn{10}{l}{\it \underline{U6923}} \\
  11 &  41 &  6 &  9 &   - &  - &  - &  41 & 65 & 341 \\
  23 &  54 &  6 &  6 &   - &  - &  - &  54 & 65 & 341 \\
  34 &  70 &  6 &  6 &  76 &  9 &  6 &  73 & 65 & 341 \\
  44 &  80 &  6 &  6 &  77 &  6 &  6 &  78 & 65 & 344 \\
  53 &   - &  - &  - &  79 &  5 &  5 &  79 & 65 & 347 \\
  61 &   - &  - &  - &  81 &  5 &  5 &  81 & 65 & 350 \\

\multicolumn{10}{l}{\it \underline{U6930}} \\
  20 &  58 & 12 & 10 &  52 & 12 & 12 &  55 & 32 &  39 \\
  40 &  88 & 10 &  7 &  83 & 12 & 10 &  85 & 32 &  39 \\
  60 &  94 &  7 &  7 &  94 & 10 & 10 &  94 & 32 &  39 \\
  80 &  98 &  7 &  7 & 100 &  7 &  7 &  99 & 32 &  39 \\
 100 & 102 &  7 &  7 & 105 &  7 &  7 & 103 & 32 &  39 \\
 120 & 105 &  7 &  7 & 109 &  7 &  7 & 107 & 32 &  39 \\
 140 & 107 &  7 &  7 & 110 &  7 &  7 & 109 & 32 &  39 \\
 150 &   - &  - &  - & 110 &  7 &  7 & 110 & 32 &  39 \\
 160 & 108 &  7 &  7 &   - &  - &  - & 108 & 32 &  39 \\
 180 & 108 &  7 &  7 &   - &  - &  - & 108 & 32 &  39 \\
 190 & 108 &  7 &  7 &   - &  - &  - & 108 & 32 &  39 \\

\multicolumn{10}{l}{\it \underline{N3992}} \\
  80 & 253 &  7 & 10 & 244 & 10 & 12 & 249 & 56 & 248 \\
 120 & 264 &  7 & 10 & 265 &  7 & 12 & 264 & 56 & 248 \\
 160 & 273 &  7 & 10 & 272 &  7 &  7 & 272 & 56 & 248 \\
 200 & 274 &  7 &  7 & 268 &  7 & 10 & 271 & 56 & 248 \\
 240 & 273 &  7 &  7 & 256 &  7 &  7 & 264 & 56 & 248 \\
 280 &   - &  - &  - & 242 &  7 &  7 & 242 & 56 & 248 \\

\hline
\hline
\end{tabular}}
\end{table}

\addtocounter{table}{-1}
\begin{table}[pht]
\caption[]{continued}
{\fontsize{9}{10}
\selectfont
\setlength{\tabcolsep}{1.7mm}
\begin{tabular}[t]{rrrrrrrrrr}
\hline
\hline
\noalign{\vspace{0.8mm}}
\multicolumn{1}{c}{Rad.} & V$^{\mbox{\scriptsize app}}_{\mbox{\scriptsize rot}}$ & \multicolumn{2}{c}{+/$-$} & V$^{\mbox{\scriptsize rec}}_{\mbox{\scriptsize rot}}$ & \multicolumn{2}{c}{+/$-$} & \multicolumn{1}{c}{V$^{\mbox{\scriptsize ave}}_{\mbox{\scriptsize rot}}$} & \multicolumn{1}{c}{i} & PA \\
\multicolumn{1}{c}{($^{\prime\prime}$)} & \multicolumn{3}{c}{ - - - km/s - - - } & \multicolumn{3}{c}{ - - - km/s - - - } & km/s & ($^\circ$) & ($^\circ$) \\
\noalign{\vspace{0.8mm}}
\hline
\noalign{\vspace{0.8mm}}

\multicolumn{10}{l}{\it \underline{N3992 (cont.)}} \\
 320 & 247 &  7 &  7 & 242 &  7 &  7 & 244 & 56 & 248 \\
 360 & 241 &  7 &  7 & 242 & 10 & 10 & 241 & 56 & 248 \\
 400 & 237 &  7 & 10 &   - &  - &  - & 237 & 56 & 248 \\

\multicolumn{10}{l}{\it \underline{U6940}} \\
  10 &  19 &  5 &  5 &  18 &  5 &  5 &  18 & 79 & 315 \\
  20 &  41 &  5 &  5 &  34 &  5 &  8 &  37 & 79 & 315 \\

\multicolumn{10}{l}{\it \underline{U6962}} \\
  10 &  50 & 20 & 15 &  75 & 12 & 10 &  62 & 37 & 359 \\
  20 & 107 & 10 &  7 & 106 & 10 &  7 & 106 & 37 & 359 \\
  30 & 129 &  7 &  7 & 126 &  7 &  7 & 128 & 37 & 359 \\
  40 & 142 &  7 &  7 & 145 & 10 &  7 & 144 & 37 & 359 \\
  50 & 155 &  7 &  7 & 163 & 10 & 10 & 159 & 37 & 359 \\
  60 & 171 &  7 &  7 &   - &  - &  - & 171 & 37 & 359 \\

\multicolumn{10}{l}{\it \underline{N4010}} \\
   0 &  34 & 15 & 15 & -34 & 15 & 15 &   0 & 90 &  66 \\
  10 &  59 & 12 & 10 &  20 &  7 &  7 &  39 & 90 &  66 \\
  20 &  66 &  7 & 10 &  43 &  7 &  7 &  55 & 90 &  66 \\
  30 &  69 &  5 & 10 &  62 & 12 & 10 &  66 & 90 &  66 \\
  40 &  80 &  5 &  5 &  88 & 10 &  7 &  84 & 90 &  66 \\
  50 &  84 &  5 &  5 & 104 & 12 & 10 &  94 & 90 &  66 \\
  60 &  96 &  5 &  5 & 113 & 10 & 10 & 104 & 90 &  66 \\
  70 & 108 &  7 &  5 & 122 &  7 &  7 & 115 & 90 &  66 \\
  80 & 125 &  7 &  5 & 129 &  7 &  7 & 127 & 90 &  66 \\
  90 & 128 &  7 &  5 & 131 &  7 &  7 & 129 & 90 &  66 \\
 100 & 123 &  7 &  5 & 131 &  7 &  7 & 127 & 90 &  66 \\
 110 & 119 &  7 &  5 & 129 &  7 &  7 & 124 & 90 &  66 \\
 120 & 119 &  5 &  5 & 125 &  5 &  7 & 122 & 90 &  66 \\

\multicolumn{10}{l}{\it \underline{U6969}} \\
  10 &   - &  - &  - &  26 &  5 &  7 &  26 & 76 & 330 \\
  20 &  34 &  5 &  7 &  44 &  7 &  7 &  39 & 76 & 330 \\
  31 &  46 &  5 &  5 &  58 &  5 &  7 &  52 & 76 & 330 \\
  41 &  60 &  5 &  5 &  69 &  5 &  7 &  65 & 76 & 330 \\
  51 &   - &  - &  - &  79 &  5 &  5 &  79 & 76 & 330 \\

\multicolumn{10}{l}{\it \underline{U6973}} \\
  20 & 162 &  5 & 10 & 179 &  5 &  7 & 170 & 71 &  41 \\
  30 & 174 &  5 &  7 & 174 &  5 &  7 & 174 & 71 &  41 \\
  40 & 170 &  5 &  7 & 170 &  5 &  7 & 170 & 71 &  42 \\
  50 & 170 &  5 &  7 & 170 &  5 &  7 & 170 & 71 &  44 \\
  61 & 171 &  5 &  7 & 172 &  7 &  7 & 171 & 71 &  45 \\
  72 & 174 &  5 &  8 & 174 &  8 &  8 & 174 & 71 &  46 \\
  78 &   - &  - &  - & 177 & 10 & 10 & 177 & 71 &  47 \\
  84 & 178 &  5 &  8 &   - &  - &  - & 178 & 71 &  47 \\
  90 & 180 &  5 & 10 &   - &  - &  - & 180 & 71 &  48 \\

\multicolumn{10}{l}{\it \underline{U6983}} \\
  20 &  58 & 10 & 10 &  56 & 10 &  7 &  57 & 49 & 270 \\
  30 &  93 &  7 &  5 &  82 &  7 &  7 &  87 & 49 & 270 \\
  40 &  87 &  7 &  7 &  97 & 10 &  7 &  92 & 49 & 270 \\
  50 &  84 &  7 &  7 & 103 &  7 &  7 &  94 & 49 & 270 \\
  60 &  93 &  5 &  5 & 103 &  7 &  7 &  98 & 49 & 270 \\
  70 &  94 &  5 &  5 & 105 &  7 &  7 & 100 & 49 & 270 \\
  80 &  95 &  5 &  5 & 108 &  5 &  5 & 102 & 49 & 270 \\
  90 & 100 &  7 &  7 & 113 &  5 &  5 & 107 & 49 & 270 \\
 100 & 105 &  7 &  7 & 112 &  5 &  5 & 108 & 49 & 270 \\
 110 & 111 &  7 &  7 & 110 &  5 &  5 & 111 & 49 & 270 \\
 120 & 113 &  7 &  7 & 112 &  5 &  5 & 113 & 49 & 270 \\
 130 & 111 &  7 &  7 & 110 &  7 &  7 & 111 & 49 & 270 \\
 140 & 107 &  7 &  7 & 109 &  7 & 10 & 108 & 49 & 270 \\
 145 & 102 & 10 & 10 &   - &  - &  - & 102 & 49 & 270 \\
 150 &   - &  - &  - & 109 &  7 & 10 & 109 & 49 & 270 \\

\hline
\hline
\end{tabular}}
\end{table}

\addtocounter{table}{-1}
\begin{table}[pht]
\caption[]{continued}
{\fontsize{9}{10}
\selectfont
\setlength{\tabcolsep}{1.7mm}
\begin{tabular}[t]{rrrrrrrrrr}
\hline
\hline
\noalign{\vspace{0.8mm}}
\multicolumn{1}{c}{Rad.} & V$^{\mbox{\scriptsize app}}_{\mbox{\scriptsize rot}}$ & \multicolumn{2}{c}{+/$-$} & V$^{\mbox{\scriptsize rec}}_{\mbox{\scriptsize rot}}$ & \multicolumn{2}{c}{+/$-$} & \multicolumn{1}{c}{V$^{\mbox{\scriptsize ave}}_{\mbox{\scriptsize rot}}$} & \multicolumn{1}{c}{i} & PA \\
\multicolumn{1}{c}{($^{\prime\prime}$)} & \multicolumn{3}{c}{ - - - km/s - - - } & \multicolumn{3}{c}{ - - - km/s - - - } & km/s & ($^\circ$) & ($^\circ$) \\
\noalign{\vspace{0.8mm}}
\hline
\noalign{\vspace{0.8mm}}

\multicolumn{10}{l}{\it \underline{U6983 (cont.)}} \\
 160 &   - &  - &  - & 108 & 10 & 10 & 108 & 49 & 270 \\
 170 &   - &  - &  - & 108 & 10 & 10 & 108 & 49 & 270 \\
 180 &   - &  - &  - & 109 & 12 & 12 & 109 & 49 & 270 \\

\multicolumn{10}{l}{\it \underline{N4051}} \\
  20 &   - &  - &  - & 121 & 15 & 15 & 121 & 49 & 310 \\
  25 & 123 & 15 & 15 &   - &  - &  - & 123 & 49 & 310 \\
  40 & 119 & 12 & 10 & 114 & 12 & 10 & 116 & 49 & 310 \\
  60 & 146 & 10 & 10 & 133 & 10 & 10 & 140 & 49 & 310 \\
  80 & 163 &  7 & 10 & 156 & 10 & 10 & 160 & 49 & 310 \\
 100 & 158 &  7 &  7 & 165 &  7 &  7 & 162 & 49 & 310 \\
 115 &   - &  - &  - & 170 &  7 &  7 & 170 & 49 & 310 \\
 120 & 154 &  7 &  7 &   - &  - &  - & 154 & 49 & 310 \\
 140 & 153 & 10 & 10 &   - &  - &  - & 153 & 49 & 310 \\

\multicolumn{10}{l}{\it \underline{N4085}} \\
  10 &  35 & 10 &  5 &  50 & 15 & 10 &  42 & 82 & 256 \\
  20 &  71 &  7 &  5 &  89 & 10 & 10 &  80 & 82 & 256 \\
  31 & 110 & 12 &  7 & 113 &  7 &  7 & 112 & 82 & 256 \\
  41 & 126 &  7 &  5 & 127 &  7 &  7 & 127 & 82 & 256 \\
  51 & 131 &  7 &  7 & 130 &  7 &  5 & 130 & 82 & 256 \\
  61 & 134 &  7 &  7 & 133 &  5 &  5 & 133 & 82 & 256 \\
  71 & 136 &  7 &  7 &   - &  - &  - & 136 & 82 & 256 \\

\multicolumn{10}{l}{\it \underline{N4088}} \\
  20 &  92 & 15 & 15 &  78 & 20 & 15 &  85 & 69 & 230 \\
  40 & 138 & 10 & 15 & 135 & 20 & 10 & 136 & 69 & 230 \\
  60 & 156 &  7 & 12 & 168 & 15 & 10 & 162 & 69 & 230 \\
  80 & 167 &  7 & 10 & 191 & 10 & 10 & 179 & 69 & 230 \\
 100 & 177 &  7 & 10 & 187 & 10 &  7 & 182 & 69 & 230 \\
 120 & 185 &  7 & 12 & 174 & 10 & 10 & 179 & 69 & 230 \\
 140 & 187 &  7 & 12 & 162 &  7 &  7 & 174 & 69 & 230 \\
 160 & 185 & 12 & 12 & 158 &  7 &  7 & 171 & 69 & 230 \\
 180 & 175 & 10 & 10 & 161 &  7 &  7 & 168 & 69 & 230 \\
 200 & 171 &  7 &  7 & 160 & 10 &  7 & 165 & 69 & 230 \\
 210 &   - &  - &  - & 156 & 10 &  7 & 156 & 69 & 229 \\
 221 & 171 & 10 &  7 &   - &  - &  - & 171 & 69 & 227 \\
 246 & 174 &  8 &  8 &   - &  - &  - & 174 & 69 & 224 \\

\multicolumn{10}{l}{\it \underline{N4100}} \\
  20 &  67 & 15 & 15 &   - &  - &  - &  67 & 73 & 345 \\
  30 & 102 & 15 & 20 & 139 & 15 &  7 & 121 & 73 & 345 \\
  40 & 138 & 12 & 12 & 159 & 10 &  7 & 148 & 73 & 345 \\
  50 & 164 &  7 & 10 & 173 & 10 &  7 & 168 & 73 & 345 \\
  60 & 177 & 10 &  7 & 188 & 10 &  7 & 182 & 73 & 345 \\
  70 & 188 &  7 &  7 & 193 &  7 &  7 & 191 & 73 & 345 \\
  80 & 193 &  7 &  7 & 195 & 10 &  7 & 194 & 73 & 345 \\
  90 & 195 & 10 &  7 & 195 & 10 &  7 & 195 & 73 & 345 \\
 100 & 193 &  5 &  7 & 194 &  7 &  7 & 193 & 73 & 345 \\
 110 & 192 &  5 &  5 & 192 &  7 &  5 & 192 & 73 & 345 \\
 120 & 193 &  5 &  5 & 191 &  5 &  5 & 192 & 73 & 345 \\
 130 & 192 &  5 &  5 & 190 &  7 &  7 & 191 & 73 & 345 \\
 140 & 188 &  7 &  7 & 189 &  7 &  5 & 189 & 73 & 345 \\
 150 & 183 &  7 & 10 & 187 &  7 &  5 & 185 & 73 & 345 \\
 160 & 180 &  7 &  7 & 185 &  5 &  5 & 182 & 73 & 345 \\
 170 & 175 & 10 & 10 & 183 & 10 &  7 & 179 & 72 & 346 \\
 180 & 172 &  7 & 10 & 181 & 10 &  7 & 177 & 71 & 346 \\
 190 & 168 & 10 & 10 & 179 & 10 &  8 & 174 & 71 & 346 \\
 200 &   - &  - &  - & 178 & 10 &  8 & 178 & 70 & 346 \\
 210 &   - &  - &  - & 177 &  8 &  5 & 177 & 70 & 347 \\
 220 & 160 &  5 &  8 & 178 & 10 & 10 & 169 & 69 & 347 \\
 230 & 158 &  5 &  8 &   - &  - &  - & 158 & 69 & 347 \\
 241 & 158 &  8 &  8 &   - &  - &  - & 158 & 68 & 348 \\

\hline
\hline
\end{tabular}}
\end{table}

\addtocounter{table}{-1}
\begin{table}[pht]
\caption[]{continued}
{\fontsize{9}{10}
\selectfont
\setlength{\tabcolsep}{1.7mm}
\begin{tabular}[t]{rrrrrrrrrr}
\hline
\hline
\noalign{\vspace{0.8mm}}
\multicolumn{1}{c}{Rad.} & V$^{\mbox{\scriptsize app}}_{\mbox{\scriptsize rot}}$ & \multicolumn{2}{c}{+/$-$} & V$^{\mbox{\scriptsize rec}}_{\mbox{\scriptsize rot}}$ & \multicolumn{2}{c}{+/$-$} & \multicolumn{1}{c}{V$^{\mbox{\scriptsize ave}}_{\mbox{\scriptsize rot}}$} & \multicolumn{1}{c}{i} & PA \\
\multicolumn{1}{c}{($^{\prime\prime}$)} & \multicolumn{3}{c}{ - - - km/s - - - } & \multicolumn{3}{c}{ - - - km/s - - - } & km/s & ($^\circ$) & ($^\circ$) \\
\noalign{\vspace{0.8mm}}
\hline
\noalign{\vspace{0.8mm}}

\multicolumn{10}{l}{\it \underline{N4100 (cont.)}} \\
 251 & 158 &  8 & 10 &   - &  - &  - & 158 & 68 & 348 \\
 261 & 159 & 10 &  8 &   - &  - &  - & 159 & 67 & 348 \\

\multicolumn{10}{l}{\it \underline{N4102}} \\
  20 & 179 & 12 & 12 & 184 & 12 & 12 & 181 & 56 &  39 \\
  30 & 177 & 12 & 10 & 181 & 10 & 12 & 179 & 56 &  39 \\
  40 & 174 & 12 & 10 & 178 & 10 & 10 & 176 & 56 &  39 \\
  50 &   - &  - &  - & 178 & 15 & 10 & 178 & 56 &  39 \\

\multicolumn{10}{l}{\it \underline{N4157}} \\
  20 &  66 & 18 & 14 & 127 & 23 & 14 &  96 & 82 &  63 \\
  40 & 142 & 18 & 14 & 173 & 14 & 14 & 157 & 82 &  63 \\
  60 & 192 &  9 & 14 & 191 & 11 & 14 & 192 & 82 &  63 \\
  80 & 202 &  7 &  9 & 201 & 11 & 11 & 201 & 82 &  63 \\
 100 & 198 & 11 & 11 & 204 &  9 &  9 & 201 & 82 &  63 \\
 120 & 192 &  9 &  9 & 197 &  9 &  9 & 195 & 82 &  63 \\
 140 & 191 &  9 &  9 & 188 &  9 &  7 & 190 & 82 &  63 \\
 160 & 191 &  9 &  9 & 181 &  9 &  9 & 186 & 82 &  63 \\
 180 & 192 &  9 &  9 & 176 &  9 &  9 & 184 & 82 &  63 \\
 200 & 191 &  9 &  9 & 173 & 11 & 11 & 182 & 82 &  63 \\
 220 & 190 &  7 &  7 & 173 & 14 &  9 & 181 & 82 &  63 \\
 240 & 189 &  9 &  7 & 177 & 11 &  9 & 183 & 82 &  63 \\
 260 & 189 & 11 &  7 & 181 &  9 &  9 & 185 & 82 &  63 \\
 280 & 186 & 11 &  7 &   - &  - &  - & 186 & 82 &  63 \\
 300 & 186 & 11 & 11 &   - &  - &  - & 186 & 82 &  63 \\
 320 & 185 & 14 & 14 &   - &  - &  - & 185 & 82 &  63 \\
 340 & 185 & 14 & 14 &   - &  - &  - & 185 & 82 &  63 \\

\multicolumn{10}{l}{\it \underline{N4183}} \\
  10 &  56 & 12 & 10 &  38 & 15 & 10 &  47 & 82 & 346 \\
  20 &  71 &  7 &  7 &  61 & 12 & 10 &  66 & 82 & 346 \\
  30 &  78 &  7 &  7 &  74 & 10 &  7 &  76 & 82 & 346 \\
  40 &  88 &  7 &  7 &  84 & 10 &  7 &  86 & 82 & 346 \\
  50 &  97 &  7 &  7 &  97 &  7 &  7 &  97 & 82 & 346 \\
  60 & 100 &  7 &  7 &  99 &  7 &  7 &  99 & 82 & 346 \\
  70 & 103 &  7 &  7 & 103 &  7 &  7 & 103 & 82 & 346 \\
  80 & 106 &  7 &  7 & 107 &  7 &  7 & 107 & 82 & 346 \\
  90 & 110 &  7 &  7 & 113 &  7 &  7 & 111 & 82 & 346 \\
 100 & 112 &  7 &  7 & 117 & 10 & 10 & 114 & 82 & 346 \\
 110 & 112 &  7 &  7 & 118 & 10 & 10 & 115 & 82 & 346 \\
 120 & 108 &  7 &  7 & 114 & 10 &  7 & 111 & 82 & 346 \\
 130 & 108 &  7 &  7 & 113 & 10 &  7 & 110 & 82 & 347 \\
 141 & 111 &  7 &  7 & 112 &  7 &  7 & 111 & 82 & 347 \\
 151 & 108 &  7 &  5 & 110 &  7 &  7 & 109 & 82 & 347 \\
 161 & 106 &  5 &  5 & 109 &  7 &  7 & 108 & 82 & 347 \\
 172 & 109 &  7 &  7 & 109 &  7 &  7 & 109 & 82 & 347 \\
 183 & 112 &  7 &  7 & 110 &  7 &  7 & 111 & 82 & 348 \\
 194 & 108 &  5 &  8 & 111 &  8 &  8 & 110 & 82 & 348 \\
 205 & 106 &  5 &  8 & 111 &  8 &  8 & 109 & 82 & 348 \\
 217 & 107 &  7 &  8 & 112 &  8 &  8 & 110 & 82 & 348 \\
 229 &   - &  - &  - & 112 & 10 & 10 & 112 & 82 & 348 \\
 241 &   - &  - &  - & 113 & 13 & 10 & 113 & 82 & 349 \\

\multicolumn{10}{l}{\it \underline{N4217}} \\
  10 &  38 & 10 & 10 &  57 & 14 & 12 &  48 & 86 & 230 \\
  20 &  82 & 10 & 12 & 116 & 14 & 10 &  99 & 86 & 230 \\
  30 & 145 & 10 & 10 & 148 & 14 & 10 & 146 & 86 & 230 \\
  40 & 162 &  7 & 10 & 165 & 10 & 10 & 164 & 86 & 230 \\
  50 & 176 &  7 & 10 & 172 &  7 & 10 & 174 & 86 & 230 \\
  60 & 176 &  7 & 10 & 175 & 10 &  7 & 175 & 86 & 230 \\
  70 & 189 &  7 & 10 & 179 & 10 & 10 & 184 & 86 & 230 \\
  80 & 188 &  7 & 10 & 182 & 10 & 10 & 185 & 86 & 230 \\
  90 & 187 &  5 & 10 & 188 & 12 & 10 & 188 & 86 & 230 \\

\hline
\hline
\end{tabular}}
\end{table}

\addtocounter{table}{-1}
\begin{table}[pht]
\caption[]{continued}
{\fontsize{9}{10}
\selectfont
\setlength{\tabcolsep}{1.7mm}
\begin{tabular}[t]{rrrrrrrrrr}
\hline
\hline
\noalign{\vspace{0.8mm}}
\multicolumn{1}{c}{Rad.} & V$^{\mbox{\scriptsize app}}_{\mbox{\scriptsize rot}}$ & \multicolumn{2}{c}{+/$-$} & V$^{\mbox{\scriptsize rec}}_{\mbox{\scriptsize rot}}$ & \multicolumn{2}{c}{+/$-$} & \multicolumn{1}{c}{V$^{\mbox{\scriptsize ave}}_{\mbox{\scriptsize rot}}$} & \multicolumn{1}{c}{i} & PA \\
\multicolumn{1}{c}{($^{\prime\prime}$)} & \multicolumn{3}{c}{ - - - km/s - - - } & \multicolumn{3}{c}{ - - - km/s - - - } & km/s & ($^\circ$) & ($^\circ$) \\
\noalign{\vspace{0.8mm}}
\hline
\noalign{\vspace{0.8mm}}

\multicolumn{10}{l}{\it \underline{N4217 (cont.)}} \\
 100 & 188 &  7 & 10 & 192 & 12 & 12 & 190 & 86 & 230 \\
 110 & 191 &  7 & 10 & 191 & 10 & 10 & 191 & 86 & 230 \\
 120 & 192 &  7 &  7 & 189 & 10 &  7 & 191 & 86 & 230 \\
 130 & 191 & 10 &  7 & 187 & 10 &  7 & 189 & 86 & 230 \\
 140 & 187 & 10 &  7 & 185 & 12 & 10 & 186 & 86 & 230 \\
 150 & 183 & 10 &  7 & 183 & 10 & 10 & 183 & 86 & 230 \\
 160 & 180 & 12 &  7 & 178 & 10 & 10 & 179 & 86 & 230 \\
 170 & 177 & 10 & 10 & 177 & 10 & 12 & 177 & 86 & 230 \\
 181 & 178 & 12 & 12 & 177 & 10 & 14 & 177 & 86 & 230 \\
 191 & 178 & 12 & 12 &   - &  - &  - & 178 & 86 & 230 \\

\multicolumn{10}{l}{\it \underline{N4389}} \\
  10 &  30 &  8 &  8 &  25 & 10 & 8 &  27 & 50 & 277 \\
  20 &  56 & 10 &  8 &  50 & 13 & 8 &  53 & 50 & 277 \\
  31 &  70 & 13 &  8 &  69 & 10 & 8 &  69 & 50 & 277 \\
  41 &  79 & 13 &  8 &  88 &  8 & 8 &  84 & 50 & 277 \\
  51 &  92 & 10 & 10 &  99 &  8 & 8 &  96 & 50 & 277 \\
  61 &   - &  - &  - & 110 &  8 & 8 & 110 & 50 & 277 \\

\hline
\hline
\noalign{\vspace{0.8mm}}

\multicolumn{10}{c}{\it Rotation curves derived from XV-diagrams only.} \\
\noalign{\vspace{0.8mm}}

\hline
\hline
\noalign{\vspace{0.8mm}}

\multicolumn{10}{l}{\it \underline{N3718}} \\
  40 & 228 & 10 & 10 & 228 & 10 & 10 & 228 & 76 & 114 \\
  80 & 228 & 10 & 10 & 228 & 10 & 10 & 228 & 80 & 130 \\
 120 & 228 & 10 & 10 & 228 & 10 & 10 & 228 & 84 & 143 \\
 160 & 228 & 10 & 10 & 228 & 10 & 10 & 228 & 90 & 162 \\
 200 & 228 & 10 & 10 & 228 & 10 & 10 & 228 & 85 & 175 \\
 240 & 220 & 10 & 10 & 235 & 10 & 10 & 228 & 80 & 186 \\
 280 & 225 & 10 & 10 & 239 & 10 & 10 & 232 & 75 & 195 \\
 320 & 240 & 10 & 10 & 245 & 10 & 10 & 242 & 70 & 196 \\
 360 & 245 & 10 & 10 & 242 & 10 & 10 & 244 & 65 & 196 \\
 400 & 235 & 10 & 10 & 240 & 10 & 10 & 237 & 65 & 194 \\
 420 & 227 & 10 & 10 &   - &  - &  - &   - & 65 & 194 \\

\multicolumn{10}{l}{\it \underline{N3729}} \\
  20 & 118 & 34 & 24 & 138 & 12 & 10 & 128 & 48 & 164 \\
  40 & 157 & 17 & 12 & 141 & 12 & 12 & 149 & 48 & 164 \\
  50 &   - &  - &  - & 144 & 10 & 10 & 144 & 48 & 164 \\
  60 & 151 & 12 & 10 &   - &  - &  - & 151 & 48 & 164 \\

\multicolumn{10}{l}{\it \underline{U6773}} \\
  10 &  28 & 10 &  7 &  34 & 10 &  7 &  31 & 60 & 341 \\
  20 &  38 &  7 &  7 &  48 &  7 &  5 &  43 & 60 & 341 \\
  30 &  46 &  5 &  5 &  44 &  7 &  7 &  45 & 60 & 341 \\
  40 &  47 &  5 &  5 &  44 & 10 &  7 &  45 & 60 & 341 \\

\multicolumn{10}{l}{\it \underline{U6818}} \\
  10 &  27 &  7 &  7 &  20 & 10 &  7 &  23 & 79 &  77 \\
  20 &  28 & 10 &  7 &  28 &  7 &  5 &  28 & 79 &  77 \\
  30 &  31 &  7 &  5 &  43 &  7 &  5 &  37 & 79 &  77 \\
  40 &  43 &  7 &  5 &  53 &  7 &  5 &  48 & 79 &  77 \\
  50 &  66 &  7 &  7 &  61 &  7 &  5 &  63 & 79 &  77 \\
  60 &  77 &  7 & 10 &  66 &  7 &  5 &  71 & 79 &  77 \\
  70 &  68 &  7 &  5 &   - &  - &  - &  68 & 79 &  77 \\
  80 &  74 &  7 &  5 &   - &  - &  - &  74 & 79 &  77 \\

\multicolumn{10}{l}{\it \underline{N3985}} \\
   0 &   8 & 15 & 10 &  -8 & 15 & 10 &   0 & 53 &  70 \\
  10 &  41 & 10 &  7 &  37 & 12 &  7 &  39 & 53 &  70 \\
  20 &  60 &  7 & 10 &  89 & 15 &  7 &  75 & 53 &  70 \\
  25 &  68 & 10 & 10 &   - &  - &  - &  68 & 53 &  70 \\
  30 &   - &  - &  - &  93 &  7 &  7 &  93 & 53 &  70 \\

\multicolumn{10}{l}{\it \underline{U6894}} \\
  10 &  28 &  7 &  7 &  28 & 12 & 10 &  28 & 89 & 269 \\
  20 &  45 &  7 &  7 &  45 &  7 &  7 &  45 & 89 & 269 \\

\hline
\hline
\end{tabular}}
\end{table}

\addtocounter{table}{-1}
\begin{table}[pht]
\caption[]{continued}
{\fontsize{9}{10}
\selectfont
\setlength{\tabcolsep}{1.7mm}
\begin{tabular}[t]{rrrrrrrrrr}
\hline
\hline
\noalign{\vspace{0.8mm}}
\multicolumn{1}{c}{Rad.} & V$^{\mbox{\scriptsize app}}_{\mbox{\scriptsize rot}}$ & \multicolumn{2}{c}{+/$-$} & V$^{\mbox{\scriptsize rec}}_{\mbox{\scriptsize rot}}$ & \multicolumn{2}{c}{+/$-$} & \multicolumn{1}{c}{V$^{\mbox{\scriptsize ave}}_{\mbox{\scriptsize rot}}$} & \multicolumn{1}{c}{i} & PA \\
\multicolumn{1}{c}{($^{\prime\prime}$)} & \multicolumn{3}{c}{ - - - km/s - - - } & \multicolumn{3}{c}{ - - - km/s - - - } & km/s & ($^\circ$) & ($^\circ$) \\
\noalign{\vspace{0.8mm}}
\hline
\noalign{\vspace{0.8mm}}

\multicolumn{10}{l}{\it \underline{U6894 (cont.)}} \\
  30 &  56 &  7 &  5 &  56 &  5 &  5 &  56 & 89 & 269 \\
  40 &  62 &  5 &  5 &  63 &  7 &  7 &  63 & 89 & 269 \\

\multicolumn{10}{l}{\it \underline{N4013}} \\
  65 &   - &  - &  - & 198 & 10 & 10 & 198 & 90 & 245 \\
  73 &   - &  - &  - & 195 &  5 &  5 & 195 & 90 & 245 \\
  82 & 193 &  5 &  5 & 195 &  3 &  3 & 194 & 90 & 245 \\
  91 & 195 &  4 &  4 & 195 &  3 &  3 & 195 & 90 & 245 \\
  99 & 195 &  3 &  3 & 195 &  3 &  3 & 195 & 90 & 245 \\
 108 & 195 &  3 &  3 & 193 &  4 &  4 & 195 & 90 & 245 \\
 117 & 196 &  3 &  3 & 185 &  5 &  5 & 192 & 90 & 245 \\
 125 & 195 &  3 &  3 & 178 &  5 &  5 & 188 & 90 & 245 \\
 134 & 190 &  4 &  4 & 178 &  8 &  8 & 186 & 90 & 245 \\
 143 & 190 &  4 &  4 &   - &  - &  - & 186 & 90 & 245 \\
 151 & 188 &  5 &  5 &   - &  - &  - & 186 & 90 & 245 \\
 160 & 187 &  6 &  6 &   - &  - &  - & 185 & 90 & 245 \\
 168 & 179 & 10 & 10 &   - &  - &  - & 180 & 90 & 243 \\
 177 & 163 &  8 &  8 &   - &  - &  - & 163 & 90 & 240 \\
 186 & 161 &  6 &  6 &   - &  - &  - & 162 & 90 & 238 \\
 194 & 162 &  5 &  5 &   - &  - &  - & 164 & 90 & 236 \\
 203 & 164 &  5 &  5 & 170 &  5 &  5 & 166 & 90 & 235 \\
 212 & 164 &  6 &  6 & 168 &  5 &  5 & 166 & 90 & 233 \\
 220 & 165 &  8 &  8 & 166 &  5 &  5 & 166 & 90 & 232 \\
 229 & 165 &  7 &  7 & 166 &  5 &  5 & 166 & 90 & 230 \\
 238 & 169 &  5 &  5 & 168 &  5 &  5 & 168 & 90 & 229 \\
 246 & 173 &  5 &  5 &   - &  - &  - & 172 & 90 & 228 \\
 255 & 173 &  5 &  5 &   - &  - &  - & 173 & 90 & 226 \\
 264 & 172 &  5 &  5 &   - &  - &  - & 171 & 90 & 225 \\
 272 & 169 &  6 &  6 & 170 &  5 &  5 & 170 & 90 & 224 \\
 281 & 162 & 10 & 10 & 172 &  5 &  5 & 172 & 90 & 224 \\
 289 &   - &  - &  - & 174 &  5 &  5 & 173 & 90 & 223 \\
 298 &   - &  - &  - & 176 &  5 &  5 & 176 & 90 & 222 \\
 307 &   - &  - &  - & 178 &  5 &  5 & 178 & 90 & 221 \\
 315 &   - &  - &  - & 180 &  6 &  6 & 180 & 90 & 221 \\
 324 &   - &  - &  - & 180 &  8 &  8 & 180 & 90 & 220 \\
 333 &   - &  - &  - & 180 &  5 &  5 & 180 & 90 & 219 \\
 341 &   - &  - &  - & 180 &  5 &  5 & 180 & 90 & 219 \\
 350 &   - &  - &  - & 178 &  5 &  5 & 178 & 90 & 218 \\
 359 &   - &  - &  - & 174 &  5 &  5 & 174 & 90 & 218 \\
 367 &   - &  - &  - & 170 & 10 & 10 & 170 & 90 & 218 \\

\multicolumn{10}{l}{\it \underline{U7089}} \\
  10 &  25 &  7 &  7 &  17 &  7 &  5 &  21 & 89 & 215 \\
  20 &  38 &  5 &  5 &  35 &  7 &  7 &  36 & 89 & 215 \\
  30 &  45 &  5 &  5 &  42 &  7 &  7 &  43 & 89 & 215 \\
  40 &  51 &  7 &  7 &  51 &  5 &  5 &  51 & 89 & 215 \\
  50 &  57 &  7 &  5 &  62 &  5 &  5 &  60 & 89 & 215 \\
  60 &  63 & 10 &  5 &  66 &  5 &  5 &  65 & 89 & 215 \\
  70 &  66 &  7 &  5 &  69 &  5 &  5 &  68 & 89 & 215 \\
  75 &   - &  - &  - &  73 &  7 &  7 &  73 & 89 & 215 \\
  80 &  70 &  7 &  5 &   - &  - &  - &  70 & 89 & 215 \\
\multicolumn{10}{l}{\it \underline{U7089 (cont.)}} \\
  90 &  74 &  7 &  5 &   - &  - &  - &  74 & 89 & 215 \\
 100 &  78 &  7 &  5 &   - &  - &  - &  78 & 89 & 215 \\
 105 &  79 &  7 &  7 &   - &  - &  - &  79 & 89 & 215 \\

\multicolumn{10}{l}{\it \underline{U7094}} \\
  20 &  32 &  5 &  5 &  32 &  5 &  5 &  32 & 72 &  39 \\
  40 &  36 &  7 &  5 &  36 &  7 &  5 &  36 & 72 &  39 \\
  60 &   - &  - &  - &  35 &  7 &  5 &  35 & 72 &  39 \\

\hline
\hline
\end{tabular}}
\end{table}

\addtocounter{table}{-1}
\begin{table}[th]
\caption[]{continued}
{\fontsize{9}{10}
\selectfont
\setlength{\tabcolsep}{1.7mm}
\begin{tabular}[t]{rrrrrrrrrr}
\hline
\hline
\noalign{\vspace{0.8mm}}
\multicolumn{1}{c}{Rad.} & V$^{\mbox{\scriptsize app}}_{\mbox{\scriptsize rot}}$ & \multicolumn{2}{c}{+/$-$} & V$^{\mbox{\scriptsize rec}}_{\mbox{\scriptsize rot}}$ & \multicolumn{2}{c}{+/$-$} & \multicolumn{1}{c}{V$^{\mbox{\scriptsize ave}}_{\mbox{\scriptsize rot}}$} & \multicolumn{1}{c}{i} & PA \\
\multicolumn{1}{c}{($^{\prime\prime}$)} & \multicolumn{3}{c}{ - - - km/s - - - } & \multicolumn{3}{c}{ - - - km/s - - - } & km/s & ($^\circ$) & ($^\circ$) \\
\noalign{\vspace{0.8mm}}
\hline
\noalign{\vspace{0.8mm}}

\multicolumn{10}{l}{\it \underline{N4138}} \\
  30 & 178 & 19 & 34 & 181 & 12 & 12 & 179 & 53 & 151 \\
  60 & 191 & 10 & 10 & 200 & 10 & 10 & 195 & 53 & 151 \\
  90 & 181 & 10 & 10 & 181 & 15 & 20 & 181 & 53 & 147 \\
 122 &   - &  - &  - & 162 & 26 & 15 & 162 & 51 & 143 \\
 154 & 145 & 14 & 14 & 145 & 14 & 14 & 145 & 48 & 140 \\
 184 &   - &  - &  - & 147 & 18 & 18 & 147 & 43 & 138 \\
 213 &   - &  - &  - & 150 & 21 & 21 & 150 & 35 & 138 \\

\multicolumn{10}{l}{\it \underline{N4218}} \\
  10 &  51 & 10 &  7 &  70 & 12 & 10 &  60 & 55 & 316 \\
  20 &  83 & 10 &  5 &  62 & 12 & 10 &  73 & 55 & 316 \\

\hline
\hline
\end{tabular}}
\end{table}

\section{Matching HI linewidths to V$_{\mbox{\scriptsize max}}$ and
V$_{\mbox{\scriptsize flat}}$}

In this section, it will be investigated how the linewidth correction
for turbulent motion can be used to match the finally corrected global
HI linewidths to the actual rotational velocities measured from the
rotation curves.

After the correction for instrumental resolution, the profile widths  
are generally corrected for broadening due to turbulent motions of the
HI gas by applying TFq's formula

\begin{center}
\begin{tabular}{rcr}
 $W^2_{R,l}$ & = & $W^2_l \; + \; W^2_{t,l} \; \left[ 1 - 2 \; e^{-\left(\frac{W_l}{W_{c,l}}\right)^2} \right]$ \\ 
\noalign{\vspace{1mm}}
             &   & $- \; 2\;W_l\;W_{t,l}\left[1 - e^{-\left(\frac{W_l}{W_{c,l}}\right)^2} \right]$ \\
\end{tabular}
\end{center}

\noindent where the subscript $l$ refers to the widths at the $l$=20\%
or the $l$=50\% level of peak flux.  This formula yields a linear
subtraction of $W_{t,l}$ if $W_l$$>$$W_{c,l}$ and a quadratic
subtraction if $W_l$$<$$W_{c,l}$.  Values of $W_{t,l}$ and $W_{c,l}$ are
different for line width corrections at the 20\% and 50\% levels.  The
values of $W_{c,l}$ indicate the profile widths where the transition
from a boxy to a Gaussian shape occurs. The amount by which a global
profile is broadened due to random motions is given by
$W_{t,l}$=2$k_l\sigma$ where, for a Gaussian velocity dispersion
$\sigma$, $k_{20}$=1.80 and $k_{50}$=1.18.

The generally adopted values for $W_{c,l}$ are \penalty-10000
$W_{c,20}$=120~km/s\ and $W_{c,50}$=100~km/s.  The more important values
of $W_{t,l}$, however, have been subject of some debate among various   
authors.  With our new HI synthesis data we can give a meaningful
contribution to this debate.

Bottinelli et al. (\cite{bottinelli83}) came up with an empirical
approach, based on a minimization of the scatter in the TF-relation. 
They assumed an anisotropic velocity dispersion of the HI gas of
$\sigma_x$=$\sigma_y$=1.5$\sigma_z$ and a velocity dispersion
perpendicular to the plane of $\sigma_z$=10~km/s.  They determined the
values of $k_l$ by minimizing the scatter in the TF-relation and found
$k_{20}$=1.89 and $k_{50}$=0.71, indicating deviations from a Gaussian
distribution (broader wings).  Due to the assumed velocity anisotropy,
$W_{t,l}$ has become a function of inclination angle and varies in the
range 45$<$$W_{t,20}$$<$57 and 17$<$$W_{t,50}$$<$21 for inclinations
ranging between 45$^\circ$$<$i$<$90$^\circ$.

The same value of $k_{20}$=1.89 was adopted by TFq but they
assumed an isotropic velocity dispersion of
$\sigma_x$=$\sigma_y$=$\sigma_z$=10~km/s\ and consequently advocate
$W_{t,20}$=2$\cdot$1.89$\cdot$10=38~km/s, independent of inclination.
They did not address the situation at the 50\% level.

Fouqu\'e et al. (\cite{fouque}) also assumed isotropy but adopted
$\sigma$=12~km/s. They determined $k_l$ in a more direct way by
comparing the corrected line width to the observed maximum rotational 
velocity V$_{\mbox{\scriptsize max}}$\ as derived from HI velocity
fields. They found $k_{20}$=1.96 and $k_{50}$=1.13, indicating a
near-Gaussian distribution, contrary to the findings of Bottinelli et
al. Consequently, Fouqu\'e et al. advocate the much larger values of
$W_{t,20}$=47~km/s\ and W$_{t,50}$=27~km/s\ respectively.

A similar procedure was followed by Broeils (\cite{broeils92}) using a
sample of 21 galaxies with well defined HI velocity fields. Broeils made
no a priori assumptions about the intrinsic velocity dispersion and did
not decouple $k_l$ and $\sigma$. He did, however, recognize that
V$_{\mbox{\scriptsize max}}$\ may exceed V$_{\mbox{\scriptsize flat}}$\
and he determined for each galaxy the values of $W^{\mbox{\scriptsize
max}}_{t,l}$ and $W^{\mbox{\scriptsize flat}}_{t,l}$ for which the
differences

\begin{center}
\begin{tabular}{llcl}
     & $\Delta W^{\mbox{\scriptsize max}}_{R,l}$  & = & $W_{R,l} - 2\mbox{V$_{\mbox{\scriptsize max}}$\ sin(i)}$ \\
 and &                                            &   & \\
     & $\Delta W^{\mbox{\scriptsize flat}}_{R,l}$ & = & $W_{R,l} - 2\mbox{V$_{\mbox{\scriptsize flat}}$\ sin(i)}$ \\
\end{tabular}
\end{center}

\noindent become zero for each galaxy.  He found mean values of

\begin{center}
\begin{tabular}{lcl}
$W^{\mbox{\scriptsize max}}_{t,20} = 21\pm2$  & , & $W^{\mbox{\scriptsize max}}_{t,50} \;= 7\pm1$ \\
\noalign{\vspace{1mm}}
$W^{\mbox{\scriptsize flat}}_{t,20} = 37\pm5$ & , & $W^{\mbox{\scriptsize flat }}_{t,50} \;= 25\pm4$ \\
\end{tabular}
\end{center}

\noindent (Note that he quoted the much larger scatters instead of the
errors in the mean quoted above.) He rejected his results, probably
discouraged by the large {\it scatters}, and adopted the values
W$_{t,20}$=38 and W$_{t,50}$=14~km/s\ which he erroneously identifies
with Bottinelli et al.'s results.

Finally, Rhee (\cite{rhee96a}) performed the same investigation using 28
galaxies, most of them in common with Broeils' (\cite{broeils92})
sample. Not surprisingly, he found

\begin{center}
\begin{tabular}{lcl}
$W^{\mbox{\scriptsize max}}_{t,20} = 20\pm2$  & , & $W^{\mbox{\scriptsize max}}_{t,50} \;= 8\pm2$ \\
\noalign{\vspace{1mm}}
$W^{\mbox{\scriptsize flat}}_{t,20} = 30\pm3$ & , & $W^{\mbox{\scriptsize flat }}_{t,50} \;= 18\pm3$ \\
\end{tabular}
\end{center}

\noindent similar to Broeils' result.
 
\begin{figure}[t]
 \resizebox{\hsize}{!}{\includegraphics{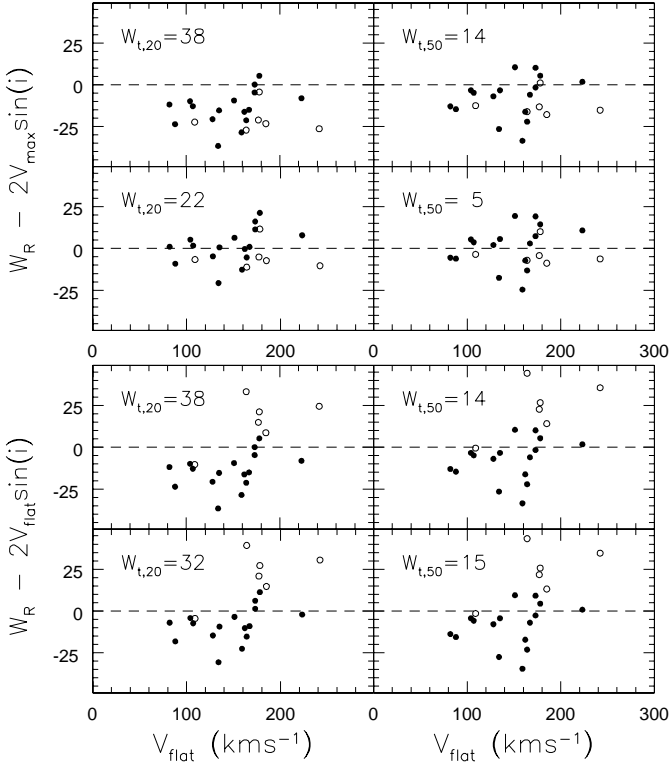}}
 \caption{Comparison of the global profile widths W$_{\mbox{\scriptsize
R},l}$, corrected for instrumental broadening and random motions, with
2V$_{\mbox{\scriptsize max}}$sin(i) (Upper panels) and with
2V$_{\mbox{\scriptsize flat}}$sin(i) (lower panels). The left panels
consider $W_{R,20}$ and the right panels $W_{R,50}$. Different values of
the random motion parameters W$_{t,l}$ are used. Open symbol indicate
galaxies with declining rotation curves (V$_{\mbox{\scriptsize
max}}$~$>$~V$_{\mbox{\scriptsize flat}}$) and filled symbols indicated
galaxies without a declining part (V$_{\mbox{\scriptsize
max}}$~=~V$_{\mbox{\scriptsize flat}}$). See section 4 for further
details.} 
  \label{UMaIVfig2}
\end{figure}

Here, with our new and independent dataset, we follow the same strategy
as Broeils and Rhee by investigating which values of $W_{t,l}$ allow an
accurate retrieval of V$_{\mbox{\scriptsize max}}$\ and
V$_{\mbox{\scriptsize flat}}$\ from the broadened global profile.  For
this purpose we will only consider those 22 galaxies in our Ursa Major
sample that show a flat part in their rotation curves (with a
significant amount of HI gas) and that are free from a major change in
inclination angle.  Of these 22, there are 6 galaxies with
V$_{\mbox{\scriptsize max}}$~$>$~V$_{\mbox{\scriptsize flat}}$.  Note
that both Broeils and Rhee used Bottinelli et al.'s prescription to
correct for instrumental broadening which we are forced to adopt here to
ensure a valid comparison between their and our results.  We calculated
the values of $W^{\mbox{\scriptsize max}}_{t,l}$ and
$W^{\mbox{\scriptsize flat}}_{t,l}$ for which the average values

\begin{center}
 $\overline{\Delta W^{\mbox{\scriptsize max}}_{R,l}} = \frac{1}{N}\sum\left( \mbox{W}_{\mbox{\scriptsize R},l} - 2\mbox{V$_{\mbox{\scriptsize max}}$sin(i)} \right)$ \\
 $\overline{\Delta W^{\mbox{\scriptsize flat}}_{R,l}} = \frac{1}{N}\sum\left( \mbox{W}_{\mbox{\scriptsize R},l} - 2\mbox{V$_{\mbox{\scriptsize flat}}$sin(i)} \right)$ \\
\end{center}

\noindent become zero.  This is done for both the entire sample of
$N$=22 galaxies and for the subsample of $N$=16 galaxies with
V$_{\mbox{\scriptsize max}}$~$=$~V$_{\mbox{\scriptsize flat}}$.  For the
entire sample we find

\begin{center}
\begin{tabular}{lcl}
$W^{\mbox{\scriptsize max}}_{t,20} = 22$  & , & $W^{\mbox{\scriptsize max}}_{t,50} \;= 5$ \\
\noalign{\vspace{1mm}}
$W^{\mbox{\scriptsize flat}}_{t,20} = 32$ & , & $W^{\mbox{\scriptsize flat }}_{t,50} \;= 15$ \\
\end{tabular}
\end{center}

\noindent These values are in good agreement with the (rejected) results
of Broeils and in excellent agreement with the results of Rhee.  The
values of $W^{\mbox{\scriptsize flat }}_{t,l}$ are larger than the
values of $W^{\mbox{\scriptsize max}}_{t,l}$ because the galaxies with
V$_{\mbox{\scriptsize max}}$~$>$~V$_{\mbox{\scriptsize flat}}$\ in our
sample have considerable amounts of HI gas at their peak velocity in the
rotation curve.  This gas, rotating at V$_{\mbox{\scriptsize max}}$\
broadens the global profile somewhat further.  If we consider only the
16 galaxies for which V$_{\mbox{\scriptsize
max}}$~$=$~V$_{\mbox{\scriptsize flat}}$\ we find

\begin{center}
\begin{tabular}{lcl}
 $W^{\mbox{\scriptsize max}}_{t,20} \; = \; W^{\mbox{\scriptsize flat }}_{t,20} \; = 23$ &
 , &
 $W^{\mbox{\scriptsize max}}_{t,50} \; = \; W^{\mbox{\scriptsize flat }}_{t,50} \; = 6$ \\
\end{tabular}
\end{center}

\noindent in agreement with the values of $W^{\mbox{\scriptsize max
}}_{t,l}$ we found when using all 22 galaxies.

Our results are illustrated in Figure 2 where we show, for each of the
22 galaxies, the deviations $\Delta W^{\mbox{\scriptsize max}}_{R,l}$
(upper panels) and $\Delta W^{\mbox{\scriptsize flat}}_{R,l}$ (lower
panels) as a function of V$_{\mbox{\scriptsize flat}}$.  Galaxies with
V$_{\mbox{\scriptsize max}}$~$=$~V$_{\mbox{\scriptsize flat}}$~ are
indicated by filled symbols, galaxies with \penalty-10000
V$_{\mbox{\scriptsize max}}$~$>$~V$_{\mbox{\scriptsize flat}}$~ are
indicated by open symbols.  The upper two panels in each block show the
results one obtains when using Broeil's adopted values of
$W_{t,20}$~=~38 and $W_{t,50}$~=~14~km/s.

From the upper panels in the upper block it is clear that the maximum
rotational velocity as derived from the corrected global profiles is
severely underestimated when using the values of W$_{t,l}$ derived by
TFq and adopted by Broeils.  This systematic underestimation disappears
when W$_{t,20}$ is decreased from 38 to 22 km/s\ and W$_{t,50}$ is
decreased from 14 to 5 km/s. The upper two panels in the lower block
show that if one is interested in the amplitude of the flat part, which
is smaller than the maximum rotational velocity for galaxies with a
declining rotation curve (open symbols), the average offset becomes less
significant simply because the open symbols scatter upward.  In this
case, to obtain an average zero offset, we find similar values for
W$_{t,l}$ as those adopted by Broeils. However, we find the curious
situation that the corrected width of the global profile systematically
overestimates V$_{\mbox{\scriptsize flat}}$~ for galaxies with a
declining rotation curve (open symbols) and systematically
underestimates V$_{\mbox{\scriptsize flat}}$~ for galaxies with a purely
flat rotation curve (filled symbols).

From this we can conclude that, in a statistical sense, the maximum
rotational velocity of a galaxy can be reasonably well retrieved from
the width of the global profile when using $W_{t,20}$ = 22 or $W_{t,50}$
= 5~km/s. The amplitude of the flat part can not be retrieved
consistently for a mixed sample containing galaxies with declining
rotation curves. Note that we have explored only a restricted range of
rotational velocities: 80--200~km/s.

Our results also indicate a non-Gaussian distribution of random
velocities in the sense that $W_{t,20}$/$W_{t,50}$~$\neq$~1.80/1.18.
Interpreting $W_{t,20}$ and $W_{t,50}$ in terms of velocity dispersions
it follows that

\begin{center}
\begin{tabular}{lclcl}
$\sigma_{20}$ & = & W$_{t,20}$/2k$_{20}$ & = & 6.1 km/s \\
\noalign{\vspace{1mm}}
$\sigma_{50}$ & = & W$_{t,50}$/2k$_{50}$ & = & 2.1 km/s \\
\end{tabular}
\end{center}

\noindent where $k_{20}$=1.80 and $k_{50}$=1.18 for a Gaussian
distribution. Recall, however, that we advocate a different correction
for instrumental broadening than Bottinelli et al.'s scheme used by
Broeils and Rhee. With our correction method for instrumental broadening
we find the somewhat smaller values of:

\begin{center}
\begin{tabular}{lcl}
$W_{t,20}$~= 22 & , & $W_{t,50}$~= 2
\end{tabular}
\end{center}

\noindent These smaller values of W$_{t,l}$ allow to retrieve
V$_{\mbox{\scriptsize flat}}$\ from the global profiles of galaxies with
purely flat rotation curves and V$_{\mbox{\scriptsize max}}$\ for
galaxies with declining rotation curves.  Applying our correction method
for instrumental resolution and the above-mentioned value of
$W_{t,20}$~$=$~22 km/s we find an rms scatter in $\Delta W_{20} =
0.5W^i_R -$ V$_{\mbox{\scriptsize max}}$\ of 6.8 km/s.

\section{A comparison of inclinations}

Present day instrumentation allows accurate measurements of the
luminosities and global HI profiles of galaxies. In general, the
observed scatter in the TF-relation is larger than can be explained by
the observational uncertainties in these measured parameters alone.     
However, the uncertainty in corrections sensitive to inclination 
contribute significantly to the observed scatter. For a sample of    
randomly oriented galaxies more inclined than 45 degrees, an uncertainty
of 1, 3 or 5 degrees in the inclination angle contributes respectively  
0.04, 0.12 or 0.19 magnitudes to the scatter due to the uncertainty in
line widths alone, assuming a slope in the TF-relation of $-10$. 
Therefore, it is important to determine the inclination angle of a
galaxy as accurate as possible and this issue deserves some special
attention.

From the photometric and HI synthesis data available, three independent
measurements of the inclination angle of a galaxy can in principle be
obtained; i$_{\mbox{\scriptsize opt}}$ from the optical axis ratio,
i$_{\mbox{\scriptsize HI}}$ from the apparent ellipticity of the HI
disk, and i$_{\mbox{\scriptsize VF}}$ from fitting tilted rings to the
HI velocity field.  Each of these methods has its own systematic
limitations which are important to recognize when estimating the actual 
inclination of a galaxy.  In the following discussion we will briefly
address those limitations and make an intercomparison of
i$_{\mbox{\scriptsize opt}}$, i$_{\mbox{\scriptsize HI}}$, and
i$_{\mbox{\scriptsize VF}}$.

\subsection{i$_{\mbox{\scriptsize opt}}$ from optical axis ratios}

The most widely used formula to infer the inclination angle from the
observed optical axis ratio (b/a)$\equiv$q was provided by Hubble (1926):

$$ \mbox{cos}^2(\mbox{i}_{\mbox{\scriptsize opt}}) = \frac{\mbox{q}^2 - \mbox{q}_0^2}{1 - \mbox{q}_0^2} $$

\noindent where q$_0$ is the intrinsic thickness of an oblate stellar
disk.  Holmberg (1946) determined an average value of q$_0$=0.20 which
is still commonly used although it is obvious from images of edge-on
systems that large variations in q$_0$ exist.  For instance, Fouqu\'e et
al. (\cite{fouque}) found q$_0$ to vary from 0.30 to 0.16 for spirals of
morphological types Sa to Sd respectively and q$_0$=0.42 for galaxies of
type Sdm-Im. Apart from the debate on the intrinsic thickness, the 
observed axis ratio q itself has limited meaning since it is often  
defined at a certain isophote around which q may still vary as a
function of radius. From images of edge-on disks in the Ursa major
cluster (see Paper I) it can often be observed that the axis ratio keeps
increasing outward until the faintest isophotes.  An extreme example is
NGC 4389, dominated by a narrow bar and surrounded by an extended faint
halo. The axis ratios presented in Table~1 were not determined at a
fixed isophote but were choosen to represent the stellar disk instead of
a bulge, lopsided structures or a faint halo.

\begin{figure}[t]
 \resizebox{\hsize}{!}{\includegraphics{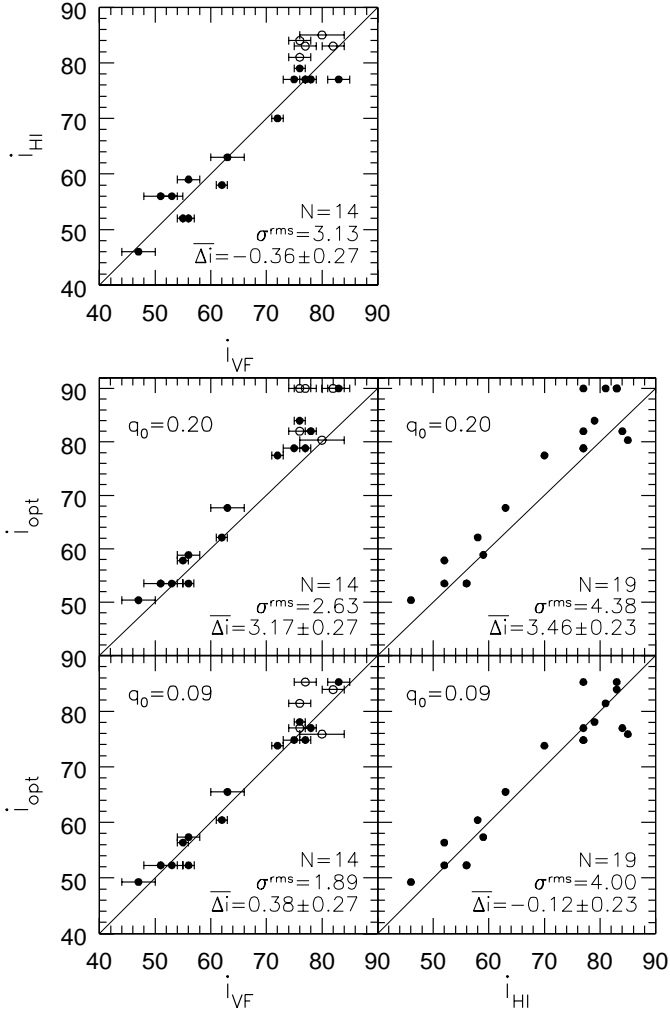}}
 \caption{Intercomparison of the three independently determined
inclination angles i$_{\mbox{\scriptsize opt}}$, i$_{\mbox{\scriptsize
HI}}$ and i$_{\mbox{\scriptsize VF}}$.  Only the filled symbols
(i$_{\mbox{\scriptsize HI}}$~$<$~80$^\circ$ when i$_{\mbox{\scriptsize
VF}}$ is involved) are considered for the unweighted quantitative
assessment.}
 \label{UMaIVfig3}
\end{figure}

\subsection{i$_{\mbox{\scriptsize HI}}$ from the inclined HI disk}

Apart from the oblate stellar disk, the HI disk can also be used to 
determine the inclination. In general, the HI disk is much thinner than
the stellar disk and its intrinsic thickness is of no concern.  However,
its patchiness, lopsidedness and the existence of warps and tidal tails
may complicate the interpretation of the results from fitting ellipses
to a certain HI isophote. Here, no correction for the intrinsic
thickness of the HI layer was applied. However, the relatively large
synthesized beams of imaging arrays at 21cm may smear the observed HI
disks to a rounder appearance. Therefore, a simple correction for beam
smearing was applied to our measurements and the inclination of the HI
disk was determined according to

 $$ \mbox{cos}^2(\mbox{i}_{\mbox{\scriptsize HI}}) = \frac{\mbox{d}^2_{\mbox{\scriptsize HI}} - \Theta^2_{\mbox{\scriptsize d}}}{\mbox{D}^2_{\mbox{\scriptsize HI}} - \Theta^2_{\mbox{\scriptsize D}}} $$

\noindent where D$_{\mbox{\scriptsize HI}}$ and d$_{\mbox{\scriptsize
HI}}$ are the observed major and minor axis diameters of the inclined HI
disk obtained by fitting an ellipse to the outer column density levels.
$\Theta_{\mbox{\scriptsize D}}$ and $\Theta_{\mbox{\scriptsize d}}$ are
the sizes of the synthesized beam in the direction of the major and
minor axis of the HI disk.

\subsection{i$_{\mbox{\scriptsize VF}}$ from HI velocity fields}

The inclination angle of an HI disk can also be measured by fitting
tilted-rings to its velocity field (Begeman \cite{begeman89}). However,
the inclination angle and the rotational velocity are strongly coupled
and reasonable results can only be obtained for inclination angles
between roughly 50 and 75 degrees. This procedure requires accurate
velocity fields with high signal-to-noise ratios as well as many
independent points along a ring. The advantage that velocity fields
offer is the possibility to identify warps and to check the kinematic
regularity of the HI disk. For instance, the optical appearance of a
galaxy may look very regular while the outer regions of the HI disk may
be strongly warped toward edge-on (e.g N3726). Such a warp would broaden
the global profile and an inclination correction based on the optical
axis ratio would lead to an overestimate of the rotational velocity when
dividing the `warp-broadened' line width by sin(i$_{\mbox{\scriptsize
opt}}$).
 
Note that the inclination measurement of a tilted ring may be affected by
non-circular motions due to spiral arms, bars and lopsidedness.

\subsection{the comparison}

For the comparison between the three differently inferred inclination   
angles we considered only those 27 galaxies with fully reduced HI data for
which the velocity fields and integrated HI maps are available.  We
excluded the interacting galaxies (N3769, N3893, U6973) because their
outer isophotes (optical and HI) are affected by tidal tails.  We also
excluded galaxies with perturbed or inadequately sampled velocity
fields\penalty-10000 (N4088, U6969, N4389), galaxies with excessively
patchy HI maps (N4102) and obviously lopsided galaxies (N4051).  These
eliminations leave us with 19 galaxies that have smooth outer isophotes,
well filled HI disks and regular HI velocity fields.

Figure~3 presents the comparison between the three differently inferred
inclination angles using two different values for q$_0$.  When
calculating mean differences and scatters using i$_{\mbox{\scriptsize
VF}}$, only galaxies with i$_{\mbox{\scriptsize HI}}$$<$80$^\circ$ are
considered because kinematic inclinations of highly inclined galaxies
are systematically underestimated.  The error bars on
i$_{\mbox{\scriptsize VF}}$ are based on the variations in
i$_{\mbox{\scriptsize VF}}$ between the various fitted rings but are not
considered any further here.

The upper most panel compares i$_{\mbox{\scriptsize VF}}$ with
i$_{\mbox{\scriptsize HI}}$. No significant offset is found for the 14
galaxies that meet the above-mentioned criteria. Assuming that
i$_{\mbox{\scriptsize VF}}$ and i$_{\mbox{\scriptsize HI}}$ contribute
equally to the scatter of 3.1 degrees implies that the inclination angle
can be determined with an accuracy of 2.2 degrees from either the
velocity fields or from the inclined HI disk. Note that the correlation
turns up for i$_{\mbox{\scriptsize HI}}$$>$80$^\circ$ due to the   
systematic underestimation of i$_{\mbox{\scriptsize VF}}$ for highly   
inclined disks.

Comparing i$_{\mbox{\scriptsize opt}}$ with i$_{\mbox{\scriptsize VF}}$
and i$_{\mbox{\scriptsize HI}}$ does show a significant offset of 
roughly 3 degrees when assuming q$_0$=0.20 (middle panels). This offset
is biggest toward edge-on as would be expected in case of an
overestimate of the intrinsic thickness. Note that there are several
galaxies with an observed optical axis ratio less than 0.20 which have
been assigned an inclination angle of 90$^\circ$.

This 3$^\circ$ offset disappears when q$_0$=0.09 is used (lower panels)
and the rms scatter is reduced to only 1.9 degrees for
i$_{\mbox{\scriptsize opt}}$ versus i$_{\mbox{\scriptsize VF}}$ but is
still 4.0 degrees in case of i$_{\mbox{\scriptsize opt}}$ versus
i$_{\mbox{\scriptsize HI}}$. In the latter case, the scatter is caused
by a few nearly edge-on systems for which the higher uncertainties have
no influence on the deprojection of the rotational velocities.

The adopted inclinations and their errors, listed in column 11 of Table~1
are best estimates based on all the information available for a
particular galaxy, including the morphology of dust lanes if present. 
For galaxies which lack fully reduced HI synthesis data, the inclination
angles were inferred from the optical axis ratios using q$_0$=0.09 for
galaxies of type Sc and later and q$_0$=0.24 for galaxies of type Sbc     
and earlier.  The latter value of q$_0$ seemed justified by the observed
axis ratios of the (nearly) edge-on systems N4013, N4026 and N4111 of  
types Sb, S0 and S0 respectively. Unfortunately, there are not enough
suitable galaxies available to determine q$_0$ as a function of
morphology.

\begin{figure}
  \resizebox{\hsize}{!}{\includegraphics{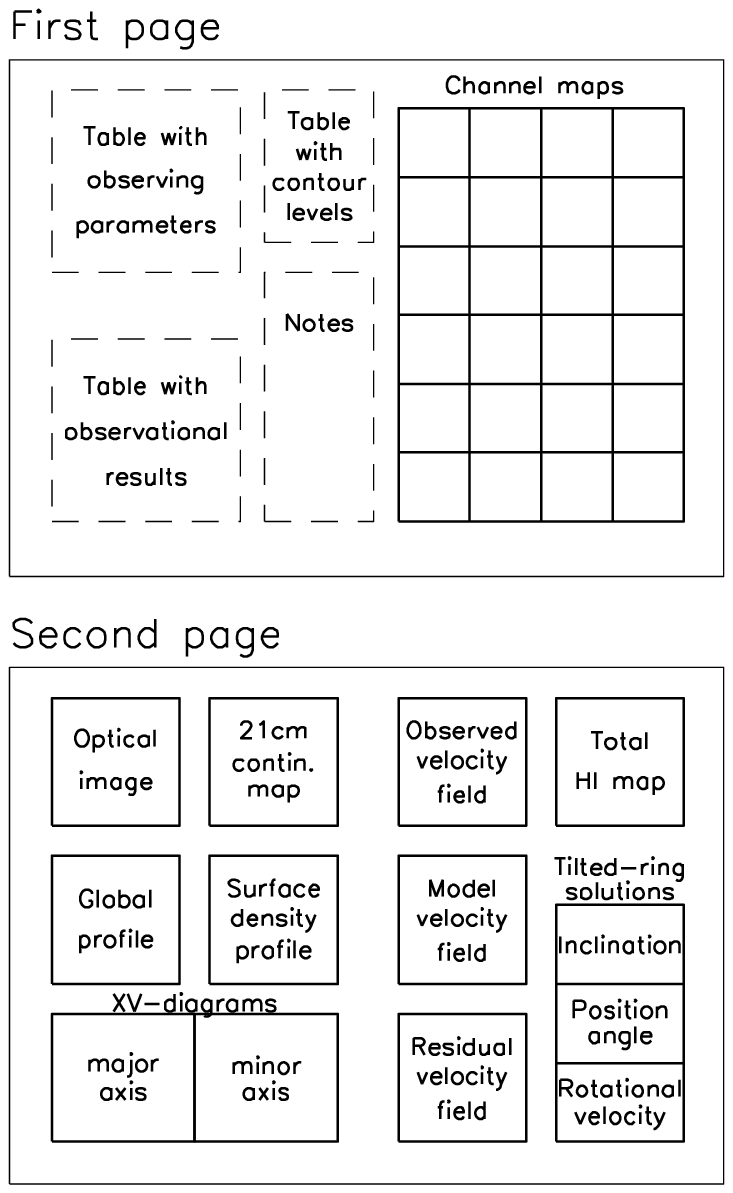}}
  \caption{Layout of the HI atlas pages for the 30 galaxies with fully
reduced data. All the data for these galaxies are presented on two
facing pages. Results for the 13 galaxies with partially reduced data
are presented on a single page per galaxy and include only the channel
maps, the global profile and the XV-diagram. The linear scale is 5.4 kpc
per arcminute.}
  \label{UMaIVfig4}
\end{figure}

\section{The atlas}

The atlas is presented in Appendix B and consists of two parts. The
first part presents the data for the 30 galaxies which have been fully
reduced and analyzed. The second part gives a less elaborate
presentation of the remaining 13 galaxies which have been only partially
reduced. 

The reduction procedures described in Section 3 were applied to the data
at all three angular resolutions.  However, to facilitate the
intercomparison of various results for a galaxy, it was decided to
present the data of a particular galaxy at the same angular resolution
as much as possible.  The rotation curves are in some cases a
combination of the rotation curves at various resolutions, the inner
parts at the highest spatial resolution and the outer parts from data of
lower resolution but higher signal-to-noise ratio.  The channel maps are
sometimes presented at a lower angular resolution than the other data. 

Figure 4 shows the graphical layout for each of the 30 galaxies in the
first part of the atlas.  The data for each of the 13 galaxies in the
second part of the atlas are presented on a single page which contains
the tables and notes as well as the mosaic of channel maps, the global
profile and the XV-diagram along the major axis.  The contents of the
various tables and panels are described below. 

\noindent\textsf{\textbf{Tables and Notes --}} There are three tables
presented for each galaxy.  The upper left table contains information on
the observations like date, integration time and correlator settings. 
The lower left table presents some of the quantities derived from the HI
data like global profile widths, integrated HI flux, systemic velocity
etc.  The upper right table provides the noise and contour levels for
the maps in the various panels.  The contours in the channel maps, 21cm
continuum maps and XV-diagrams are always drawn at levels which are
related to the rms noise. 

The notes contain information about specific aspects of a particular
galaxy like optical and HI appearance. 

\noindent\textsf{\textbf{Channel maps --}} The channel maps show how the
HI emission behaves as a function of velocity.  From these maps it is
possible to recognize the presence of warps, non-circular motions and
HI-bridges between interacting galaxies like N3769/1135+48 and
N3893/N3896.  The contours are drawn at levels of -3, -1.5 (dashed),
1.5, 3, 4.5, 6, 9, 12, 15,...$\sigma$.  The value of the rms noise level
$\sigma$ is given in the upper right table.  In each panel, the cross
indicates the adopted dynamical center of the galaxy.  The ellipse in
the upper left panel is centered on this position and the position angle
is set at the kinematic major axis of the inner regions.  The
ellipticity represents the inclination as derived from the optical axis
ratio and the major axis diameter is equal to D$^{b,i}_{25}$.  The
synthesized beam at half power is shown in the lower left corners of the
left panels.  The panel at the lower right shows the subtracted,
`dirty' continuum map. 

\noindent\textsf{\textbf{Optical image --}} The optical image of a
galaxy was scanned from the blue POSS plates.  In the upper left corner,
the morphological type according to the RC3 is given.  CCD images of far
superior quality can be found in Paper I. An example is given for N3726
in which case the CCD image is pasted into the POSS image.

\noindent\textsf{\textbf{Radio continuum map --}} The cleaned 21cm radio
continuum map is plotted at the same scale as the optical image.  The
contours are drawn at levels of -4, -2 (dashed), 2, 4, 8, 16, 32,
64,...$\sigma$.  The value of the rms noise $\sigma$ is given in the
upper right table.  The ellipse represents the optical image as in the
upper left panel in the channels maps.  The synthesized beam is plotted
in the lower left corner and the small cross indicates the adopted
position of the center of rotation. 

\noindent\textsf{\textbf{Global profile --}} Each point in the global HI
profile gives the primary-beam corrected, total HI flux density
integrated over a single channel map.  The errors are derived as
explained in Section 3.2.  The vertical arrow indicates the systemic
velocity (lower left table on the first page) as derived from the HI
velocity field and does not necessarily indicate the center of the
global HI profile.  A significant offset from the center could indicate
non-circular motions or a strong lopsidedness.  These anomalies can
often be traced in the individual channel maps. 

\noindent\textsf{\textbf{Surface density profile --}} The open and
filled symbols indicate the azimuthally averaged radial surface
densities for the approaching and receding sides.  The solid line
follows the average value.  At the adopted distance of 18.6 Mpc, 1
arcmin corresponds to 5.4 kpc.  The vertical arrow indicates
R$_{25}^{b,i}$.  The profile becomes unreliable for highly inclined
systems since no correction for beam smearing was applied. 

\noindent\textsf{\textbf{XV-diagrams --}} The position-velocity diagrams
are shown for two orthogonal cuts through the adopted center of rotation
along the kinematic major (left) and minor (right) axes. The position
angles of these two axes are printed in the upper right corner of each
panel. Note that the quoted position angles refer to the positive
offset axes. Consequently, the position angle of the major axis refers
to the receding side which also can be inferred from the channel maps.
The vertical dashed line indicates the position of the center of
rotation. The horizontal dashed line indicates the systemic velocity as
derived from either the velocity field for galaxies with fully reduced
data, or from the XV-diagram for galaxies in the second part of the
atlas. The two vertical arrows show where the ellipse with major axis
diameter D$_{25}^{b,i}$ intersects the XV-slice. The horizontal arrows
in the left panel show the systemic velocity V$_{\mbox{\scriptsize
sys}}^{\mbox{\scriptsize GP}}$ derived from the global profile and
V$_{\mbox{\scriptsize sys}}^{\mbox{\scriptsize
GP}}$~$\pm$~$\frac{1}{2}$W$_{R,I}$ where W$_{R,I}$ is the width of the
global profile at the 20\% level of peak flux, corrected for
instrumental broadening (see section 3.2) and turbulent motions
according to TFq. 

Contours are at levels of -3, -1.5 (dashed), 1.5, 3, 4.5, 6, 9, 12
, 15,...$\sigma$.  The value of the rms noise $\sigma$ is given in the
upper right table.  The cross in the lower left corners indicates the
angular and velocity resolutions.  An offset of 1 arcmin corresponds to
a projected distance of 5.4 kpc from the center. 

The crosses give the projected rotational velocities as derived from the
tilted rings fit.  In some cases, at large radii, a cross can be found
without any underlying signal in the XV-diagram.  In such cases, the
rotational velocity at that radius is defined by points in the velocity
field away from the major axis.  The open and filled circles indicate
the projected rotational velocity estimated directly from these
XV-diagrams.  These points must be deprojected using the appropriate
inclination and position angles to obtain the actual rotation curve for
both halves of the galaxy. 

\noindent\textsf{\textbf{Total HI map --}} All pixels in the total HI
map with a positive signal have a greyscale value assigned.  Because the
signal-to-noise ratio along a contour is not constant, the
`3$\sigma$-contour' is not defined. Section 3.3 and the appendix explain
why and how the noise varies across an integrated HI map. 

The second contour in the total HI maps corresponds to the {\em average}
value of all pixels with a signal-to-noise ratio between 2.75 and 3.25
and this contour can thus be considered as a pseudo 3$\sigma$-contour. 
Wherever a contour goes through an area with $(\frac{S}{N})<3$, the
contour is plotted much thinner.  Consequently, the lowest contour,
plotted at the `1.5$\sigma$' level, is plotted thin over most of its
stretch. The various contour levels in atoms$\;$cm$^{-2}$ are given in
the upper right table. The size of the synthesized beam is plotted in
the lower left corner. The beamwidths are the same as in the channel
maps unless specified otherwise in the note. The small cross indicates
the adopted position of the center of rotation (1 arcmin corresponds to
5.4 kpc).

\noindent\textsf{\textbf{Velocity fields --}} The greyscales indicate
the pixels where a radial velocity was measured.  Darker greyscales and
white isovelocity contours indicate the receding side.  The thick first
black contour adjacent to the white ones indicates the adopted systemic
velocity.  In the ideal case of circular motion and no noise, this thick
contour should be a straight line through the center and coinciding with
the kinematic minor axis of the galaxy.  The isovelocity contours are
plotted with constant velocity intervals as given by the upper right
table. The synthesized beam is plotted in the lower left corner. 

The observed velocity field was modeled by fitting tilted rings to it. 
The orientation and rotational velocity of each ring were then used to
construct the model velocity field.  The model velocity field is plotted
with the same orientation and on the same scale as the observed velocity
field.  The isovelocity contours are plotted at the same velocities in
the observed as in the model velocity fields.  For nearly edge-on
systems, the model velocity field is only one or two pixels wide in
which case no contours could be drawn. 

The residual velocity field was made by subtracting the model from the
observed velocity field.  White contours indicate positive residuals,
black contours indicate negative residuals.  The contour levels are
...,-15, -10, -5, 5, 10, 15,...  km$\;$s$^{-1}$.

\noindent\textsf{\textbf{Tilted-ring fits --}} The three combined panels
show the results from the tilted-ring fits to the observed velocity
field.  The upper panel shows the inclination angle, the middle panel
the position angle and the lower panel the rotational velocity. 

The crosses with errorbars in the panels for inclination and position
angle are the results from the second step of the fitting procedure as
explained in section 3.6.1.  The dashed lines, mostly coinciding with
the solid lines,

\begin{landscape}

\begin{table}[p]

\begin{center}
\caption[]{Results from the HI synthesis observations.}

{\fontsize{8.5}{9.5}
\selectfont

\begin{tabular}[t]{lrrrrrrrrrrrrrrrlrrrr}

\hline
\noalign{\vspace{0.8mm}}
Name      &
\multicolumn{2}{c}{W$_{20}$ $\pm$} &
\multicolumn{2}{c}{W$_{50}$ $\pm$} & Res. &
\multicolumn{2}{c}{V$_{\mbox{\scriptsize hel}}$ $\pm$} &
\multicolumn{2}{c}{$\int$Sdv $\pm$} &
\multicolumn{2}{c}{$F_\nu$ $\pm$} &
R$_{\mbox{\scriptsize HI}}$  &
R$^{\mbox{\scriptsize lmp}}$ & 
\multicolumn{2}{c}{V$^{\mbox{\scriptsize lmp}}_{\mbox{\scriptsize rot}}$ $\pm$} & 
\multicolumn{1}{c}{shape} &
\multicolumn{2}{c}{V$_{\mbox{\scriptsize max}}$  $\pm$} &
\multicolumn{2}{c}{V$_{\mbox{\scriptsize flat}}$ $\pm$} \\
\noalign{\vspace{0.8mm}}
 & 
\multicolumn{7}{c}{-- -- -- -- -- -- -- -- km$\;$s$^{-1}$ -- -- -- -- -- -- -- --} &
\multicolumn{2}{c}{Jy$\;$km$\;$s$^{-1}$} &
\multicolumn{2}{c}{-- mJy --} &            
($^\prime$) & ($^\prime$) &
\multicolumn{2}{c}{-- km$\;$s$^{-1}$ --} &
 &
\multicolumn{4}{c}{-- -- -- -- km$\;$s$^{-1}$ -- -- -- --} \\
\noalign{\vspace{0.8mm}}

\multicolumn{1}{c}{(1)}&
\multicolumn{1}{c}{(2)}&
\multicolumn{1}{c}{(3)}&
\multicolumn{1}{c}{(4)}&
\multicolumn{1}{c}{(5)}&
\multicolumn{1}{c}{(6)}&
\multicolumn{1}{c}{(7)}&
\multicolumn{1}{c}{(8)}&
\multicolumn{1}{c}{(9)}&
\multicolumn{1}{c}{(10)}&
\multicolumn{1}{c}{(11)}&
\multicolumn{1}{c}{(12)}&
\multicolumn{1}{c}{(13)}&
\multicolumn{1}{c}{(14)}&
\multicolumn{1}{c}{(15)}&
\multicolumn{1}{c}{(16)}&
\multicolumn{1}{c}{(17)}&
\multicolumn{1}{c}{(18)}&
\multicolumn{1}{c}{(19)}&
\multicolumn{1}{c}{(20)}&
\multicolumn{1}{c}{(21)} \\
\noalign{\vspace{0.8mm}}

\hline
\hline

\noalign{\vspace{0.8mm}}

\multicolumn{21}{l}{\hspace{-2mm}{\em \underline{Galaxies with fully reduced HI data:}}} \\
\noalign{\vspace{0.8mm}}
U6399    & 188.1 & 1.4 & 172.5 & 2.9 &  8.3 &  791.5 & 0.6 &  10.5 & 0.3 &$<$2.5 &     & 1.68 & 1.50 &  88 &  5 & R/F            &  88 &  5 &  88 &  5 \\
U6446    & 154.1 & 1.0 & 131.9 & 1.2 &  5.0 &  644.3 & 0.8 &  40.6 & 0.5 &$<$7.2 &     & 2.96 & 2.93 &  80 & 11 & F     \hfill L &  82 &  4 &  82 &  4 \\
N3726    & 286.5 & 1.6 & 260.6 & 1.8 &  5.0 &  865.6 & 0.9 &  89.8 & 0.8 &  49.7 & 5.0 & 4.24 & 6.22 & 167 & 15 & F/(D)          & 162 &  9 & 162 &  9 \\
N3769    & 265.3 & 6.7 & 230.5 & 3.6 &  8.3 &  737.3 & 1.8 &  62.3 & 0.6 &  12.1 & 2.9 & 4.31 & 7.10 & 113 & 11 & F/(D)          & 122 &  8 & 122 &  8 \\
U6667    & 187.5 & 1.4 & 178.1 & 1.9 &  5.0 &  973.2 & 1.2 &  11.0 & 0.4 &$<$2.7 &     & 1.64 & 1.50 &  86 &  3 & R     \hfill L &  86 &  3 &  86 &  3 \\
N3877    & 373.4 & 5.0 & 344.5 & 6.2 & 33.2 &  895.4 & 3.8 &  19.5 & 0.6 &  35.6 & 2.4 & 2.19 & 2.17 & 169 &  7 & F     \hfill L & 167 & 11 & 167 & 11 \\
N3893    & 310.9 & 1.0 & 277.9 & 4.1 &  5.0 &  967.2 & 1.0 &  69.9 & 0.5 & 137.4 & 2.9 & 3.98 & 3.88 & 148 & 19 & F/(D)          & 188 & 11 & 188 & 11 \\
N3917    & 294.5 & 1.9 & 279.1 & 2.1 &  8.3 &  964.6 & 1.4 &  24.9 & 0.6 &$<$7.2 &     & 2.69 & 2.83 & 137 &  8 & F              & 135 &  3 & 135 &  3 \\
N3949    & 286.5 & 1.4 & 258.3 & 1.7 &  8.3 &  800.2 & 1.2 &  44.8 & 0.4 & 134.1 & 3.6 & 2.62 & 1.35 & 169 &  8 & F     \hfill L & 164 &  7 & 164 &  7 \\
N3953    & 441.9 & 2.4 & 413.9 & 3.2 & 33.1 & 1052.3 & 2.0 &  39.3 & 0.8 &  50.9 & 2.5 & 3.32 & 3.00 & 215 & 10 & F              & 223 &  5 & 223 &  5 \\
N3972    & 281.2 & 1.4 & 260.7 & 5.5 &  8.3 &  852.2 & 1.4 &  16.6 & 0.4 &$<$5.8 &     & 1.92 & 1.67 & 134 &  5 & R     \hfill L & 134 &  5 & ... & .. \\
U6917    & 208.9 & 3.2 & 189.6 & 1.6 &  8.3 &  910.7 & 1.4 &  26.2 & 0.3 &$<$4.4 &     & 2.42 & 2.00 & 111 &  7 & R/F            & 104 &  4 & 104 &  4 \\
U6923    & 166.8 & 2.4 & 147.1 & 4.5 & 10.0 & 1061.6 & 2.2 &  10.7 & 0.6 &$<$2.6 &     & 1.29 & 1.02 &  81 &  5 & R     \hfill L &  81 &  5 & ... & .. \\
U6930    & 136.5 & 0.5 & 122.1 & 0.7 &  8.3 &  777.2 & 0.4 &  42.7 & 0.3 &$<$5.8 &     & 3.20 & 3.17 & 108 &  7 & R/F            & 107 &  4 & 107 &  4 \\
N3992    & 478.5 & 1.4 & 461.4 & 2.4 & 10.0 & 1048.2 & 1.2 &  74.6 & 1.5 &  30.2 & 7.6 & 4.75 & 6.67 & 237 &  9 & F/D            & 272 &  6 & 242 &  5 \\
U6940    &  59.3 & 3.8 &  40.6 & 7.8 & 10.0 & 1118.0 & 1.7 &   2.1 & 0.3 &$<$1.3 &     & 0.61 & 0.33 &  37 &  4 & R              &  37 &  4 & ... & .. \\
U6962    & 220.3 & 6.6 & 182.4 & 3.7 &  8.3 &  807.4 & 3.2 &  10.0 & 0.3 &  13.4 & 1.7 & 1.38 & 1.00 & 171 &  7 & R     \hfill L & 171 &  7 & ... & .. \\
N4010    & 277.7 & 1.0 & 264.1 & 1.2 &  8.3 &  901.9 & 0.8 &  38.2 & 0.3 &  16.9 & 1.6 & 3.36 & 2.00 & 122 &  2 & (R)/F \hfill L & 128 &  9 & 128 &  9 \\
U6969    & 132.1 & 6.4 & 123.5 & 2.9 & 10.0 & 1118.5 & 2.4 &   6.1 & 0.5 &$<$3.8 &     & 0.95 & 0.85 &  79 &  5 & R              &  79 &  5 & ... & .. \\
U6973    & 367.8 & 1.8 & 350.4 & 1.2 &  8.3 &  700.5 & 1.0 &  22.9 & 0.2 & 127.5 & 2.1 & 2.21 & 1.50 & 180 &  8 & F/(D)          & 173 & 10 & 173 & 10 \\
U6983    & 188.4 & 1.3 & 173.0 & 1.1 &  5.0 & 1081.9 & 0.8 &  38.5 & 0.6 &$<$5.4 &     & 3.07 & 3.00 & 109 & 12 & F              & 107 &  7 & 107 &  7 \\
N4051    & 255.4 & 1.8 & 224.6 & 1.5 &  5.0 &  700.3 & 1.2 &  35.6 & 0.8 &  26.5 & 2.6 & 2.89 & 2.33 & 153 & 10 & R/F   \hfill L & 159 & 13 & 159 & 13 \\
N4085    & 277.4 & 6.6 & 255.4 & 7.8 & 19.8 &  745.7 & 5.0 &  14.6 & 0.9 &  44.1 & 1.3 & 1.94 & 1.18 & 136 &  7 & R/F   \hfill L & 134 &  6 & 134 &  6 \\
N4088    & 371.4 & 1.7 & 342.1 & 1.9 & 19.8 &  756.7 & 1.2 & 102.9 & 1.1 & 222.3 & 1.9 & 4.25 & 4.10 & 174 &  8 & F/(D) \hfill L & 173 & 14 & 173 & 14 \\
N4100    & 401.8 & 2.0 & 380.5 & 1.8 & 19.9 & 1074.4 & 1.3 &  41.6 & 0.7 &  54.3 & 1.7 & 3.45 & 4.35 & 159 &  9 & F/D            & 195 &  7 & 164 & 13 \\
N4102    & 349.8 & 2.0 & 322.4 & 8.5 &  8.3 &  846.3 & 2.0 &   8.0 & 0.2 & 276.0 & 1.5 & 1.16 & 0.83 & 178 & 12 & F              & 178 & 11 & 178 & 11 \\
N4157    & 427.6 & 2.2 & 400.7 & 3.1 & 19.9 &  774.4 & 1.8 & 107.4 & 1.6 & 179.6 & 2.3 & 4.60 & 5.67 & 185 & 14 & F/D            & 201 &  7 & 185 & 10 \\
N4183    & 249.6 & 1.2 & 232.5 & 1.5 &  8.3 &  930.1 & 1.0 &  48.9 & 0.7 &$<$5.8 &     & 3.07 & 4.02 & 113 & 11 & F/D   \hfill L & 115 &  6 & 109 &  4 \\
N4217    & 428.1 & 5.1 & 395.6 & 3.8 & 33.2 & 1027.0 & 3.0 &  33.8 & 0.7 & 115.6 & 2.2 & 3.19 & 3.17 & 178 & 12 & F/D            & 191 &  6 & 178 &  5 \\
N4389    & 184.0 & 1.5 & 164.9 & 1.6 &  8.3 &  718.4 & 1.2 &   7.6 & 0.2 &  23.3 & 1.2 & 1.30 & 1.02 & 110 &  8 & R              & 110 &  8 & ... & .. \\
\noalign{\vspace{0.5mm}}
\multicolumn{21}{l}{\hspace{-2mm}{\em \underline{Galaxies with partially reduced HI data:}}} \\
\noalign{\vspace{0.8mm}}
N3718    & 492.8 & 1.0 & 465.7 & 1.0 & 33.2 &  993.0 & 0.8 & 140.9 & 0.9 &  11.4 & 0.4 &      & 6.67 & 223 & 12 & F              & 232 & 11 & 232 & 11 \\
N3729    & 270.8 & 1.5 & 253.2 & 3.9 & 33.2 & 1059.8 & 1.4 &   5.5 & 0.3 &  18.0 & 0.9 &      & 1.00 & 151 & 11 & F              & 151 & 11 & 151 & 11 \\
U6773    & 110.4 & 2.3 &  91.1 & 2.2 &  8.3 &  923.6 & 1.6 &   5.6 & 0.4 &$<$2.6 &     &      & 0.67 &  45 &  5 & R     \hfill L &  45 &  5 & ... & .. \\
U6818    & 166.9 & 2.3 & 141.9 & 5.7 &  8.3 &  808.1 & 2.1 &  13.9 & 0.2 &   2.4 & 1.0 &      & 1.33 &  74 &  7 & R/(F) \hfill L &  73 &  5 &  73 &  5 \\
U6894    & 141.8 & 1.1 & 132.2 & 1.5 &  8.3 &  848.6 & 1.8 &   5.8 & 0.2 &$<$2.7 &     &      & 0.67 &  63 &  5 & R              &  63 &  5 & ... & .. \\
N3985    & 160.2 & 3.7 &  88.0 & 2.4 &  8.3 &  948.2 & 2.0 &  15.7 & 0.6 &   9.7 & 1.4 &      & 0.50 &  93 &  7 & R              &  93 &  7 & ... & .. \\
N4013    & 425.0 & 0.9 & 395.0 & 0.8 & 33.0 &  831.3 & 0.6 &  41.5 & 0.2 &  36.3 & 0.8 &      & 6.12 & 170 & 10 & F/D            & 195 &  3 & 177 &  6 \\
U7089    & 156.7 & 1.7 &  97.7 & 3.0 & 10.0 &  770.0 & 1.5 &  17.0 & 0.6 &$<$3.4 &     &      & 1.75 &  79 &  7 & R     \hfill L &  79 &  7 & ... & .. \\
U7094    &  83.7 & 1.7 &  71.9 & 5.5 & 10.0 &  779.6 & 1.6 &   2.9 & 0.2 &$<$2.6 &     &      & 1.00 &  35 &  6 & R     \hfill L &  35 &  6 & ... & .. \\
N4117    & 289.4 & 7.5 & 260.3 & 5.2 & 10.0 &  934.0 & 1.5 &   6.9 & 1.1 &   3.7 & 1.2 &      &      &     &    & ?              &     &    &     &    \\
N4138    & 331.6 & 4.5 & 266.0 & 7.8 & 19.9 &  893.8 & 3.9 &  19.2 & 0.7 &  16.7 & 4.6 &      & 3.55 & 150 & 21 & F/D            & 195 &  7 & 147 & 12 \\
N4218    & 138.0 & 5.0 &  79.9 & 1.9 &  8.3 &  729.9 & 1.7 &   7.8 & 0.2 &   6.3 & 0.8 &      & 0.33 &  73 &  7 & R              &  73 &  7 & ... & .. \\
N4220    & 438.1 & 1.3 & 423.3 & 3.3 & 33.1 &  914.2 & 1.2 &   4.4 & 0.3 &$<$4.9 &     &      &      &     &    & ?              &     &    &     &    \\
\hline
\hline
\end{tabular}
}
\end{center}
\end{table}
\end{landscape}

\noindent in these upper two panels indicate the final values of
the inclination and position angles kept fixed when the rotational
velocity was fitted.  The resulting rotation curve is shown by crosses
with errorbars in the lower panel.  The errorbars indicate the formal
errors, as given by the least squares minimization algorithm. 

The horizontal arrows in the upper two panels indicate the inclination
and position angles as derived from the optical isophotes in the outer
regions.  The diamonds indicate the inclination and position angles as
determined from the total HI maps.  When the total HI maps are very
patchy, these diamonds are very uncertain.  The horizontal arrow in the
lower panel indicates the rotational velocity as derived from the width
of the global HI profile corrected for instrumental broadening,
turbulent motion and inclination. The adopted inclination is
representative for the outer parts. The vertical arrow in the lower
panel indicates R$_{25}^{b,i}$.

\begin{figure}[t]
  \resizebox{\hsize}{!}{\includegraphics{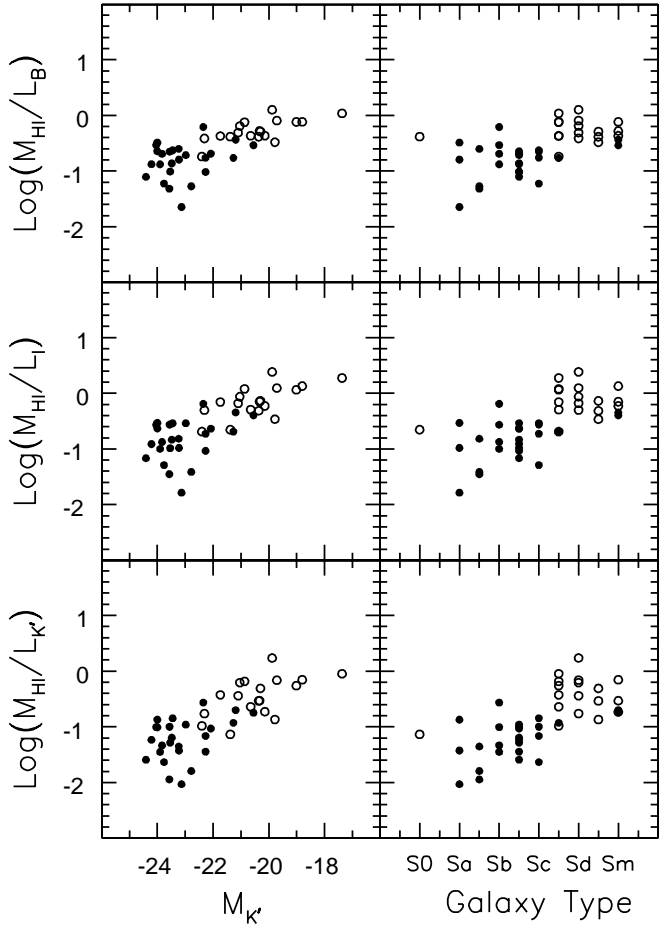}}
  \caption{Correlations between HI mass-to-light ratios and absolute
$K^\prime$-band magnitudes (left column) and morphological type (right
column).  Solid symbols indicate HSB galaxies and open symbols denote
galaxies of the LSB type.}
  \label{UMaIVfig5}
\end{figure}

The solid lines in the upper and middle panels show the inclination and
position angles that were adopted to deproject the radial velocities
determined from the XV-diagrams.  This deprojection results in the
rotation curves plotted as open and filled circles in the lower panel
(same symbols as in the XV-diagrams).  Note that although the rotational
velocities at a certain radius may be different for the approaching and
receding sides, both sides were assumed to have the same inclination and
position angles at that radius.  The solid line in the lower panel shows
the mean rotation curve derived from the XV-diagram.  1 arcmin on the
horizontal axis corresponds to 5.4 kpc. 

\begin{figure*}
  \resizebox{\hsize}{!}{\includegraphics{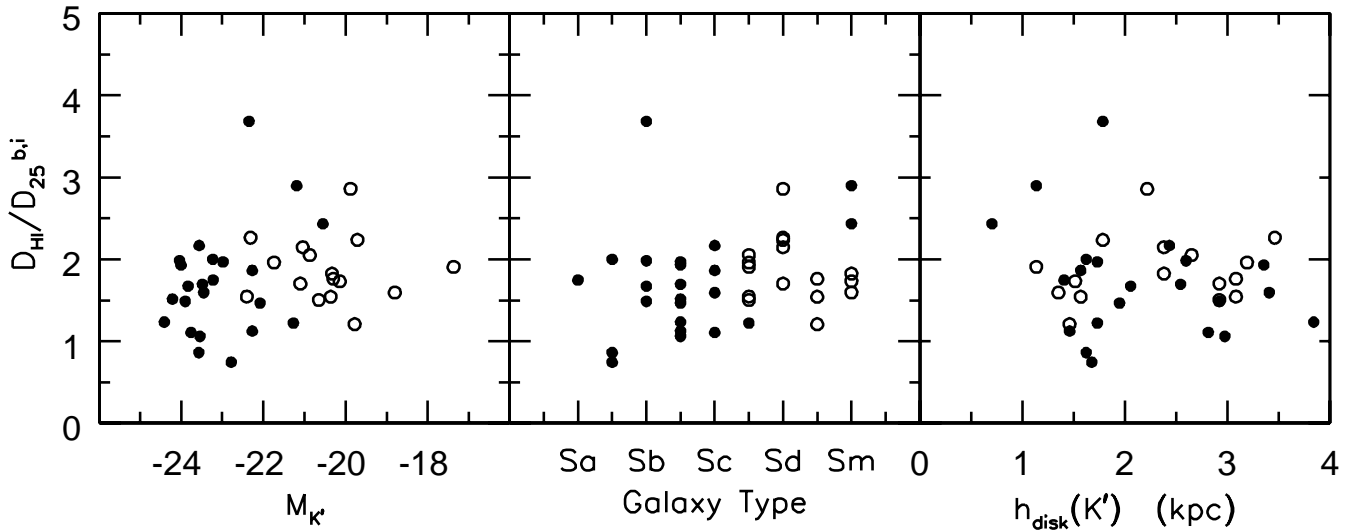}}
  \caption{Ratio of HI-to-optical diameter as a function of $K^\prime$
magnitudes, morphological type and disk scale length.  The HI diameters
were measured at the 1 M$_\odot$pc$^{-2}$ isodensity contour.  Solid
symbols indicate HSB galaxies and open symbols denote galaxies of the LSB
type.}
  \label{UMaIVfig6}
\end{figure*}

\noindent\textsf{\textbf{Tabulated data --}} The various parameters
derived from the HI data presented here are summarized in Table~5.

 \noindent {\it Column} (1) gives the NGC or UGC numbers.  \\
 \noindent {\it Columns} (2)-(5) give the uncorrected widths with formal
errors of the global profiles at 20\% and 50\% of the peak flux.  \\
 \noindent {\it Column} (6) gives the instrumental velocity resolution at
which the global profiles were observed.  \\
 \noindent {\it Columns} (7) and 8 contain the heliocentric systemic
velocities and their uncertainties as derived from the global profiles.
\\
 \noindent {\it Columns} (9) and (10) provide the integrated HI flux and
the uncertainty in Jy$\;$km$\;$s$^{-1}$.  \\
 \noindent {\it Columns} (11) and (12) contain the 21cm continuum flux
density and its uncertainty in mJy.  In case no continuum flux was
detected, a 3$\sigma$ upper limit for extended emission is given.  \\
 \noindent {\it Column} (13) gives the radius of the HI disk,
R$_{\mbox{\scriptsize HI}}$ in arcmin, at the azimuthally averaged
surface density of 1 M$_{\odot}$pc$^{-2}$, measured from the radial
surface density profiles. \\
 \noindent {\it Column} (14) gives the radius R$^{\mbox{\scriptsize
lmp}}$ of the last measured point of the rotation curve in arcmin.  The
differences between R$^{\mbox{\scriptsize lmp}}$ and
R$_{\mbox{\scriptsize HI}}$ depend on the sensitivity of the measurement
and the distribution of the HI gas along the kinematic major axis.  \\
 \noindent {\it Columns} (15) and (16) give the rotational velocity of
the last measured point V$^{\mbox{\scriptsize lmp}}$ and its
uncertainty.  \\
 \noindent {\it Column} (17) contains information on the overal shape of
the rotation curve; R: rising rotation curve, F: the rotation curve
shows a flat part, D: the rotation curve shows a declining part, L:
lopsided. \\
 \noindent {\it Columns} (18) and (19) give the maximum observed
rotational velocity V$_{\mbox{\scriptsize max}}$ and its uncertainty. 
For galaxies with a rising rotation curve (R) V$_{\mbox{\scriptsize
max}}$~=~V$^{\mbox{\scriptsize lmp}}$. \\
 \noindent {\it Columns} (20) and (21) give the average rotational
velocity of the flat part of the rotation curve V$_{\mbox{\scriptsize
flat}}$ and its uncertainty.  For galaxies with a flat rotation curve
(F) V$_{\mbox{\scriptsize flat}}$~=~V$_{\mbox{\scriptsize max}}$ may
deviate from V$^{\mbox{\scriptsize lmp}}$ because V$_{\mbox{\scriptsize
flat}}$ was averaged over the flat part of the rotation curve while
V$^{\mbox{\scriptsize lmp}}$ was measured at a single point.  \\

\section{HI properties of spirals}

The HI survey of the Ursa Major cluster presented here provides not only
the kinematical information necessary for the study of the Tully-Fisher
relation and of the dark and luminous matter for a well defined sample
of galaxies. It also serves to investigate the general HI properties of
disks and to make a comparison with galaxies in the field and with
galaxies in denser environments. The HI studies of the Virgo cluster
galaxies by Warmels (\cite{warmels88a}, \cite{warmels88b}) and
especially by Cayatte et al. (\cite{cayatte90}) have shown that the
spiral galaxies in the central parts of the cluster have smaller HI
disks of lower surface density. In the Hydra Cluster McMahon
(\cite{mcmahon}) did not find any such significant HI deficiency.  She
did find, however, a surprisingly large number of isolated HI-rich dwarf
galaxies near the center of the cluster.  Dickey (\cite{dickey97})
surveyed the more distant Hercules SuperCluster and found a similar HI
deficiency of spirals near the X-ray gas as in the case of Virgo. 

The Ursa Major cluster differs from those just mentioned. It has no 
central concentration, no X-ray emitting gas and contains mainly spirals
of  late morphological types. In many respects its conditions are very
similar to  those of a field environment. For this reason it is useful
to compare the  properties of the Ursa Major spirals not only with those
of galaxies in  dense cluster environments but also with those of field
galaxies as found  in various recent studies (see e.g. Broeils
\cite{broeils92}, Puche \& Carignan \cite{puche},  Rhee \cite{rhee96a},
Swaters \cite{swaters}). 

Here we give only a brief description of the global parameters and of the 
main properties of the HI disks of the Ursa Major galaxies. A more detailed
discussion and a comparison with results from previous work is beyond
the scope if this data paper.

\subsection{Global parameters}

Integral properties and global parameters of spiral galaxies have been
derived for a large number of objects from single-dish observations (cf.
 Roberts and Haynes \cite{roberts}).  In recent years also synthesis
observations (Broeils \cite{broeils92}, Rhee \cite{rhee96a}) have been
used to obtain similar information for smaller samples of galaxies. 

The M$_{\mbox{\scriptsize HI}}$/L ratios obtained for the galaxies of
the Ursa Major sample listed in Table~5 are shown here in Figure~5 as a
function of absolute magnitude and of morphological type.  It is well
known (see refs.  above) that the M$_{\mbox{\scriptsize HI}}$/L ratio of
galaxies depends on luminosity and morphological type.  The present
sample of galaxies shows a clear increase of the HI mass fraction with
decreasing luminosity and from early to late morphological types.  The
correlation is clearly stronger for the $K^\prime$-band magnitudes which
is a better tracer of the stellar mass.

\subsection{Sizes of HI disks and radial surface density profiles}

Detailed information on the sizes and radial distributions of HI disks
has been obtained recently from synthesis observations of limited
samples of field and cluster galaxies (Broeils and Van Woerden
\cite{broeils94}, Cayatte et al. \cite{cayatte94}, Rhee \cite{rhee96a}
and \cite{rhee96b}).  Here we present only some of the main results on
the comparison of HI and optical diameters, on the relation between HI
mass and diameter and on the radial density profiles for the Ursa Major
sample. 

Figure~6 shows the ratio of the HI diameter D$_{\mbox{\scriptsize HI}}$
(defined at an HI surface density of 1 M$_\odot$pc$^{-2}$) to the
optical diameter D$_{25}^{\mbox{\scriptsize b,i}}$ as a function of
luminosity, morphological type and disk scale-length.  The diagrams do
not indicate any clear trend or dependence of the diameter ratio on any
of those quantities.  The spread is large.  There may be a hint of a
slight increase of the ratio from early to later types and from more
luminous to less luminous systems.  For almost all galaxies
D$_{\mbox{\scriptsize HI}}$ is larger than D$^{b,i}_{25}$. 

As shown in previous investigations (see refs. above), there is a tight
correlation between HI mass and HI diameter as illustrated in Figure~ 7.
This implies a nearly constant mean HI surface density regardless of
size. The HI mass correlates also with the optical diameter, but, as in
previous work, with a much larger scatter.

\begin{figure}
  \resizebox{\hsize}{!}{\includegraphics{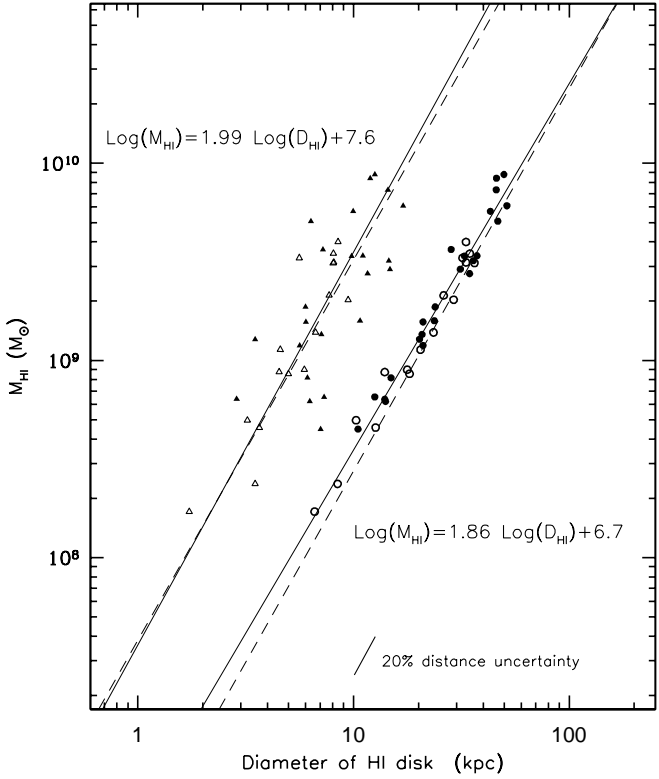}}
  \caption{Correlations between HI mass and the isophotal diameters of
the stellar (triangles) and HI disks (circles). Solid lines indicate
fits to the plotted data points while the dashed lines represent the
fits found by Broeils (\cite{broeils92}). Filled symbols indicate HSB
galaxies and open symbols denote galaxies of the LSB type. Triangles
are offset by 0.3 dex to the left.}
  \label{UMaIVfig7}
\end{figure}

\subsection{Warps, asymmetries and interactions}

\begin{table}
\begin{center}

\caption[]{Warps, asymmetries and interactions.}

\begin{tabular}[t]{lccccc}
\hline
\noalign{\vspace{0.8mm}}
Name & Type & Warped & \multicolumn{2}{c}{Lopsided} & Inter$-$ \\
     &      &        &    HI distr.   &   HI kin.   & acting   \\
\noalign{\vspace{0.8mm}}
\hline
\hline
\noalign{\vspace{0.8mm}}

U6399  &  Sm   &          & $\times$ &          &   \\ 
U6446  &  Sd   &          &          & $\times$ &   \\
N3718  &  Sa   & $\times$ &          &          &   \\   
N3726  &  SBc  & $\times$ &          &          &   \\
N3729  &  SBab &          &          &          &   \\   
\hline
N3769  &  SBb  & $\times$ &          &          & $\times$ \\
U6667  &  Scd  &          &          & $\times$ &   \\
N3877  &  Sc   &          &          & $\times$ &   \\
U6773  &  Sm   &          &          & $\times$ &   \\   
N3893  &  Sc   &          &          &          & $\times$ \\
\hline
N3917  &  Scd  &          & $\times$ &          &   \\
U6818  &  Sd   &          &          & $\times$ &   \\   
N3949  &  Sbc  &          & $\times$ & $\times$ &   \\
N3953  &  SBbc &          &          &          &   \\
U6894  &  Scd  &          &          &          &   \\   
\hline
N3972  &  Sbc  &          &          & $\times$ &   \\
U6917  &  SBd  & $\times$ &          &          &   \\
N3985  &  Sm   & $\times$ &          &          &   \\   
U6923  &  Sdm  & $\times$ &          & $\times$ &   \\
U6930  &  SBd  &          & $\times$ &          &   \\
\hline
N3992  &  SBbc &          &          &          &   \\
U6940  &  Scd  &          &          &          &   \\
N4013  &  Sb   & $\times$ &          &          &   \\
U6962  &  SBcd &          &          & $\times$ & $\times$ \\
N4010  &  SBd  & $\times$ &          & $\times$ &   \\
\hline
U6969  &  Sm   &          &          &          &   \\
U6973  &  Sab  & $\times$ &          &          & $\times$ \\
U6983  &  SBcd &          &          &          &   \\
N4051  &  SBbc &          &          & $\times$ &   \\
N4085  &  Sc   &          &          & $\times$ &   \\
\hline
N4088  &  Sbc  & $\times$ & $\times$ & $\times$ &   \\
U7089  &  Sdm  &          & $\times$ &          &   \\
N4100  &  Sbc  & $\times$ &          &          &   \\
U7094  &  Sdm  &          & $\times$ &          &   \\
N4102  &  SBab &          &          &          &   \\
\hline
N4117  &  S0   &          &          &          &   \\
N4138  &  Sa   & $\times$ & $\times$ &          &   \\
N4157  &  Sb   & $\times$ &          &          &   \\
N4183  &  Scd  & $\times$ &          &          &   \\
N4218  &  Sm   &          &          &          &   \\
\hline
N4217  &  Sb   &          &          &          &   \\
N4220  &  Sa   &          &          &          &   \\
N4389  &  SBbc &          &          &          &   \\
\hline
\hline

\end{tabular}
\end{center}
\end{table}

The radial distributions of the HI surface densities are shown in
Figure~8. Only galaxies with fully reduced data, more inclined than 80
degrees and with R$_{\mbox{\scriptsize HI}}<1$ arcmin are considered in
order to avoid the most severe cases of beam smearing. There is clearly
a considerable diversity of shapes and intensities. The upper row shows
the profiles grouped for galaxies of similar morphological types. The
dotted lines represent low surface brightness galaxies. No obvious trend
with morphological type or surface brightness can be discerned. However,
in the lower row, the profiles are grouped according to the galaxy
properties as listed in Table~6. In this case, a clear trend is visible
in the sense that galaxies with high HI surface densities in their inner
regions are either involved in interactions, are lopsided or display a
warped HI disk. Especially in the case of interacting and strongly
lopsided systems, no longlived stable gas orbits can be expected and
evidently, the cold gas becomes concentrated toward the inner regions of
the disk. Note that some galaxies appear in more than one panel.

\begin{figure*} 
  \resizebox{\hsize}{!}{\includegraphics{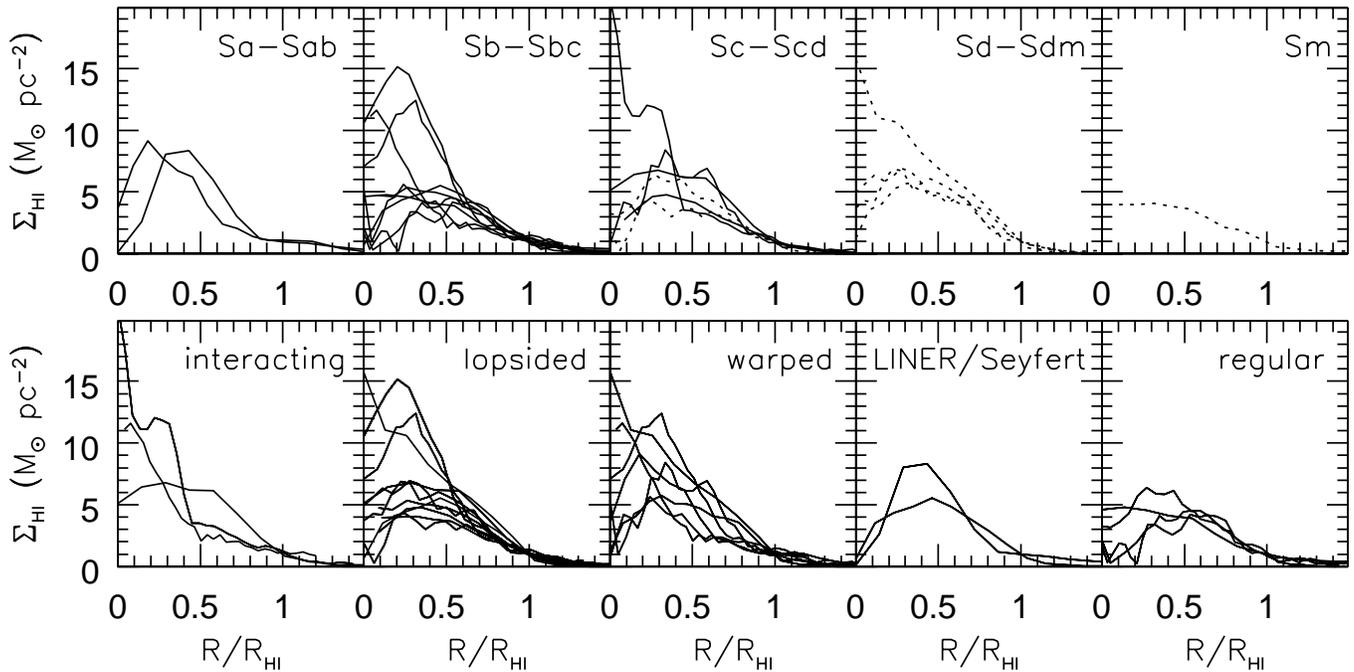}}
  \caption{Azimuthally averaged deprojected radial HI surface density
profiles of galaxies with fully reduced data, less inclined than 80
degrees and with R$_{\mbox{\scriptsize HI}}>1$ arcmin. The profiles were
scaled in radius by the radius of the HI disk measured at the 1
M$_\odot$pc$^{-2}$ isodensity contour. Upper row:
$\Sigma_{\mbox{\scriptsize HI}}$ as a function of morphological type.
The dashed lines indicate low surface brightness galaxies. Lower row:
$\Sigma_{\mbox{\scriptsize HI}}$ as a function of the kinematic state,
according to Table~6, and nuclear activity.}
  \label{UMaIVfig8}
\end{figure*}

Warps are thought to be a quite common feature of the outer HI layers of
spiral galaxies.  But, in spite of attempts made in recent years (Bosma,
1991), a good statistics on their occurrence does not exist yet.  The
information on HI warps provided by the present survey of the Ursa Major
sample is somewhat limited mainly because of the already noted small
radial extent of the HI layers with respect to the optical.  There are
pronounced HI warps like the well-known ones found in NGC 3718 (Schwarz
1985) and NGC 4013 (Bottema 1995) and those seen in other galaxies
(N3726, N3985 N4010, N4157, N4183).  In most cases the warps are visible
in the outer parts beyond the optical bright disk and are present in
normal, regular, not interacting disks.  But there are also systems,
like NGC 4088, which are strongly distorted in their optical appearance
and also in their kinematics.  Similar distortions are found in clearly
interacting systems like NGC 3769.  In the sample of 43 galaxies
(Table~6) there are at least 13 objects with clear indications of
warping. 

Also asymmetries are thought to occur frequently in field spirals
(Richter and Sancisi, 1994).  In the present sample we see a large
number of objects (at least half, see Table~6) with a lopsided HI
distribution and/or kinematics.  The majority shows kinematical
asymmetries: on one side of the disk the rotation curve rises more
slowly and reaches its flat part at a larger radius than on the other
side (see for example NGC 3877, 3949, 4051).  Note that this occurs in
the inner parts of non-interacting, regular, normal systems. 

Finally, there are four galaxies in this sample of 43 which have close
companions and show clear HI signs of tidal interactions.  A few more
objects (examples NGC 3718, 4088) show distortions or peculiar
structures in their density and velocity maps.

\section{Concluding remarks}

This data paper has presented the results of an extensive HI synthesis
imaging survey with the WSRT of a well-defined complete equidistant
sample of spiral galaxies in the nearby Ursa Major cluster. Figures B.1
and B.2 show a compilation of all the available HI maps.  Individual
galaxies are at their proper position on the sky but they are
individually four times enlarged. Some galaxies had to be shifted to
avoid overlapping maps. Rotation curves have been derived for most
galaxies as well as detailed information on the kinematical state of the
galaxy disks as indicated by the presence of global perturbations,
warps, interactions and lopsidedness. Since the galaxies were not
selected on the basis of their HI size or content, the quality of the
kinematical data varies widely from galaxy to galaxy. Nevertheless,
these data will be useful for an analysis of the statistical properties
of the Tully-Fisher relation and the rotation curves may be decomposed
into contributions from the main dynamical constituents like the stellar
and gaseous disks, the bulge and the dark matter halo. These issues will
be addressed in forthcoming papers. 

Those who would like to use the rotation curves for their own purposes
are advised not to take the rotation curves at face value but to check
their validity against the actual data and to take notice of the
comments provided on the atlas pages. It should be realized that the
rotation curves presented here are derived from the kinematics of the HI
gas which is assumed to be a tracer of a galaxy's potential. There are
some caveats to be aware of like the projection effects of streaming
motions and partially filled HI disks of edge-on systems. 

Finally, optical long-slit spectroscopy is already available for nearly
all galaxies in the sample. These high resolution optical rotation
curves will be used to supplement the HI rotation curves in the inner
regions, allowing for better decompositions and maximum-disk
constraints.

\begin{acknowledgements}

 The Westerbork Synthesis Radio Telescope is operated by the
Netherlands Foundation for Research in Astronomy (NFRA/ASTRON), with
financial support by the Netherlands Organization for Scientific
Research (NWO).  This research has made use of the NASA/IPAC
Extragalactic Database (NED) which is operated by the Jet Propulsion
Laboratory, Caltech, under agreement with the National Aeronautics and
Space Association. The results presented in this paper were obtained
during MV's thesis research at the Kapteyn Institute of the University
of Groningen, The Netherlands. This paper was finalized at the
National Radio Astronomy Observatory which is a facility of the
National Science Foundation operated under cooperative agreement by
Associated Universities, Inc.

\end{acknowledgements}

\appendix

\clearpage

\section{The noise in an integrated HI map.}

This appendix explains how the noise in a total HI map can be
calculated.  A total HI map is usually constructed from a 3 dimensional
datacube containing a number of so called channelmaps.  Each channelmap
shows an HI image of the galaxy at a certain velocity.  A total HI map
is made by adding those channelmaps which contain the HI signal.  Before
adding the channel maps the signal in each channelmap should be
isolated.  When the signal is not isolated one merely adds noise to the
total HI map because the location of the signal in a channel map varies
with velocity due to the galactic rotation.  The signals can be isolated
interactively by blotting away the surrounding noise or in a more
objective way by taking a certain contour level in the smoothed maps as a
mask.  As a consequence of adding channel maps with isolated regions, the
noise in the total HI map is not constant but varies from pixel to
pixel.  The noise at a certain pixel in the total HI map depends on the
number N of non-blank pixels at the same position in the individual
channel maps that were added. 

In case the data cube was obtained with an uniform taper during the
observation, the noise $\sigma^{u}$ in two channelmaps will be
independent.  The noise equivalent bandwidth B$^{u}$ in a uniform
tapered spectrum is equal to the channel separation b.  When adding N
uniform tapered channelmaps at a certain pixel the noise
$\sigma^{u}_{N}$ at the same pixel position in the total HI map will be
increased by a factor $\sqrt{N}$:

$$ \sigma^{u}_{N} = \sqrt{N} \; \sigma^{u} $$

Usually, the observations are made using a hanning taper in which case the
noise in two adjacent channelmaps is no longer independent. A hanning
taper effectively smooths the data in velocity by convolving the velocity
profile at each pixel. If U$_{i}$ is the pixel value in the i$^{th}$
uniform tapered channel map, the value H$_{i}$ in the i$^{th}$ hanning
tapered channelmap is given by 

$$ H_{i} = {\textstyle\frac{1}{4}} U_{i-1} + {\textstyle\frac{1}{2}} U_{i} + {\textstyle\frac{1}{4}} U_{i+1} $$

Since the $\sigma^{u}_{i}$'s are independent and all equal to $\sigma^{u}$,
the noise $\sigma^{h}_{i}$ in the i$^{th}$ hanning tapered channelmap
can be calculated according to

\begin{eqnarray*}
\sigma^{h}_{i} & = & \left[ ( {\textstyle\frac{1}{4}} \sigma^{u}_{i-1} )^{2} + 
                            ( {\textstyle\frac{1}{2}} \sigma^{u}_{i}   )^{2} +
                            ( {\textstyle\frac{1}{4}} \sigma^{u}_{i+1} )^{2} \right]
                            ^{ \frac{1}{2} }                              \\
               & = & \left[ 
                      {\textstyle\frac{1}{16}} + {\textstyle\frac{1}{4}} + {\textstyle\frac{1}{16}} 
                     \right] ^{ \frac{1}{2} } \sigma^{u}                  \\
               & = & {\textstyle\frac{\sqrt{6}}{4}} \; \sigma^{u} = 0.61 \; \sigma^{u}
\end{eqnarray*}

\noindent In this case the noise equivalent bandwidth B$^{h}$ for a
hanning tapered spectrum is given by

 $$ B^{h} = {\textstyle\frac{16}{6}} B^{u} = 2.67 \; B^{u} $$

As a consequence, the noise in two hanning tapered channelmaps may be
correlated depending on their separation.  Two channelmaps separated by
one channelmap are correlated because both contain a quarter of the flux
from the channel map between them.  Only channel maps separated by more
than one channel are independent.  This will be shown in the following
three cases in which two hanning tapered channel maps at different
separations will be added. 

\vspace{0.5cm}   

\begin{enumerate}

\item Adding two adjacent hanning tapered channelmaps i and (i+1) gives
a signal H$_{(i)+(i+1)}$ of

\begin{eqnarray*}
H_{(i)+(i+1)} & = & H_{i} + H_{i+1}                                      \\
              & = & \left(
                  {\textstyle\frac{1}{4}} U_{i-1} + {\textstyle\frac{1}{2}} U_{i} + {\textstyle\frac{1}{4}} U_{i+1}
                  \right)                                                \\
              &   & + \left(
                  {\textstyle\frac{1}{4}} U_{i} + {\textstyle\frac{1}{2}} U_{i+1} + {\textstyle\frac{1}{4}} U_{i+2}
                  \right)                                                \\
              & = &
                  {\textstyle\frac{1}{4}} U_{i-1} + {\textstyle\frac{3}{4}} U_{i} + 
                  {\textstyle\frac{3}{4}} U_{i+1} + {\textstyle\frac{1}{4}} U_{i+2} 
\end{eqnarray*}

and the noise $\sigma^{h}_{(i)+(i+1)}$ in that map will be

\begin{eqnarray*}
\sigma^{h}_{(i)+(i+1)} & = & [ \; ( {\textstyle\frac{1}{4}} \sigma^{u}_{i-1} )^{2} +
                                    ( {\textstyle\frac{3}{4}} \sigma^{u}_{i}   )^{2}   \\
                       &   & + ( {\textstyle\frac{3}{4}} \sigma^{u}_{i+1} )^{2} +
                                    ( {\textstyle\frac{1}{4}} \sigma^{u}_{i+2} )^{2} 
                             \; ]^{ \frac{1}{2} }                      \\
                       & = & {\textstyle\frac{ \sqrt{20} }{4}} \; \sigma^{u} = 
                             {\textstyle\frac{ \sqrt{20} }{4}} 
                             {\textstyle\frac{4}{ \sqrt{6} }} \; \sigma^{h}          \\
                       & = & \sqrt{ 3 {\textstyle\frac{1}{3}} } \; \sigma^{h} = 
                             1.83 \; \sigma^{h}
\end{eqnarray*}

\item Adding the hanning tapered channels i and (i+2) gives

\begin{eqnarray*}
H_{(i)+(i+2)} & = & H_{i} + H_{i+2} \\
        & = & \left(
              {\textstyle\frac{1}{4}} U_{i-1} + {\textstyle\frac{1}{2}} U_{i} + {\textstyle\frac{1}{4}} U_{i+1}
              \right)               \\
        &   & + \left(
              {\textstyle\frac{1}{4}} U_{i+1} + {\textstyle\frac{1}{2}} U_{i+2} + {\textstyle\frac{1}{4}} U_{i+3}
              \right) \\
        & = & {\textstyle\frac{1}{4}} U_{i-1} + {\textstyle\frac{1}{2}} U_{i} + {\textstyle\frac{1}{2}} U_{i+1} \\
        &   & + {\textstyle\frac{1}{2}} U_{i+2} + {\textstyle\frac{1}{4}} U_{i+3}
\end{eqnarray*}

and the noise becomes

\begin{eqnarray*}
\sigma^{h}_{(i)+(i+2)} & = & [ \; ( {\textstyle\frac{1}{4}} \sigma^{u}_{i-1} )^{2} +
                                  ( {\textstyle\frac{1}{2}} \sigma^{u}_{i}   )^{2} +
                                  ( {\textstyle\frac{1}{2}} \sigma^{u}_{i+1} )^{2}  \\
                       &   &  +   ( {\textstyle\frac{1}{2}} \sigma^{u}_{i+2} )^{2} +
                                  ( {\textstyle\frac{1}{4}} \sigma^{u}_{i+3} )^{2} 
                             \; ]^{ \frac{1}{2} }                          \\
                       & = & {\textstyle\frac{ \sqrt{14} }{4}} \; \sigma^{u} =
                             {\textstyle\frac{ \sqrt{14} }{4}}
                             {\textstyle\frac{4}{ \sqrt{6} }} \; \sigma^{h}           \\
                       & = & \sqrt{ 2 {\textstyle\frac{1}{3}} } \; \sigma^{h} = 
                             1.53 \; \sigma^{h}
\end{eqnarray*}

\item Adding the hanning tapered channels i and (i+3) gives

\begin{eqnarray*}
H_{(i)+(i+3)} & = & H_{i} + H_{i+3} \\
        & = & \left(
              {\textstyle\frac{1}{4}} U_{i-1} + {\textstyle\frac{1}{2}} U_{i} + {\textstyle\frac{1}{4}} U_{i+1}
              \right)               \\
        &   & + \left(
              {\textstyle\frac{1}{4}} U_{i+2} + {\textstyle\frac{1}{2}} U_{i+3} + {\textstyle\frac{1}{4}} U_{i+4}
              \right) 
\end{eqnarray*}

with a resulting noise of

\begin{eqnarray*}
\sigma^{h}_{(i)+(i+3)} & = & [ \; ( {\textstyle\frac{1}{4}} \sigma^{u}_{i-1} )^{2} +
                                  ( {\textstyle\frac{1}{2}} \sigma^{u}_{i}   )^{2} +
                                  ( {\textstyle\frac{1}{4}} \sigma^{u}_{i+1} )^{2}   \\
                       &   &    + ( {\textstyle\frac{1}{4}} \sigma^{u}_{i+2} )^{2} +
                                  ( {\textstyle\frac{1}{2}} \sigma^{u}_{i+3} )^{2} +
                                  ( {\textstyle\frac{1}{4}} \sigma^{u}_{i+4} )^{2}
                             \; ]^{ \frac{1}{2} }                        \\
                       & = & {\textstyle\frac{ \sqrt{12} }{4}} \; \sigma^{u} =
                             {\textstyle\frac{ \sqrt{12} }{4}}
                             {\textstyle\frac{4}{ \sqrt{6} }} \; \sigma^{h}            \\
                       & = & \sqrt{2} \; \sigma^{h} = 
                             1.41 \; \sigma^{h}
\end{eqnarray*}

\noindent So, channelmaps i and (i+3) are independent. 

\end{enumerate}

\noindent Because the noise is correlated, adding N {\em adjacent}
hanning tapered channelmaps does not give an increase of the noise with
a factor $\sqrt{N}$ but with a factor $\sqrt{ N - \frac{3}{4} } \cdot
\frac{4}{\sqrt{6}}$ as is shown below.  First the total signal H$_N$ is
calculated. 

\begin{center}
{\scriptsize
\begin{tabular}{c|p{5mm}p{2mm}p{4mm}p{3mm}p{8mm}p{9mm}p{6mm}p{2mm}}
Channel  &   U$_{i-1}$   &   U$_{i}$     &   U$_{i+1}$   & $\cdots$ &  U$_{i+N-2}$  &  U$_{i+N-1}$  &  U$_{i+N}$    &   \\
\hline
i        &\centering 1/4 &\centering 1/2 &\centering 1/4 &          &               &               &               &   \\
i+1      &               &\centering 1/4 &\centering 1/2 & $\cdots$ &               &               &               &   \\
i+2      &               &               &\centering 1/4 & $\cdots$ &               &               &               &   \\
$\cdots$ &               &               &               & $\cdots$ &               &               &               &   \\
i+N-3    &               &               &               & $\cdots$ &\centering 1/4 &               &               &   \\
i+N-2    &               &               &               & $\cdots$ &\centering 1/2 &\centering 1/4 &               &   \\
i+N-1    &               &               &               &          &\centering 1/4 &\centering 1/2 &\centering 1/4 & + \\
\hline
         & $\frac{1}{4}$U$_{i-1}$ & $\frac{3}{4}$U$_i$ & U$_{i+1}$ & $\cdots$  
         & U$_{i+N-2}$ & $\frac{3}{4}$U$_{i+N-1}$ & $\frac{1}{4}$U$_{i+N}$ & \\ 
\end{tabular}}
\end{center}

\noindent and thus

\begin{center}
H$_N$ =  $\frac{1}{4}$U$_{i-1}$ + $\frac{3}{4}$U$_i$ + U$_{i+1}$ + $\cdots$ +
         U$_{i+N-2}$ + $\frac{3}{4}$U$_{i+N-1}$ + $\frac{1}{4}$U$_{i+N}$
\end{center}

\noindent From this it follows that the noise $\sigma^h_N$ is given by

\begin{eqnarray*}
 \sigma^h_N & = & \left[ ({\textstyle\frac{1}{4}})^2 + ({\textstyle\frac{3}{4}})^2 + (N-2) \cdot 1^2 +
                   ({\textstyle\frac{3}{4}})^2 + ({\textstyle\frac{1}{4}})^2 \right]^{ \frac{1}{2} } \sigma^u \\
            & = & \left[ N - {\textstyle\frac{3}{4}} \right]^{ \frac{1}{2} } \; \sigma^u =
                  \left[ N - {\textstyle\frac{3}{4}} \right]^{ \frac{1}{2} } {\textstyle\frac{4}{ \sqrt{6} }} \; \sigma^h \\
            & = & \sqrt{ (N - {\textstyle\frac{3}{4}} ) \mbox{B}^h } \; \sigma^h
\end{eqnarray*}

However, before the hanning tapered channelmaps are added to form a
total HI map, the continuum must be subtracted.  This operation
introduces extra noise in the channelmaps which doesn't behave like a
hanning tapered correlation.  Here, it will be assumed that the average
continuum map is formed by averaging N$_1$ line free channels at the low
velocity end of the datacube and N$_2$ channels at the high velocity end
which gives

$$ C_{low} = \frac{1}{N_1} \sum\limits_{j=1}^{N_1}H_j 
   \mbox{ \hspace{0.6cm} and \hspace{0.6cm} }
   C_{high} = \frac{1}{N_2} \sum\limits_{j=1}^{N_2}H_j  $$

\noindent Since all channels are hanning tapered the noise in these maps
can be calculated according to

$$ \sigma_{C_{low}} = \frac{1}{N_1} \sqrt{ \left( N_1 - {\textstyle\frac{3}{4}} \right) } \; \sigma^u $$

and

$$ \sigma_{C_{high}} = \frac{1}{N_2} \sqrt{ \left( N_2 - {\textstyle\frac{3}{4}} \right) } \; \sigma^u $$

\noindent The average continuum map to be subtracted is then formed by 

$$ <C> = {\textstyle\frac{1}{2}} ( C_{low} + C_{high} ) $$

\noindent Since $\sigma_{C_{low}}$ and $\sigma_{C_{high}}$ are
independent it follows that the noise $\sigma_{<C>}$ in the finally
averaged continuum map is given by

\begin{eqnarray*}
 \sigma_{<C>} & = & {\textstyle\frac{1}{2}} \sqrt{ \sigma^2_{C_{low}} + \sigma^2_{C_{high}} } \\
              & = & \sigma^u \;\; \sqrt{ \left( \frac{ N_1 - \frac{3}{4} }{ 4N^2_1 } + 
                           \frac{ N_2 - \frac{3}{4} }{ 4N^2_2 } \right) } \\
              & \equiv & \sigma^u \; {\cal N}
\end{eqnarray*}

\noindent After subtraction of the continuum the channelmaps only
contain signal from the HI emission line.  The signal in the channelmaps
containing the line emission is now given by

\begin{eqnarray*}
 L_i & = & H_i - <C> \\
     & = & {\textstyle\frac{1}{4}} U_{i-1} + {\textstyle\frac{1}{2}} U_{i} + {\textstyle\frac{1}{4}} U_{i+1} - <C> 
\end{eqnarray*}

\noindent Because $\sigma_{<C>}$ is independent from $\sigma^{u}_{i}$ in
the velocity range which is not used to form the averaged continuum map,
it can be written

\begin{eqnarray*}
 \sigma^l_i & = & \left[ ( {\textstyle\frac{1}{4}} \sigma^{u}_{i-1} )^{2} +
                         ( {\textstyle\frac{1}{2}} \sigma^{u}_{i}   )^{2} +
                         ( {\textstyle\frac{1}{4}} \sigma^{u}_{i+1} )^{2} +
                           \sigma_{<C>}^{2} \right] ^{ \frac{1}{2} } \\
            & = & \left[ {\textstyle\frac{1}{16}} + {\textstyle\frac{4}{16}} + {\textstyle\frac{1}{16}} +
                         {\cal N}^2 \right] ^{ \frac{1}{2} } \; \sigma^{u} \\
            & = & \sqrt{ \left( {\textstyle\frac{3}{8}} + {\cal N}^2 \right) } \; \sigma^{u}
\end{eqnarray*}

\noindent When adding N {\em adjacent} hanning tapered and continuum
subtracted channelmaps containing the line emission, the signal L$_N$
will be

\begin{eqnarray*}
L_N & = & {\textstyle\frac{1}{4}}U_{i-1} + {\textstyle\frac{3}{4}}U_i + U_{i+1} + \cdots \\
    &   & \cdots + U_{i+N-2} + {\textstyle\frac{3}{4}}U_{i+N-1} + {\textstyle\frac{1}{4}}U_{i+N} \\
    &   & - N \cdot <C>
\end{eqnarray*}

The noise $\sigma^l_N$ at each pixel in the final map can be derived
analogous to the calculation of $\sigma^h_N$ and is given by

\begin{eqnarray*}
 \sigma^l_N & = & \left[ (N - {\textstyle\frac{3}{4}} ) + N^2{\cal N}^2 \right]
                  ^{ \frac{1}{2} } \; \sigma^u \\
            & = & B^h \; \sqrt{ \left( \frac{ ( N - \frac{3}{4} ) +
                  N^2{\cal N}^2 }{B^h} \right) } \; \sigma^h  
\end{eqnarray*}

\clearpage
\newpage

\onecolumn

\begin{center}
\section{The HI Atlas}
\end{center}

\begin{figure*}[bh]
  \resizebox{\hsize}{!}{\includegraphics{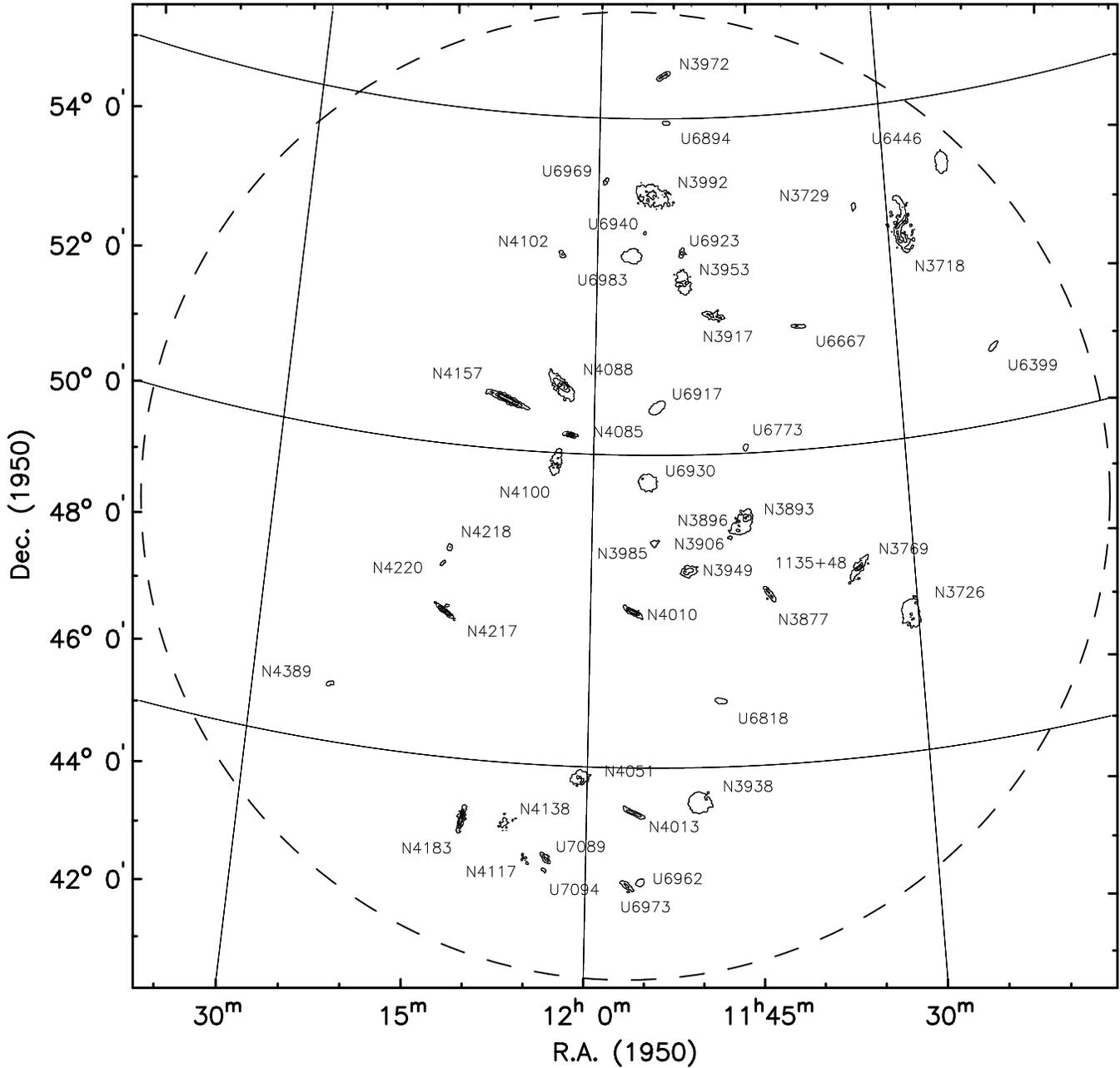}}
  \caption{Integrated HI maps of all galaxies in the cluster more
inclined than 45 degrees and brighter than M(B)=-16.5.  The angular
resolution is 30$\times$30 arcsec.  Contour levels are at column
densities of 0.5, 2.0, 3.5 and 5.0 $\times$10$^{21}$ atoms$\;$cm$^{-2}$.
 Individual galaxies are four times enlarged.  To avoid overlap, some
galaxies are slightly displaced from their actual position.  N3906,
N3938, U6962 and U6930 are more face-on than 45 degrees.  U6940 is
fainter than the complete sample limit.  The dashed circle indicates the
adopted boundary of the cluster.}
  \label{fig7}
\end{figure*}

\begin{figure*}[h]
  \caption{Same as figure 1, displayed here in grayscale.  Note the
interacting pairs N3769/1135+48, N3893/N3896 and U6962/U6973.  The
galaxies N3718, N3726, N4010, N4013, N4088 and N4138 are strongly
warped.  It is clear that the integrated column density strongly depends
on inclination.  Many of the more face-on galaxies show a depletion of
the HI gas in their inner regions.}
  \label{fig8}
\end{figure*}

\twocolumn

\begin{figure*}[h]
  \begin{center}
  \resizebox{!}{\vsize}{\includegraphics{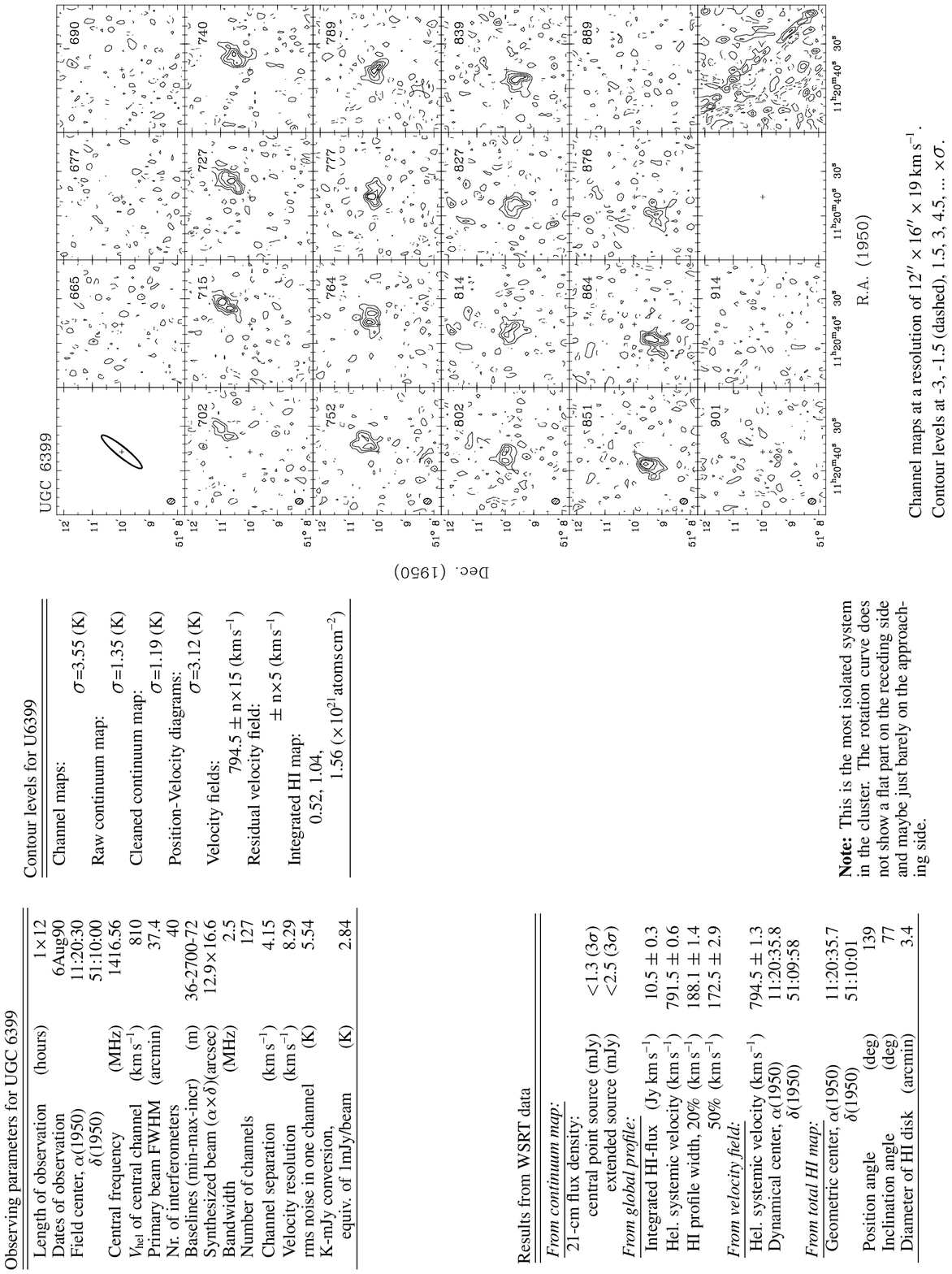}}
  \end{center}   
  \label{figA01a}
\end{figure*}

\begin{figure*}[h]
  \begin{center}
  \resizebox{!}{\vsize}{\includegraphics{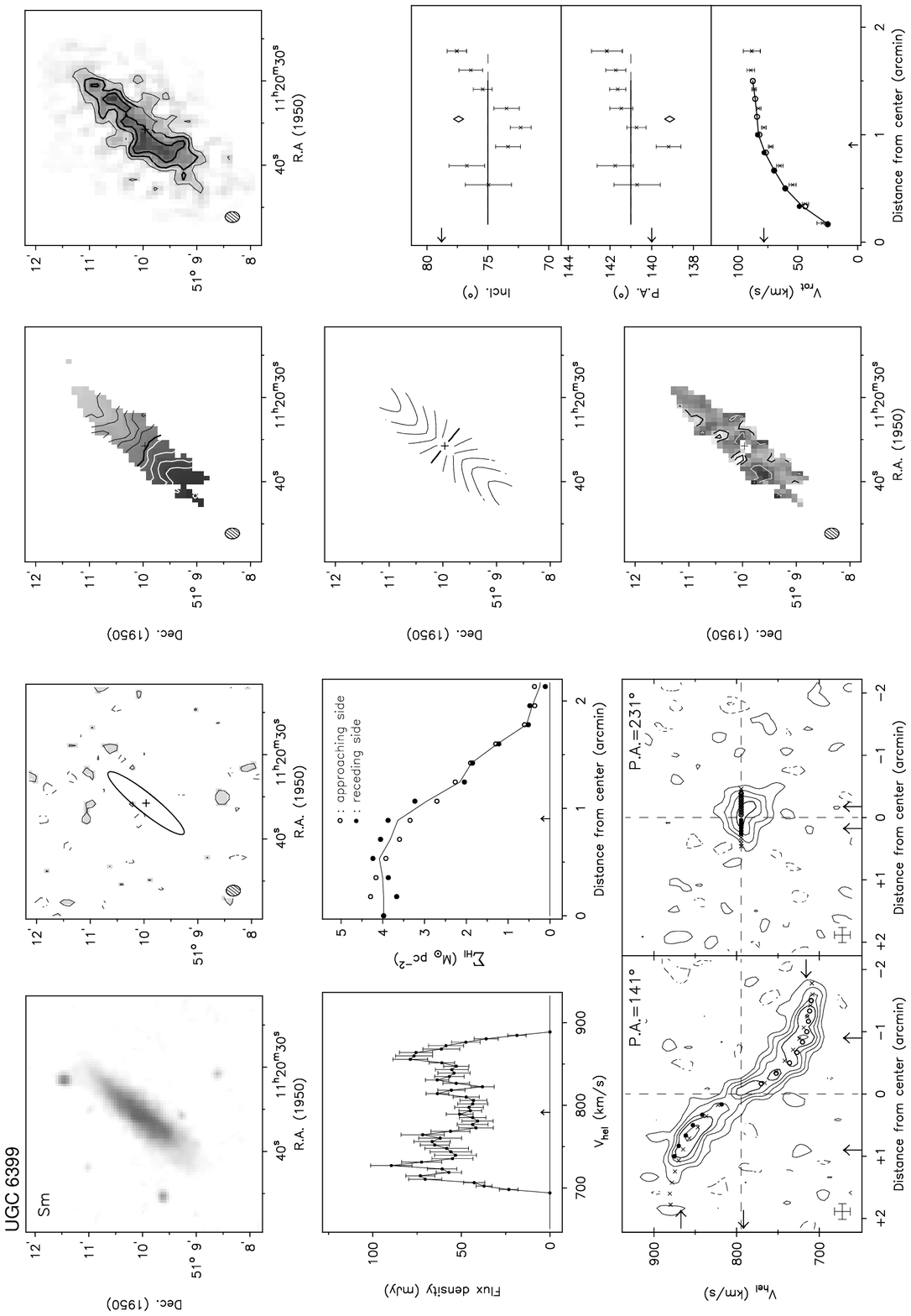}}
  \end{center}
  \label{figA01b}
\end{figure*}

  

  

  

\end{document}